\documentstyle[12pt,dina4]{article}

\def\title #1 {
   \headsep=1.0in
   \baselineskip=30pt
			\begin{center}
   {\titlebold #1}
   \end{center}
			\vskip .75in }
\def\author #1 {
   \baselineskip=30pt
   \begin{center}
   {\timeslarge #1}
   \end{center}
			\vskip .25in }
\def\address #1 {
   \baselineskip=24pt
   \begin{center}
   {\timesitalic #1}
   \end{center}
   \vskip 1.0in }

\def\a {\mbox{$\alpha$}}

\def\L {\mbox{$\Lambda$}}

\def\Tr {\mathop{\rm Tr}}

\def\Im {\mathop{\rm Im}}

\def\Re {\mathop{\rm Re}}

\def\disp {\displaystyle}

\def\conj #1 {\overline #1}

\def\be {\begin{equation}}
\def\ee {\end{equation}}

\def\ba {\begin{array}}
\def\ea {\end{array}}
\def\bea {\begin{eqnarray}}
\def\eea {\end{eqnarray}}

\def\et {$$}

\def\etn {$$}

\def\ett {$$}

\def\ettn{$$}

\def\eqn#1 {(\ref{#1}) }

\newdimen\twoeqncolwidth
\newdimen\twoeqncolwidtha
\newdimen\twoeqncolwidthb
\newdimen\twoeqncolsep
\newdimen\twoeqnlinset
\def\twoeqn#1&#2\et{
   \hbox to\twoeqnlinset{\hfil}
   \hbox to\twoeqncolwidth{$\disp#1$\hfil}
   \hbox to\twoeqncolsep{\hfil}
   \hbox to\twoeqncolwidth{$\disp#2$\hfil}\eqno{\rm (\theequation)}$$}
\def\twoeqnt#1&#2\ett{
   \hbox to\twoeqnlinset{\hfil}
   \hbox to\twoeqncolwidtha{$\disp#1$\hfil}
   \hbox to\twoeqncolsep{\hfil}
   \hbox to\twoeqncolwidthb{$\disp#2$\hfil}\eqno{\rm (\theequation)}$$}
\def\twoeqnn#1&#2\etn{
   \hbox to\twoeqnlinset{\hfil}
   \hbox to\twoeqncolwidth{$\disp#1$\hfil}
   \hbox to\twoeqncolsep{\hfil}
   \hbox to\twoeqncolwidth{$\disp#2$\hfil}\eqno\phantom{\rm
(\theequation)}$$}
\def\twoeqntn#1&#2\ettn{
   \hbox to\twoeqnlinset{\hfil}
   \hbox to\twoeqncolwidtha{$\disp#1$\hfil}
   \hbox to\twoeqncolsep{\hfil}
   \hbox to\twoeqncolwidthb{$\disp#2$\hfil}\eqno\phantom{\rm
(\theequation)}$$}

\def\rawpicture #1 by #2 (#3){
  \vbox to #2{
    \hrule width #1 height 0pt depth 0pt
    \vfill
    \special{picture #3} % this is the low-level interface
    }
  }
\def\scaledpicture #1 by #2 (#3 scaled #4){{
  \dimen0=#1 \dimen1=#2
  \divide\dimen0 by 1000 \multiply\dimen0 by #4
  \divide\dimen1 by 1000 \multiply\dimen1 by #4
  \rawpicture \dimen0 by \dimen1 (#3 scaled #4)}
  }
\def\beginparmode{\endmode
  \begingroup \def\endmode{\par\endgroup}}
\let\endmode=\par
{\obeylines\gdef\
{}}

\def\body			% Begin text body;  can be used to end
  {\beginparmode}		% \title, \author, \affil, \abstract,
\def\head#1{			% Head;  NOTE enclose the text in {}
  \goodbreak\vskip 0.5truein	%  e.g., \head{I. Introduction}
  {\immediate\write16{#1}
   \uppercase{#1}\par}
   \nobreak\vskip 0.25truein\nobreak}

\def\itemj{\par\hang\textindent}
\def\beginitems{
\par\medskip\bgroup\def\i##1 {\itemj{##1}}\def\ii##1 {\itemitem{##1}}
\leftskip=36pt\parskip=0pt}
\def\enditems{\par\egroup}

\def\beneathrel#1\under#2{\mathrel{\mathop{#2}\limits_{#1}}}

\def\refto#1{[#1]}		% For references in text as superscript

\def\references			% Begin references -- basic format is Phys Rev
  {\head{REFERENCES}		% I.e., volume, page, year (space after commas).
   \beginparmode
   \frenchspacing \parindent=0pt \leftskip=1truecm
   \parskip=8pt plus 3pt \everypar{\hangindent=\parindent}}

\gdef\refis#1{\itemj{#1.\ }}			% Ref list numbers.

\gdef\journal#1, #2, #3, 1#4#5#6{		% Journal reference.  Comma sets
    {\sl #1~}{\bf #2} (1#4#5#6), #3 }		% off: name, vol, page, year

\def\annp{\journal Ann. Phys. (N.Y.), }

\def\aspm{\journal Advanced Studies in Pure Mathematics, }

\def\cmp{\journal Comm. Math. Phys., }

\def\eurolett{\journal Europhysics Lett., }

\def\ijmpa{\journal Int. J. Mod. Phys. A, }

\def\ijmpb{\journal Int. J. Mod. Phys. B, }

\def\jappp{\journal J. Appl. Phys., }

\def\jphc{\journal J. Physique C, }

\def\jcr{\journal J. Chem. Res., }

\def\jetp{\journal Sov. Phys. JETP, }

\def\jpj{\journal J. Phys. Soc. Japan, }

\def\jmp{\journal J. Math. Phys., }

\def\jpa{\journal J. Phys. A, }

\def\jpc{\journal J. Phys. C, }

\def\jpcon{\journal J. Phys.: Condens. Matter, }

\def\ptp{\journal Prog. Theor. Phys., }

\def\jetp{\journal Sov. Phys. JETP, }

\def\jsp{\journal J. Stat. Phys., }

\def\lmp{\journal Lett. Math. Phys., }

\def\lnp{\journal Lecture Notes in Physics, }

\def\mpla{\journal Mod. Phys. Lett. A, }

\def\npb{\journal Nucl. Phys. B, }

\def\physica{\journal Physica, }

\def\pla{\journal Phys. Lett. A, }

\def\plb{\journal Phys. Lett. B, }

\def\prep{\journal Physics Reports, }

\def\pra{\journal Phys. Rev. A, }

\def\prb{\journal Phys. Rev. B, }

\def\prl{\journal Phys. Rev. Lett., }

\def\prs{\journal Proc. Roy. Soc. (London) A, }

\def\pr{\journal Phys. Rev., }

\def\rmp{\journal Rev. Mod. Phys., }

\def\sjnp{\journal Sov. J. Nucl. Phys., }

\def\tmp{\journal Theor. Math. Phys., }

\def\zpb{\journal Z. Phys. B, }

\def\zp{\journal Z. Phys., }

\def\reff#1{Ref.~#1}			% 	for inline references
\def\Reff#1{Ref.~#1}			% 	ditto
\def\[#1]{[\refcite{#1}]}
\def\refcite#1{{#1}}
\def\(#1){(\call{#1})}
\def\call#1{{#1}}
\def\taghead#1{}
\def\frac#1#2{{#1 \over #2}}

\def\sla{\raise.15ex\hbox{$/$}\kern-.57em}
\def\leaderfill{\leaders\hbox to 1em{\hss.\hss}\hfill}
\def\twiddle{\lower.9ex\rlap{$\kern-.1em\scriptstyle\sim$}}
\def\bigtwiddle{\lower1.ex\rlap{$\sim$}}
\def\gtwid{\mathrel{\raise.3ex\hbox{$>$\kern-.75em\lower1ex\hbox{$\sim$}}}}
\def\ltwid{\mathrel{\raise.3ex\hbox{$<$\kern-.75em\lower1ex\hbox{$\sim$}}}}
\def\square{\kern1pt\vbox{\hrule height 1.2pt\hbox{\vrule width 1.2pt\hskip
3pt
   \vbox{\vskip 6pt}\hskip 3pt\vrule width 0.6pt}\hrule height
0.6pt}\kern1pt}
\def\tdot#1{\mathord{\mathop{#1}\limits^{\kern2pt\ldots}}}

\def\pmb#1{\setbox0=\hbox{#1}%
  \kern-.025em\copy0\kern-\wd0
  \kern  .05em\copy0\kern-\wd0
  \kern-.025em\raise.0433em\box0 }

\def\refto#1{[#1]}		% For references in text as superscript

\def\references			% Begin references -- basic format is Phys Rev
  {\section*{References}		% I.e., volume, page, year (space after commas).
   \beginparmode
   \frenchspacing \parindent=0pt \leftskip=1truecm
   \parskip=8pt plus 3pt \everypar{\hangindent=\parindent}}

\def\endreferences{\body}

\def\reff#1{Ref.~#1}			% 	for inline references
\def\Reff#1{Ref.~#1}			% 	ditto
\def\[#1]{[\refcite{#1}]}
\def\refcite#1{{#1}}
\def\refeq#1{(\ref{#1})}

\catcode`@=11
\newcount\r@fcount \r@fcount=0
\newcount\r@fcurr
\immediate\newwrite\reffile
\newif\ifr@ffile\r@ffilefalse
\def\w@rnwrite#1{\ifr@ffile\immediate\write\reffile{#1}\fi\message{#1}}

\def\writer@f#1>>{}
\def\referencefile{%			  Stuff to write .REF file
  \r@ffiletrue\immediate\openout\reffile=\jobname.ref%
  \def\writer@f##1>>{\ifr@ffile\immediate\write\reffile%
    {\noexpand\refis{##1} = \csname r@fnum##1\endcsname = %
     \expandafter\expandafter\expandafter\strip@t\expandafter%
     \meaning\csname r@ftext\csname r@fnum##1\endcsname\endcsname}\fi}%
  \def\strip@t##1>>{}}

\def\citeall#1{\xdef#1##1{#1{\noexpand\refcite{##1}}}}
\def\refcite#1{\each@rg\citer@nge{#1}}
\def\each@rg#1#2{{\let\thecsname=#1\expandafter\first@rg#2,\end,}}
\def\first@rg#1,{\thecsname{#1}\apply@rg}	% each@ag is a general purpose
\def\apply@rg#1,{\ifx\end#1\let\next=\relax%	  variable no. of arg. macro.
\else,\thecsname{#1}\let\next=\apply@rg\fi\next}% args separated by commas

\def\citer@nge#1{\citedor@nge#1-\end-}
\def\citer@ngeat#1\end-{#1}
\def\citedor@nge#1-#2-{\ifx\end#2\r@featspace#1 % Single argument
  \else\citel@@p{#1}{#2}\citer@ngeat\fi}	% M-N range of arguments
\def\citel@@p#1#2{\ifnum#1>#2{\errmessage{Reference range #1-#2\space is
bad.}
    \errhelp{If you cite a series of references by the notation M-N, then M
and
    N must be integers, and N must be greater than or equal to M.}}\else%
 {\count0=#1\count1=#2\advance\count1
by1\relax\expandafter\r@fcite\the\count0,%
  \loop\advance\count0 by1\relax%	  Loop from M to N
    \ifnum\count0<\count1,\expandafter\r@fcite\the\count0,%
  \repeat}\fi}

\def\r@featspace#1#2 {\r@fcite#1#2,}	% Eat spaces at beginning or end of arg
\def\r@fcite#1,{\ifuncit@d{#1}%		  Cite individual reference
    \newr@f{#1}%
    \expandafter\gdef\csname r@ftext\number\r@fcount\endcsname%
                     {\message{Reference #1 to be supplied.}%
                      \writer@f#1>>#1 to be supplied.\par}%
 \fi%
 \csname r@fnum#1\endcsname}
\def\ifuncit@d#1{\expandafter\ifx\csname r@fnum#1\endcsname\relax}%
\def\newr@f#1{\global\advance\r@fcount by1%
    \expandafter\xdef\csname r@fnum#1\endcsname{\number\r@fcount}}

\let\r@fis=\refis			% Save old \refis, redefine
\def\refis#1#2#3\par{\ifuncit@d{#1}%      Use two params #2 #3 to strip blank
   \newr@f{#1}%
   \w@rnwrite{Reference #1=\number\r@fcount\space is not cited up to
now.}\fi%
  \expandafter\gdef\csname r@ftext\csname r@fnum#1\endcsname\endcsname%
  {\writer@f#1>>#2#3\par}}

\def\ignoreuncited{%   redefine \refis if ignoring uncited references
   \def\refis##1##2##3\par{\ifuncit@d{##1}%
     \else\expandafter\gdef\csname r@ftext\csname
r@fnum##1\endcsname\endcsname%
     {\writer@f##1>>##2##3\par}\fi}}

\def\r@ferr{\endreferences\errmessage{I was expecting to see
\noexpand\endreferences before now;  I have inserted it here.}}
\let\r@ferences=\references
\def\references{\r@ferences\def\endmode{\r@ferr\par\endgroup}}

\let\endr@ferences=\endreferences
\def\endreferences{\r@fcurr=0%		  Save old \endreferences, redefine
  {\loop\ifnum\r@fcurr<\r@fcount%	  Loop over refnum and produce text
    \advance\r@fcurr by
1\relax\expandafter\r@fis\expandafter{\number\r@fcurr}%
    \csname r@ftext\number\r@fcurr\endcsname%
  \repeat}\gdef\r@ferr{}\endr@ferences}

\let\r@fend=\endpaper\gdef\endpaper{\ifr@ffile
\immediate\write16{Cross References written on []\jobname.REF.}\fi\r@fend}

\catcode`@=12

\citeall\refto		% These macros will generate citations
\citeall\reff		%
\citeall\Reff		%

\ignoreuncited
\renewcommand{\theequation}{\thesection.\arabic{equation}}
\renewcommand{\title}[1]{\large\bf \mbox{}\\ \mbox{}\\ \mbox{}\\ \mbox{}\\
     #1\bigskip\medskip\\}
\renewcommand{\author}[1]{\large #1\\ \smallskip}

\renewcommand{\address}[1]{{\narrower\normalsize\it #1\\}\bigskip}
\renewenvironment{abstract}{\narrower\small}{\par\normalsize\bigskip}

\catcode`\@=11

\font\twelvemsx=msxm10 scaled \magstep1
\font\tenmsx=msxm10
\font\sevenmsx=msxm7

\font\twelvemsy=msym10 scaled \magstep1
\font\tenmsy=msym10
\font\sevenmsy=msym7

\newfam\msxfam
\newfam\msyfam
\textfont\msxfam=\twelvemsx
\scriptfont\msxfam=\tenmsx
  \scriptscriptfont\msxfam=\sevenmsx
\textfont\msyfam=\twelvemsy
\scriptfont\msyfam=\tenmsy
  \scriptscriptfont\msyfam=\sevenmsy

\def\hexnumber@#1{\ifcase#1 0\or1\or2\or3\or4\or5\or6\or7\or8\or9\or
	A\or B\or C\or D\or E\or F\fi }

\font\twelveeuf=eufm10 scaled \magstep1
\font\teneuf=eufm10
\font\seveneuf=eufm7

\newfam\euffam
\textfont\euffam=\twelveeuf
\scriptfont\euffam=\teneuf
\scriptscriptfont\euffam=\seveneuf
\def\frak{\relaxnext@\ifmmode\let\next\frak@\else
 \def\next{\Err@{Use \string\frak\space only in math mode}}\fi\next}
\def\goth{\relaxnext@\ifmmode\let\next\frak@\else
 \def\next{\Err@{Use \string\goth\space only in math mode}}\fi\next}
\def\frak@#1{{\frak@@{#1}}}
\def\frak@@#1{\noaccents@\fam\euffam#1}
\edef\msx@{\hexnumber@\msxfam}
\edef\msy@{\hexnumber@\msyfam}

\mathchardef\boxdot="2\msx@00
\mathchardef\boxplus="2\msx@01
\mathchardef\boxtimes="2\msx@02
\mathchardef\square="0\msx@03
\mathchardef\blacksquare="0\msx@04
\mathchardef\centerdot="2\msx@05
\mathchardef\lozenge="0\msx@06
\mathchardef\blacklozenge="0\msx@07
\mathchardef\circlearrowright="3\msx@08
\mathchardef\circlearrowleft="3\msx@09
\mathchardef\rightleftharpoons="3\msx@0A
\mathchardef\leftrightharpoons="3\msx@0B
\mathchardef\boxminus="2\msx@0C
\mathchardef\Vdash="3\msx@0D
\mathchardef\Vvdash="3\msx@0E
\mathchardef\vDash="3\msx@0F
\mathchardef\twoheadrightarrow="3\msx@10
\mathchardef\twoheadleftarrow="3\msx@11
\mathchardef\leftleftarrows="3\msx@12
\mathchardef\rightrightarrows="3\msx@13
\mathchardef\upuparrows="3\msx@14
\mathchardef\downdownarrows="3\msx@15
\mathchardef\upharpoonright="3\msx@16

\mathchardef\downharpoonright="3\msx@17
\mathchardef\upharpoonleft="3\msx@18
\mathchardef\downharpoonleft="3\msx@19
\mathchardef\rightarrowtail="3\msx@1A
\mathchardef\leftarrowtail="3\msx@1B
\mathchardef\leftrightarrows="3\msx@1C
\mathchardef\rightleftarrows="3\msx@1D
\mathchardef\Lsh="3\msx@1E
\mathchardef\Rsh="3\msx@1F
\mathchardef\rightsquigarrow="3\msx@20
\mathchardef\leftrightsquigarrow="3\msx@21
\mathchardef\looparrowleft="3\msx@22
\mathchardef\looparrowright="3\msx@23
\mathchardef\circeq="3\msx@24
\mathchardef\succsim="3\msx@25
\mathchardef\gtrsim="3\msx@26
\mathchardef\gtrapprox="3\msx@27
\mathchardef\multimap="3\msx@28
\mathchardef\therefore="3\msx@29
\mathchardef\because="3\msx@2A
\mathchardef\doteqdot="3\msx@2B

\mathchardef\triangleq="3\msx@2C
\mathchardef\precsim="3\msx@2D
\mathchardef\lesssim="3\msx@2E
\mathchardef\lessapprox="3\msx@2F
\mathchardef\eqslantless="3\msx@30
\mathchardef\eqslantgtr="3\msx@31
\mathchardef\curlyeqprec="3\msx@32
\mathchardef\curlyeqsucc="3\msx@33
\mathchardef\preccurlyeq="3\msx@34
\mathchardef\leqq="3\msx@35
\mathchardef\leqslant="3\msx@36
\mathchardef\lessgtr="3\msx@37
\mathchardef\backprime="0\msx@38
\mathchardef\risingdotseq="3\msx@3A
\mathchardef\fallingdotseq="3\msx@3B
\mathchardef\succcurlyeq="3\msx@3C
\mathchardef\geqq="3\msx@3D
\mathchardef\geqslant="3\msx@3E
\mathchardef\gtrless="3\msx@3F
\mathchardef\sqsubset="3\msx@40
\mathchardef\sqsupset="3\msx@41
\mathchardef\vartriangleright="3\msx@42
\mathchardef\vartriangleleft="3\msx@43
\mathchardef\trianglerighteq="3\msx@44
\mathchardef\trianglelefteq="3\msx@45
\mathchardef\bigstar="0\msx@46
\mathchardef\between="3\msx@47
\mathchardef\blacktriangledown="0\msx@48
\mathchardef\blacktriangleright="3\msx@49
\mathchardef\blacktriangleleft="3\msx@4A
\mathchardef\vartriangle="0\msx@4D
\mathchardef\blacktriangle="0\msx@4E
\mathchardef\triangledown="0\msx@4F
\mathchardef\eqcirc="3\msx@50
\mathchardef\lesseqgtr="3\msx@51
\mathchardef\gtreqless="3\msx@52
\mathchardef\lesseqqgtr="3\msx@53
\mathchardef\gtreqqless="3\msx@54
\mathchardef\Rrightarrow="3\msx@56
\mathchardef\Lleftarrow="3\msx@57
\mathchardef\veebar="2\msx@59
\mathchardef\barwedge="2\msx@5A
\mathchardef\doublebarwedge="2\msx@5B
\mathchardef\angle="0\msx@5C
\mathchardef\measuredangle="0\msx@5D
\mathchardef\sphericalangle="0\msx@5E
\mathchardef\varpropto="3\msx@5F
\mathchardef\smallsmile="3\msx@60
\mathchardef\smallfrown="3\msx@61
\mathchardef\Subset="3\msx@62
\mathchardef\Supset="3\msx@63
\mathchardef\Cup="2\msx@64

\mathchardef\Cap="2\msx@65

\mathchardef\curlywedge="2\msx@66
\mathchardef\curlyvee="2\msx@67
\mathchardef\leftthreetimes="2\msx@68
\mathchardef\rightthreetimes="2\msx@69
\mathchardef\subseteqq="3\msx@6A
\mathchardef\supseteqq="3\msx@6B
\mathchardef\bumpeq="3\msx@6C
\mathchardef\Bumpeq="3\msx@6D
\mathchardef\lll="3\msx@6E

\mathchardef\ggg="3\msx@6F

\mathchardef\circledS="0\msx@73
\mathchardef\pitchfork="3\msx@74
\mathchardef\dotplus="2\msx@75
\mathchardef\backsim="3\msx@76
\mathchardef\backsimeq="3\msx@77
\mathchardef\complement="0\msx@7B
\mathchardef\intercal="2\msx@7C
\mathchardef\circledcirc="2\msx@7D
\mathchardef\circledast="2\msx@7E
\mathchardef\circleddash="2\msx@7F
\def\ulcorner{\delimiter"4\msx@70\msx@70 }
\def\urcorner{\delimiter"5\msx@71\msx@71 }
\def\llcorner{\delimiter"4\msx@78\msx@78 }
\def\lrcorner{\delimiter"5\msx@79\msx@79 }
\def\yen{\mathhexbox\msx@55 }
\def\checkmark{\mathhexbox\msx@58 }
\def\circledR{\mathhexbox\msx@72 }
\def\maltese{\mathhexbox\msx@7A }
\mathchardef\lvertneqq="3\msy@00
\mathchardef\gvertneqq="3\msy@01
\mathchardef\nleq="3\msy@02
\mathchardef\ngeq="3\msy@03
\mathchardef\nless="3\msy@04
\mathchardef\ngtr="3\msy@05
\mathchardef\nprec="3\msy@06
\mathchardef\nsucc="3\msy@07
\mathchardef\lneqq="3\msy@08
\mathchardef\gneqq="3\msy@09
\mathchardef\nleqslant="3\msy@0A
\mathchardef\ngeqslant="3\msy@0B
\mathchardef\lneq="3\msy@0C
\mathchardef\gneq="3\msy@0D
\mathchardef\npreceq="3\msy@0E
\mathchardef\nsucceq="3\msy@0F
\mathchardef\precnsim="3\msy@10
\mathchardef\succnsim="3\msy@11
\mathchardef\lnsim="3\msy@12
\mathchardef\gnsim="3\msy@13
\mathchardef\nleqq="3\msy@14
\mathchardef\ngeqq="3\msy@15
\mathchardef\precneqq="3\msy@16
\mathchardef\succneqq="3\msy@17
\mathchardef\precnapprox="3\msy@18
\mathchardef\succnapprox="3\msy@19
\mathchardef\lnapprox="3\msy@1A
\mathchardef\gnapprox="3\msy@1B
\mathchardef\nsim="3\msy@1C
\mathchardef\ncong="3\msy@1D

\mathchardef\varsubsetneq="3\msy@20
\mathchardef\varsupsetneq="3\msy@21
\mathchardef\nsubseteqq="3\msy@22
\mathchardef\nsupseteqq="3\msy@23
\mathchardef\subsetneqq="3\msy@24
\mathchardef\supsetneqq="3\msy@25
\mathchardef\varsubsetneqq="3\msy@26
\mathchardef\varsupsetneqq="3\msy@27
\mathchardef\subsetneq="3\msy@28
\mathchardef\supsetneq="3\msy@29
\mathchardef\nsubseteq="3\msy@2A
\mathchardef\nsupseteq="3\msy@2B
\mathchardef\nparallel="3\msy@2C
\mathchardef\nmid="3\msy@2D
\mathchardef\nshortmid="3\msy@2E
\mathchardef\nshortparallel="3\msy@2F
\mathchardef\nvdash="3\msy@30
\mathchardef\nVdash="3\msy@31
\mathchardef\nvDash="3\msy@32
\mathchardef\nVDash="3\msy@33
\mathchardef\ntrianglerighteq="3\msy@34
\mathchardef\ntrianglelefteq="3\msy@35
\mathchardef\ntriangleleft="3\msy@36
\mathchardef\ntriangleright="3\msy@37
\mathchardef\nleftarrow="3\msy@38
\mathchardef\nrightarrow="3\msy@39
\mathchardef\nLeftarrow="3\msy@3A
\mathchardef\nRightarrow="3\msy@3B
\mathchardef\nLeftrightarrow="3\msy@3C
\mathchardef\nleftrightarrow="3\msy@3D
\mathchardef\divideontimes="2\msy@3E
\mathchardef\varnothing="0\msy@3F
\mathchardef\nexists="0\msy@40
\mathchardef\mho="0\msy@66
\mathchardef\eth="0\msy@67
\mathchardef\eqsim="3\msy@68
\mathchardef\beth="0\msy@69
\mathchardef\gimel="0\msy@6A
\mathchardef\daleth="0\msy@6B
\mathchardef\lessdot="3\msy@6C
\mathchardef\gtrdot="3\msy@6D
\mathchardef\ltimes="2\msy@6E
\mathchardef\rtimes="2\msy@6F
\mathchardef\shortmid="3\msy@70
\mathchardef\shortparallel="3\msy@71
\mathchardef\smallsetminus="2\msy@72
\mathchardef\thicksim="3\msy@73
\mathchardef\thickapprox="3\msy@74
\mathchardef\approxeq="3\msy@75
\mathchardef\succapprox="3\msy@76
\mathchardef\precapprox="3\msy@77
\mathchardef\curvearrowleft="3\msy@78
\mathchardef\curvearrowright="3\msy@79
\mathchardef\digamma="0\msy@7A
\mathchardef\varkappa="0\msy@7B
\mathchardef\hslash="0\msy@7D
\mathchardef\hbar="0\msy@7E
\mathchardef\backepsilon="3\msy@7F
\def\Bbb@@#1{\fam\msyfam#1}
\def\Bbb@#1{{\Bbb@@{#1}}}
\def\Bbb{\Bbb@}

\catcode`\@=12

\font\twelvemsx=msxm10 scaled \magstep1
\font\twelvemsy=msym10 scaled \magstep1
\font\twelveeuf=eufm10 scaled \magstep1
\textfont\msxfam=\twelvemsx
\textfont\msyfam=\twelvemsy
\textfont\euffam=\twelveeuf
\def\frak#1{\mbox{\twelveeuf #1}}

\def\a{\frak a}
\def\A{\frak A}

\def\a{{\frak a}}
\def\A{{\frak A}}

\def\e{e}

\def\e{{\rm e}}
\def\L{\Lambda}
\def\a{{\frak a}}
\def\A{{\frak A}}
\def\ba{{\overline{\frak a}}}
\def\bA{{\overline{\frak A}}}
\def\P{\Phi}
\def\lam{\lambda}
\def\dint{\int_{-2K}^{2K}}

\def\lint{\int_{\cal L}}
\def\ksumx{\sum_{{k=-\infty}\atop{k\neq 0}}^{\infty}}
\def\ksum{\sum_{{k=-\infty}}^{\infty}}
\def\ch{{\rm ch\,}}
\def\sh{{\rm sh\,}}
\def\sn{{\rm sn\,}}
\def\cn{{\rm cn\,}}
\def\dn{{\rm dn\,}}
\def\snh{{\rm snh\,}}
\def\cnh{{\rm cnh\,}}
\def\dnh{{\rm dnh\,}}
\def\jprod{\prod_{j=1}^{N/2}}
\def\jlprod{\prod_{j=1}^{L}}
\def\ihalf{{i\over 2}}
\def\T{\mbox{\boldmath $T$}}
\def\Tr{{\rm Tr}}
\def\Nlim{\lim_{N\to\infty}}
\def\max{{\rm max}}

\def\ksumo{\sum_{{k=1}}^{\infty}}
\def\jsum{\sum_{j=1}^{N/2}}
\def\lhs{\hbox{left-hand-side }}
\def\rhs{\hbox{right-hand-side }}
\def\Im{{\rm Im\,}}
\def\Re{{\rm Re\,}}
\def\bc{\overline c}
\def\bx{\overline x}
\def\Re{\hbox{Re\,}}
\def\dify{\hbox{d}y}
\def\difx{\hbox{d}x}
\def\difk{\hbox{d}k}
\def\difp{\hbox{d}p}
\begin{document}
\begin{center}
\titlepage
\title{THERMODYNAMICS OF THE ANISOTROPIC SPIN-1/2 HEISENBERG
CHAIN AND RELATED QUANTUM CHAINS\footnote[1]{
Work performed within the research program of the
Sonderforschungsbereich 341, K\"oln-Aachen-J\"ulich}}
\author{Andreas Kl\"umper}
\address{Institut f\"ur
Theoretische Physik,
Universit\"at zu K\"oln, Z\"ulpicher Str. 77,\\
D-5000 K\"oln 41, Germany.
\footnote[7]{Email: kluemper@thp.uni-koeln.de,
Tel: [0221] 470-4312, FAX: -5159}}
\end{center}
\begin{abstract}
\noindent
The free energy and correlation lengths of the spin-1/2 $XYZ$ chain are studied
at finite temperature. We use the quantum transfer matrix approach and
derive non-linear integral equations for all eigenvalues. Analytic results
are presented for the low-temperature asymptotics, in particular for the
critical $XXZ$ chain in an external
magnetic field. These results are compared to predictions by conformal
field theory. The integral equations are solved numerically for the
non-critical
$XXZ$ chain and the related spin-1 biquadratic chain at arbitrary temperature.

\vspace{.3in}
\noindent
Keywords: integrable quantum chains, thermodynamic Bethe Ansatz,
quantum transfer matrix
\end{abstract}

%\vspace{1in}

\begin{center}
\today

\noindent
to be published in \it Z. Phys. B: Condensed Matter\rm \\
\end{center}

\vspace{.3in}

\bibliographystyle{alpha}

%\newpage

\section{Introduction}

\setcounter{equation}{0}
One of the simplest models of magnetism is the Heisenberg model
describing the exchange interaction of spins. Despite its simplicity,
in general only approximate methods are available for its study.
An exception to this situation is the one-dimensional
spin-1/2 case. Very early Bethe \refto{Bethe31}
constructed the eigenstates for the
isotropic spin-1/2 Heisenberg chain. Much later also the
fully anisotropic spin-1/2 $XYZ$ chain was discovered to be integrable
\refto{Suth70,Baxt71b} as it is related to the exactly solvable eight-vertex
model \refto{Baxt82b}. In this way the spectrum of the $XYZ$ chain and the
ground state correlation lengths are known, cf. \refto{JohnKM73,KlumZ8VM}.
The thermodynamics of this model were studied in
\refto{TakTBA} by an elaborate version of the method used in \refto{Yang69}.
Unfortunately, only the free energy of the $XYZ$ chain could be studied
within this approach, the correlation functions at finite temperature
remained out of reach.

In this paper we apply an alternative method to the thermodynamics
of integrable quantum chains giving the free energy and correlation
lengths. Our treatment will follow somewhat the approach of
\refto{Koma,Tak91}
where the Suzuki-Trotter formula was employed leading to a mapping
of quantum chains at finite temperature to classical two-dimensional
lattice models.
The quantum transfer matrix of the Heisenberg
model was identified as the diagonal-to-diagonal transfer matrix of
the exactly solvable six-vertex model which is identical
to the row-to-row transfer matrix of an inhomogeneous
six-vertex model. The eigenvalues are known in terms of a Bethe
ansatz \refto{Bariev82,TruS83,SuzukiAW90},
the largest one yielding the free energy and the next-leading
ones the correlation lengths.
The study of the limit of infinite Trotter number poses a certain problem.
In \refto{Koma} this limit
was taken numerically by extrapolation of the eigenvalues, in
\refto{SuzukiAW90,SuzukiNW92} it was done analytically for low temperatures.
In \refto{Tak91} the limit of infinite Trotter number was
taken analytically at the Bethe ansatz level
for any finite temperature. The derived equations are not of the
integral equation type as in \refto{TakTBA}. However, they are very useful for
numerical calculations. Unfortunately, only the
zero temperature limit of them could be studied analytically \refto{Tak91}.
This is the main motivation for us to choose another approach to the
study of the quantum transfer matrix.

In section 2 we review the two-dimensional eight-vertex model and its
relation to the spin-1/2 $XYZ$ chain. The quantum transfer matrix is
introduced by a direct mapping of the $XYZ$ chain at finite temperature
to an inhomogeneous eight-vertex model.
In section 3 the eigenvalue
problem for the quantum transfer matrix is investigated. The Bethe ansatz
equations are analyzed within the 'non-linear integral equation' approach
\refto{KlumB90,KlumBP91,PearK91,Klum92}.
This allows for a very simple treatment of the
infinite Trotter number limit. For the partially isotropic $XXZ$ chains,
also the thermodynamics in an external magnetic field can be studied.
As a result we obtain integral equations for the largest eigenvalue and
the next-leading eigenvalues of the quantum transfer matrix. In section 4
certain limiting cases are investigated analytically, in particular the
low-temperature limit. The critical $XXZ$ chain in a magnetic field
is studied in section 5.
The low-temperature asymptotics of the specific heat and correlation lengths
are derived and compared to predictions by conformal field theory.
The non-critical $XXZ$ chain and the related
spin-1 purely biquadratic chain are investigated analytically as well
as numerically in section 6. Several technical details and calculations
are deferred to three appendices.

%%%%%%%%%%%%%%%%%%%%%%%%%%%%%%%%%%%%%
%quanten transfer
%%%%%%%%%%%%%%%%%%%%%%%%%%%%%%%%%%%%%

\section{The quantum transfer matrix of the $XYZ$ chain}

We first consider the symmetric eight-vertex model on a square lattice
with periodic boundary conditions. Each bond of the lattice is occupied
by an arrow which is pointing either up/down or right/left,
respectively. The only allowed arrow configurations are such that there is
an even number of arrows pointing into (and out of) each vertex of the lattice.
There are eight-vertex configurations with associated Boltzmann weights
$a$, $b$, $c$ and $d$, cf. Fig. 1. This model is exactly solvable
for all Boltzmann weights \refto{Baxt82b}
which are conveniently parametrized by elliptic
functions
\bea
a &=& -i \rho \Theta(i \lambda)H\biggl[{1 \over 2}i( \lambda-v)\biggr]
\Theta\biggl[{1 \over 2}i( \lambda+v)\biggr],\cr
b &=& -i \rho \Theta(i \lambda) \Theta\biggl[{1 \over 2}i(
\lambda-v)\biggr]H\biggl[{1 \over 2}i( \lambda+v)\biggr],\cr
c &=& -i \rho H(i \lambda) \Theta\biggl[{1 \over 2}i( \lambda-v)\biggr]
\Theta\biggl[{1 \over 2}i( \lambda+v)\biggr],\cr
d &=& +i \rho H(i \lambda)H\biggl[{1 \over 2}i( \lambda-v)\biggr]H\biggl[{1
\over 2}i( \lambda+v)\biggr],
\label{weights}\eea
with the elliptic modulus $k$, the crossing parameter $\lam$ and the
spectral variable $v$ as new parameters. The theta functions $H$, $\Theta$,
and the quarter-periods $K$, $K'$ are defined in appendix A.
The 'antiferroelectric' regime of the eight-vertex model in the new parameters
is given by the restriction
\be
\quad 0 < k < 1,\quad 0 < \lambda < K',\quad -\lambda < v < \lambda.
\ee
For convenience we set the trivial scale of the problem to the value
\be
\rho={1\over h(\lam)},
\ee
where the function $h$ is defined in appendix A. The Boltzmann weights
\refeq{weights} possess the socalled 'crossing' symmetry. A rotation of the
vertices by $90^o$ leads to an interchange of configurations 1, 2 with 3, 4
thus amounting to an interchange of $a$ and $b$ or a substitution of $v$
by $-v$. This relation is depicted graphically in Fig. 2 and will play
an important role below.

We next consider the row-to-row transfer matrix $\T(v)$ as a function of
the spectral parameter $v$, holding $k$ and $\lam$ fixed. $\T(v)$ is a
family of commuting matrices \refto{Baxt82b}
with two special points $v=\pm\lam$ at which
$\T$ reduces to left and right shift operators which are exponentials of the
momentum operator $P$. At these points the Hamiltonian limit of the model
can be taken by the logarithmic derivative of the transfer matrix
\be
H_{XYZ} = \hbox{ const. } \pm (\ln\T) '  (\pm\lambda),\label{HamLim}
\ee
leading to the $XYZ$ chain describing the exchange interaction
of 1/2 spins
\be
H_{XYZ} = -{1 \over 2} \sum_{j=1}^L {(J_X \sigma_j^X \sigma_{j+1}^X
+J_Y \sigma_j^Y \sigma_{j+1}^Y+J_Z \sigma_j^Z \sigma_{j+1}^Z)}.
\ee
$\sigma^{X,Y,Z}$ are the Pauli spin matrices and we have cyclic boundary
conditions $\sigma_{L+1} = \sigma_1$. The interaction coefficients are
functions of $k$ and $\lam$
\be
J_X={1+k\,\snh^2\lambda \over 2 \snh  \lambda},\qquad
{{J_Y}} =  {{1 - k\,\snh^2 \lambda} \over 2 \snh  \lambda},\qquad
{{J_Z}} = - {{\cnh  \lambda   \dnh  \lambda} \over 2 \snh  \lambda},
\label{InterCoeff}
\ee
where $\snh$, $\cnh$, $\dnh$ are related to standard elliptic functions, cf.
appendix A.
In place of \refeq{HamLim} we shall utilize more directly the relation
\be
\T(\mp(\lam-v))=\e^{\hbox{$\pm i P-v H+O(v^2)$}},
\ee
with $H=H_{XYZ}$ (up to an additive constant)
from which we immediately conclude
\be
\left[\T\left(-\lam+{\beta\over N}\right)
      \T\left(\lam-{\beta\over N}\right)\right]^{N/2}=\e^{\hbox{
$-\beta H+O(1/N)$}}.
\label{Trotter}
\ee
This equation relates a product of certain row-to-row transfer matrices
to the exponential $\exp(-\beta H)$ where $\beta$ is the inverse of the
temperature $T$ and $N$ is a large even 'Trotter' number. The partition
function of the Hamiltonian can now be expressed in terms of the transfer
matrices
\be
Z=\Nlim\Tr\left[\T\left(-\lam+{\beta\over N}\right)
      \T\left(\lam-{\beta\over N}\right)\right]^{N/2}.
\ee
So, the partition function of the quantum chain at finite temperature is
given by the partition
function of an inhomogeneous eight-vertex model with alternating rows, cf.
Fig. 3.
For the calculation of this partition function the column-to-column transfer
matrix is best adapted. This 'vertical' transfer matrix (or quantum transfer
matrix) is nothing but a row-to-row transfer matrix of an inhomogeneous
eight-vertex model with alternating spectral parameters $\pm(\lam-\beta/N)$.
This is easily understood from the crossing symmetry, cf. Fig. 2.
Thus we obtain the representation
\be
Z=\Nlim\Tr\,\T^L\left(\lam-{\beta\over N},-\lam+{\beta\over N}\right),
\ee
where $\T(.,.)$ denotes the inhomogeneous transfer matrix with length $N$
whose eigenvalues satisfy Bethe ansatz like equations \refto{Baxt82b}.
The free energy per site is
$f=\lim_{L\to\infty}F/L=-1/\beta\lim_{L\to\infty}\ln Z/L$
where first the limit $N\to\infty$ has to be taken and then
${L\to\infty}$. We are allowed to interchange these limits due to
the theorems in \refto{Suzuki85,SuzukiI87}. Standard reasoning then yields
\be
f=-{1\over\beta}\Nlim\ln\L_\max,\label{freeEn}
\ee
where $\L_\max$ denotes the largest eigenvalue of the inhomogeneous transfer
matrix $\T(.,.)$. All other eigenvalues of $\T(.,.)$ are separated by a gap
which is finite for finite temperature even in the limit $N\to\infty$. This
is an important difference to and an
advantage over the transfer matrices on the
\lhs of \refeq{Trotter}. The next-leading eigenvalues give the
exponential correlation lengths\footnote[1]{The
derivation is standard involving a spectral representation and the observation
that correlation functions of operators acting on vertical bonds are identical
to correlations of operators on horizontal bonds (for spectral parameter
$v\to\pm\lam$).} of the equal time correlators at finite
temperature
\be
{1\over\xi}=-\Nlim\ln\left|{\L\over\L_\max}\right|.\label{length}
\ee

Lastly we want to comment on the study of thermodynamics of the quantum
chain in the presence of an external magnetic field $h$ coupling to the
spin $S=\sum_{j=1}^LS_j$, where $S_j$ denotes a certain component of the
$j$th spin for instance $S_j^z$. Of course this changes \refeq{Trotter}
only trivially
\be
\left[\T\left(-\lam+{\beta\over N}\right)
      \T\left(\lam-{\beta\over N}\right)\right]^{N/2}
\cdot\e^{\beta h S}
=\e^{-\beta (H-h S)+O(1/N)}.
\ee
On the lattice,
the equivalent two-dimensional model is modified in a simple way
by a horizontal seam. Each vertical bond of this seam carries an individual
Boltzmann weight $\e^{\pm\beta h/2}$ if $S_j=\pm1/2$ which indeed describes the
action of the operator
\be
\e^{\beta h S}=\jlprod\e^{\beta h S_j}.
\ee
Consequently, the vertical transfer matrix is modified by an $h$ dependent
boundary condition. It will turn out that these modifications can still be
treated exactly if the additional operators acting on the bonds belong
to symmetries of the model. Therefore, a magnetic field cannot
be studied unless limiting cases of the $XYZ$ chain are investigated such as
the $XXZ$ chain.

%%%%%%%%%%%%%%%%%%%%%%%%%%%%%%%%%%%%%
%nicht-lineare integralgl (nicht kritisch)
%%%%%%%%%%%%%%%%%%%%%%%%%%%%%%%%%%%%%

\section{Bethe ansatz equations and non-linear integral equations}

\setcounter{equation}{0}

In \refto{Baxt82b} the treatment of the eigenvalue problem for
inhomogeneous transfer matrices of the
eight-vertex model was described. We content ourselves with quoting the
relevant
formulae. Each eigenvalue of $\T(.,.)$ is given by a function $\L(x)$ at $x=0$,
where $\L(x)$ satisfies the equation
\be
\L(x)q(x)=s\left[{1\over\omega}\P(x-i\lam)q(x+i2\lam)+
\omega\P(x+i\lam)q(x-i2\lam)\right].\label{eigen}
\ee
$\P(x)$ is a 'known' function with definition
\be
\P(x)=\P_1(x)\P_2(x),\qquad\P_{1/2}(x)=
\left[\rho h\left(\ihalf(x\mp i x_0)\right)\right]^{N/2},
\ee
where $h(x)$ is defined in appendix A, $\omega$ is a constant, $s=\pm 1$,
and $x_0$ involves the temperature via
\be
x_0=\lam-{\beta\over N}.
\ee
The function $q(x)$ is 'unknown' with functional form
\be
q(x)=\exp(-i\tau x)\jprod h\left(\ihalf(x-y_j)\right),\label{qfunc}
\ee
and zeros $y_j$ henceforth called Bethe ansatz numbers which have to
be determined under the condition that $\L(x)$ is analytic for all $x$
in the complex plane. This imposes that each zero $y_j$ of the
\lhs of \refeq{eigen} is also a zero of its \rhs
thus yielding the well-known Bethe ansatz equations.

For the $XYZ$ chain with zero magnetic field we have $\omega=1$ and
a sum rule
\be
\sum_{j=1}^{N/2} {y_j} = (1-rs+N)K +{i \over 2}(s-1+N)K' +4pK+ i2p'K',
\ee
where $s=\pm 1$ and $r=\pm 1$ are quantum numbers of the system and
$p$, $p'$ some integers related to $\tau$
\be
\tau = {\pi \over 8K}(s-1+N+4p').
\ee

For the critical and non-critical $XXZ$ models the corresponding limiting
cases of modulus $k\to 1$ and $k\to 0$ describe the problem. In any case
$\tau=0$, but there may be less than $N/2$ Bethe ansatz numbers such that
the product on the \rhs of \refeq{qfunc} runs over less than $N/2$ factors.
Now an external magnetic field $h$ may be introduced leading to a non-trivial
$\omega$ namely
\be
\omega=\e^{\beta h /2}.\label{omeg}
\ee

A compact way for writing down the Bethe ansatz equations is $p(y_j)=-1$
where $p(x)$ is the ratio of the two summands on the \rhs of \refeq{eigen}
\be
p(x)={1\over\omega^2}{\P(x-i\lam)q(x+i2\lam)\over\P(x+i\lam)q(x-i2\lam)}.
\ee
Instead of dealing with coupled non-linear equations for the Bethe ansatz
numbers we work with \refeq{eigen} on the basis of functional equations
in analogy to the approach in \refto{KlumBP91}. Shortly, the usefulness of the
following functions will become obvious
\bea
\a(x)={1\over p(x-i\lam)},\qquad \A(x)=1+\a(x),\cr
\ba(x)={p(x+i\lam)},\qquad \bA(x)=1+\ba(x),\label{SubCond}
\eea
where $\a$, $\ba$ etc. denote independent functions which are related
by complex conjugation actually only if ${\overline\omega}=1/\omega$.
Quite generally $\a(x)$, $\ba(x)$ are very small for
real $x$ in the limit of large values of $N$. The functions $\A(x)$, $\bA(x)$
in turn are close to $1$. In terms of these functions $\L(x)$ can be written
in two ways
\bea
\L(x-i\lam)&=&{s\over\omega}\P(x-i2\lam){q(x+i\lam)\over q(x-i\lam)}\A(x),\cr
\L(x+i\lam)&=&{s\omega}\P(x+i2\lam){q(x-i\lam)\over q(x+i\lam)}\bA(x).
\label{diffRep}
\eea
Now we recall that all eigenvalues $\L(x)$ are analytic in the whole
complex plane. For a while we focus our study on the two largest eigenvalues
with only real Bethe ansatz numbers. Each of these functions $\L(x)$
is non-zero in a strip $-2\lam<\Im x <2\lam$ where the logarithm
defines an analytic function. The Fourier transform of $\ln\L(x)$
can therefore be calculated in two different ways
by integrals with integration paths along $\Im x=-\lam$
or $\Im x=\lam$ using the two different representations \refeq{diffRep}
of $\L$ in terms of the functions $q$, $\A$ and $\bA$. Of course both
calculations yield the same result due to Cauchy's theorem
thus imposing a non-trivial relation
on the Fourier transforms of $\ln q(x)$, $\ln\A(x)$ and $\ln\bA(x)$.
In other words, the Fourier transform of $\ln q(x)$ is given by the transforms
of $\ln\A(x)$ and $\ln\bA(x)$. Next we observe that the transform of $\ln\a(x)$
is given in terms of the transforms of $\ln q(x)$ or, alternatively by
$\ln\A(x)$, $\ln\bA(x)$. (In fact this reasoning should be applied to
the derivatives of all the mentioned functions, because
$\ln q(x)$ does not admit a Fourier transform in contrast to $[\ln q(x)]'$.
For more details particularly concerning the regimes of validity of the
Fourier transforms see appendix B or the exposition of similar
calculations in \refto{KlumBP91,KlumWZ93}.)
To the final equation for the Fourier transform of $\ln\a(x)$ the
inverse transform can be applied resulting in the non-linear integral
equation
\bea
\ln\a(x)&=&
{N\over 2}\ln\left[k_1\sn{K_1\over 2K}(x-x_0)\sn{K_1\over 2K}(x+x_0)\right]\cr
&&+\dint k(x-y)\ln\A(y)\dify-\dint k(x-y-i2\lam+i\epsilon)\ln\bA(y)\dify\cr
&&+D,\label{IntEq}
\eea
where $\epsilon$ is an infinitesimally small number.
The function $\sn$ is the standard elliptic sn function (cf. appendix A),
here however to a new modulus $k_1$ which is defined by the requirement
that the ratio of the corresponding quarterperiods $K_1$, $K_1'$ is
\be
{K_1'\over K_1}={\lam\over K}.\label{k1}
\ee
The kernel of the integral equation $k(x)$ (not to be confused with the
modulus) is defined by the series
\be
k(x)={1\over 4K}\left[{1\over 2}+\sum_{k=1}^\infty{\sh(k(K'-2\lam))\ch(i k
x)\over
\ch(k\lam)\sh(k(K'-\lam))}\right],\label{kernel}
\ee
where we introduced the short-hand notations
\be
\ch(x)=\cosh{\pi\over 2K}x,\qquad\sh(x)=\sinh{\pi\over 2K}x.
\ee
The remaining integration constant $D$ has the value
\be
D=\ln r\omega -2\tau\lambda,
\ee
where $r=\pm 1$. For $\ba(x)$ we obtain an equation very similar to
\refeq{IntEq}. Both equations close since $\a$ and $\ba$ are related to
$\A$ and $\bA$ by \refeq{SubCond}. The integral equation \refeq{IntEq}
is valid for the $XYZ$ model with repulsive ($0<\lam<K'/2$) and attractive
($K'/2<\lam<K'$) interaction. In the latter case the integration paths of
the first and second integral have to be shifted into the upper and lower
half plane by $\pm i(2\lam-K')$. For this situation an alternative set of
equations is more suitable, cf. appendix C.

Lastly we want to find an equation for $\L(x)$ in terms of $\A(x)$ and
$\bA(x)$. For this purpose we may use any of the two equations \refeq{diffRep}
as we can calculate $\ln q(x)$ from $\A(x)$ and $\bA(x)$. However, it is more
convenient to multiply both equations \refeq{diffRep} with the result
\be
\L(x-i\lam)\L(x+i\lam)=\P(x-i2\lam)\P(x+i2\lam)\A(x)\bA(x),
\ee
where the \rhs does not involve $q(x)$ anymore. This equation is the
socalled 'inversion identity' for finite $N$. Taking the
logarithm, it is solved immediately by Fourier transforms
\bea
\ln\L(x)&=&\ln r+{N}\ksumo{\ch(k\lam)-\ch(ik x)\ch(k x_0)\over
k\,\ch(k \lam)\sh(k K')}\ch(k(K'-2\lam))\cr
&&+\dint c(x-y)\ln[\A\bA(y)]\dify,\label{LambEx}
\eea
where $c(x)$ is defined by
\be
c(x)={1\over 4K}\left[{1\over 2}+\ksumo{\ch(ik x)\over \ch(k\lam)}\right].
\label{Defc}
\ee
This function is related to the $\sn$ function with modulus $k_1$ \refeq{Repc1}
and other elliptic functions \refeq{Repc2}.

It remains to study the limit of infinite Trotter number $N\to\infty$ in
\refeq{IntEq}, \refeq{LambEx}. Using \refeq{funcsn} and \refeq{Repc1}
we first obtain
\bea
\ln\a(x)&=&-{2\pi\beta}c(x)+\ln r\omega\cr
&&+\dint k(x-y)\ln\A(y)\dify-\dint k(x-y-i2\lam+i\epsilon)\ln\bA(y)\dify,\cr
&&\label{IntEqT}
\eea
where we have used $\tau=0$ for the two largest eigenvalues. The analogous
equation for $\ba$ is obtained from \refeq{IntEqT} by the replacement
of $\a$, $\A$ by $\ba$, $\bA$, and $i$, $\ln r\omega$ by $-i$, $-\ln r\omega$,
respectively.
{}From \refeq{LambEx} we find (as we are only interested in $\L=\L(0)$
we set $x=0$)
\be
\ln\L=\ln r+{\pi\beta\over 2K}\ksumo{\sh(k\lam)\ch(k(K'-2\lam))\over
\ch(k\lam)\sh(k K')}+\dint c(x)\ln[\A\bA(x)]\difx.\label{LambExT}
\ee
Equations \refeq{IntEqT} and \refeq{LambExT} determine the two largest
eigenvalues of the quantum transfer matrix corresponding to $r=\pm 1$.
The leading eigenvalue ($r=+1$) yields the free energy according to
\refeq{freeEn}
\be
\beta f=\beta e_0-\dint c(x)\ln[\A\bA(x)]\difx,\label{ResultFree}
\ee
where the ground state energy $e_0$ is given by the series in
\refeq{LambExT}.
The next-leading eigenvalue ($r=-1$) determines the length \refeq{length}
of the leading correlation function. Numerical solutions to these equations
compare well to numerical results in \refto{Tak91}. However, the mathematical
equations therein are not related in an obvious way to our ones.

All other solutions to \refeq{eigen} are obtained by allowing for complex
Bethe ansatz numbers. For the next-largest eigenvalues of the quantum
transfer matrix most of the Bethe ansatz
numbers are real apart from one number on the horizontal line
$\Im x=K'$ (1-string) or two
complex conjugate numbers at an approximate distance of $2\lam$ (2-string)
\refto{KlumZ8VM}. The main modification of the properties of the eigenvalue
function $\L(x)$ is the occurrence of two zeros $\theta_1$, $\theta_2$ in
the strip $-2\lam<\Im x <2\lam$.
For these and also more general situations \refeq{IntEqT}
and \refeq{LambExT} are still valid, however with a modified integration path
$\cal L$ encircling certain singularities of the involved functions
\refto{KlumWZ93}.

For a 1-string consisting of a Bethe ansatz number $y_0$ ($\Im y_0=K'$) the
path
$\cal L$ follows the real axis with loops in the upper half plane encircling
$\theta_1+i\lam$, $\theta_2+i\lam$ (simple zeros of $\A$) and
$y_0-i\lam$ (simple pole/zero of $\A/\bA$) clockwise. The loops can be
removed using Cauchy's theorem. Technically this is achieved with the
differentiated version of \refeq{IntEqT} (with integration path
$\cal L$). This ensures that the integrands $[\ln\A]'$, $[\ln\bA]'$ possess
simple poles at the before mentioned points whereas $\ln\A$, $\ln\bA$ have
non-isolated singularities. The application of Cauchy's theorem produces
contributions by the kernel $k(x)$. Integrating again we obtain
\bea
\ln\a(x)&=&-{2\pi\beta}c(x)+\ln r\omega-{\pi\lam\over 2K}\cr
&&-K(x-(\theta_1+i\lam))-K(x-(\theta_2+i\lam))
+\ln{h_2(x-(y_0+i\lam))\over h_2(x-(y_0-i\lam))}\cr
&&+\dint k(x-y)\ln\A(y)\dify-\dint k(x-y-i2\lam+i\epsilon)\ln\bA(y)\dify,\cr
&&\label{IntEq1}
\eea
where
\be
K(x)={\pi i\over 4K}x
+\sum_{k=1}^\infty{\sh(k(K'-2\lam))\sh(i k x)\over
k\ch(k\lam)\sh(k(K'-\lam))},
\ee
satisfying $K'(x)={2\pi i}\,k(x)$, and we have used $\tau=\pi/4K$.
The function $h_2$ is defined by
\be
h_2(x)=h\left(i{K_2\over 2K}x,k_2\right),\label{defh2}
\ee
in terms of the function $h$ (cf. Appendix A), now however with an elliptic
modulus $k_2$ with quarterperiods
\be
{K_2'\over K_2}={K'-\lam\over K}.\label{k2}
\ee
The function $h_2$ enters \refeq{IntEq1} because of an application of
\refeq{funcEqk1}. The integral equation \refeq{IntEq1}
has to be solved under the subsidiary conditions
$\a(\theta_{1,2}+i\lam)=\a(y_{0}+i\lam)=-1$.

In order to derive the counterpart to \refeq{LambExT} we have to go back to
\refeq{LambEx} (with integration path $\cal L$). Taking the derivative,
removing the loops of $\cal L$, collecting the contributions due to the
residues at $\theta_1+i\lam$, $\theta_2+i\lam$, integrating again we find
\bea
\ln\L_1(x)&=&\ln\L_{\rm b}(x)+\ln (-r)+
\ln\left[k_1\sn{K_1\over 2K}(x-\theta_1)
\sn{K_1\over 2K}(x-\theta_2)\right]\cr
&&+\dint c(x-y)\ln[\A\bA(y)]\dify,\label{LambEx1}
\eea
where $\ln\L_{\rm b}$ denotes the 'bulk' term in \refeq{LambEx}.
Taking now $x=0$ and $N\to\infty$ we get
\be
\ln\L_1=-\beta e_0+\ln r+
\ln\left[k_1\sn{K_1\over 2K}\theta_1\sn{K_1\over 2K}\theta_2\right]
+\dint c(x)\ln[\A\bA(x)]\difx.
\ee

Quite similarly we proceed with a 2-string consisting of two complex Bethe
ansatz numbers $y_\pm$ with $\Im y_\pm$ close to $\pm\lam$.
Here $\cal L$ follows the real axis with loops encircling
$\theta_1+i\lam$, $\theta_2+i\lam$ (simple zeros of $\A$),
$y_+-i\lam$ (simple pole/zero of $\A/\bA$) clockwise, and
$y_-+i\lam$ (simple zero/pole of $\A/\bA$) anticlockwise. The loops
around $\theta_1+i\lam$, $\theta_2+i\lam$ and $y_+-i\lam$
are removed. Integration leads to
\bea
\ln\a(x)&=&-{2\pi\beta}c(x)+\ln (-r\omega)\cr
&&-K(x-(\theta_1+i\lam))-K(x-(\theta_2+i\lam))
+\ln{h_2(x-(y_++i\lam))\over h_2(x-(y_+-i\lam))}\cr
&&+\lint k(x-y)\ln\A(y)\dify-\lint k(x-y-i2\lam+i\epsilon)\ln\bA(y)\dify,
\label{IntEq2}
\eea
where $\cal L$ is an integration path along the real axis below
$y_\pm\mp i\lam$, and we have used $\tau=0$.
This integral equation has to be solved
under the subsidiary conditions
$\a(\theta_{1,2}+i\lam)=\a(y_{\pm}+i\lam)=-1$.
The corresponding eigenvalue of the quantum transfer matrix
is
\be
\ln\L_2=-\beta e_0+\ln r+
\ln\left[k_1\sn{K_1\over 2K}\theta_1\sn{K_1\over 2K}\theta_2\right]
+\lint c(x)\ln[\A\bA(x)]\difx.\label{LambEx2}
\ee
%

%%%%%%%%%%%%%%%%%%%%%%%%%%%%%%%%%%%%%
%Grenzfaelle
%%%%%%%%%%%%%%%%%%%%%%%%%%%%%%%%%%%%%
\section{Limiting cases}

\setcounter{equation}{0}

In this section we study several cases of the $XYZ$ chain which can be
treated analytically.
First we consider the $XY$ chain which is equivalent
to free fermions. Second we investigate the low- and high-temperature
behaviour of the general $XYZ$ chain.

\begin{itemize}
\item[a)] The $XY$ chain
\end{itemize}

In this case the crossing parameter is given by the decoupling condition
$\lam=K'/2$ where $J_Y=0$ \refeq{InterCoeff}. For this value \refeq{IntEqT},
\refeq{IntEq1}
and \refeq{IntEq2} simplify considerably as they are no integral equations
anymore ($k(x)\equiv$ const.), but explicit expressions for the function $\a$
\be
\ln\a(x)=-{2\pi\beta}c(x)+\ln rs.\label{1order}
\ee
The two largest eigenvalues of the quantum transfer matrix are then exactly
given by
\be
\ln\L=-\beta e_0+\ln r+\dint c(x)\ln[\A\bA(x)]\difx.
\ee
Observing that the elementary excitation energy $\epsilon(x)$ and momentum
$p(x)$ (as functions of the rapidity $x$) are related to $c(x)$ by
\be
\epsilon(x)={\hbox{d}\over \difx}p(x)={2\pi}c(x),
\ee
we find the familiar expression
\be
\beta f=\beta e_0-{1\over\pi}\int_{-\pi/2}^{\pi/2}\difp
\ln[1+\e^{-\beta\epsilon(p)}].\label{freeEnAs}
\ee
The correlation length of the $\langle \sigma^z \sigma^z \rangle$ correlation
function is
\be
{1\over\xi_z}={1\over\pi}\int_{-\pi/2}^{\pi/2}\difp
\ln\left[{1+\e^{-\beta\epsilon(p)}\over 1-\e^{-\beta\epsilon(p)}}\right].
\ee
Of course, also the results for 1- and 2-string excitations
\refeq{LambEx1} and \refeq{LambEx2} are exact with \refeq{1order} (under the
subsidiary condition $\a(\theta_{1,2}+i\lam)=-1$). These eigenvalues
yield the correlation lengths of the $\langle \sigma^x \sigma^x \rangle$
and $\langle \sigma^y \sigma^y \rangle$ functions.

\begin{itemize}
\item[b)] Low and high-temperature asymptotics
\end{itemize}

Next we study the general $XYZ$ chain at low temperatures. The integral
equation \refeq{IntEqT} does not admit an analytic solution, however an
iterative procedure is conceivable. For such an approch \refeq{1order}
takes into account just the lowest order in temperature. In this approximation
the asymptotics of the free energy is given by an equation similar to
\refeq{freeEnAs}. Performing
a saddle point integration we find explicitly
\be
\beta f=\beta e_0-\e^{-\beta\epsilon_0}[B_1T^{1/2}+B_2T^{3/2}+O(T^{5/2})].
\label{AsympF}
\ee
where
\be
\epsilon_0={K_1\over 2K}(1-k_1),\qquad
B_1=\left({K\over\pi K_1}{1-k_1\over k_1}\right)^{1/2},\qquad
B_2={1-k_1^3\over 4\pi^{1/2}}\left(K\over K_1k_1(1-k_1)\right)^{3/2}.
\ee
Within the same approximation we obtain the correlation length
of the $\langle \sigma^z \sigma^z \rangle$ function
\be
{1\over\xi_z}=2\e^{-\beta\epsilon_0}[B_1T^{1/2}+B_2T^{3/2}+O(T^{5/2})].
\label{AsympXi}
\ee
In addition to the terms in \refeq{AsympF} and \refeq{AsympXi} there will be
$O(\e^{-2\beta\epsilon_0})$ contributions from the first iteration of
the integral equations. This is not studied here. However, for the $XYZ$ chain
with attractive interaction ($K'/2<\lam<K'$) higher order iterations may lead
to contributions which become dominant for sufficiently strong attraction.
For $\Im x\simeq 2\lam-K'$ we find
\be
\ln\a(x)=-{2\pi\beta}c(x)+\exp\left(-{2\pi\beta}
c[x-i2(2\lam-K')]\right)+...\ .
\ee
Within this accuracy we get
\bea
\beta f&=&\beta e_0-2\dint c(x)\exp\left(-{2\pi\beta}c(x)\right)
-\dint c_b(x)\exp\left(-{2\pi\beta}c_b(x)\right)\cr
&=&-{1\over\pi}\int_{-\pi/2}^{\pi/2}\difp\e^{-\beta\epsilon(p)}
-{1\over 2\pi}\int_{-\pi}^{\pi}\difp\e^{-\beta\epsilon_b(p)},\label{AsympFAttr}
\eea
with
\bea
c_b(x)&=&c(x-i(2\lam-K'))+c(x+i(2\lam-K')),\cr
\epsilon_b(x)&=&{\hbox{d}\over\difx}p_b(x)={2\pi}c_b(x),
\eea
where $\epsilon_b(p)$ is the energy-momentum dispersion of the lowest
bound states \refto{KlumZ8VM}. The result of \refeq{AsympFAttr} is physically
very intuitive. For certain interaction parameters the gap of
the bound states may be lower than that of the free states thus becoming
the thermodynamically relevant one, cf. also \refto{TakTBA,Tak91}.

Without derivation we communicate our result
for the low-temperature asymptotics
of the correlation length of the $\langle \sigma^x \sigma^x \rangle$
and $\langle \sigma^y \sigma^y \rangle$ functions
\be
{1\over\xi_{x,y}}=-\ln k_1+\left({2\pi K\over K_1k_1'}\right)^2T^2+O(T^3).
\ee
The second term in the asymptotic expansion of the
$\langle \sigma^z \sigma^z \rangle$ function decays exponentially with length
$\xi_z'$
\be
{1\over\xi_{z}'}=-\ln k_1+2\left({2\pi K\over K_1k_1'}\right)^2T^2+O(T^3).
\ee
Of course, these results are only valid in the repulsive regime
$0<\lam<K'/2$. In the
attractive regime the lengths are dominated by bound states.
For the general $XYZ$ chain ($\epsilon_0>0$) the length $\xi_z$ diverges at
$T=0$ with an essential singularity while all other lengths $\xi_{x,y}$,
$\xi_z'$ remain finite. This implies long-range order at zero temperature.

We conclude this section with a comment on the high-temperature asymptotics
of the free energy. For small values of $\beta$ (and $r=+1$, $\omega=1$)
the function
$\a(x)$ becomes independent of $x$ and \refeq{IntEqT} turns into a simple
algebraic equation. The
result is $\a=\ba=1$. The integral in \refeq{ResultFree} can be done,
yielding
\be
f\sim -T \ln 2,
\ee
with a high-temperature entropy $\ln 2$ as it should be for a model
with two states per site.

%%%%%%%%%%%%%%%%%%%%%%%%%%%%%%%%%%%%%
%kritischer Fall + tieftemp
%%%%%%%%%%%%%%%%%%%%%%%%%%%%%%%%%%%%%
%
\section{The critical $XXZ$ chain}

\setcounter{equation}{0}

Next we study the critical limit of the $XYZ$ chain with modulus $k\to 1$
(entailing $K\to\infty$, $K'\to\pi/2$) where \refeq{InterCoeff} reads
\be
J_X={1\over \sin\gamma},\qquad
{{J_Y}} =  {\cos\gamma \over \sin\gamma},\qquad
{{J_Z}} = - {1\over \sin\gamma},
\label{InterCoeffxxz}
\ee
with $\gamma=2\lam\in(0,\pi)$. As we are allowed to change any two signs of the
interaction coefficients due to a unitary transformation \refto{Baxt82b}
we see that the limiting case $k\to 1$ corresponds
to the $XXZ$ chain with equal $J_X$ and $J_Z$ coefficients and a coefficient
$J_Y$ which is smaller in absolute value. The limiting cases of $\gamma\to 0$
and $\pi$ yield the antiferromagnetic and ferromagnetic Heisenberg chain,
respectively. For $0<\gamma<\pi/2$ the excitations are due to free states only,
for $\pi/2<\gamma<\pi$ there are also bound states. The dispersion relation
of the free states is
\be
\epsilon(k)={\pi\over\gamma} \left|\sin k\right|,
\ee
with an obvious sound velocity of
\be
v={\pi\over\gamma}.
\ee
At zero temperature the model is critical, all correlation lengths diverge
like $\xi\sim 1/T$ which will be shown below.

The $k\to 1$ limit of \refeq{IntEqT} and \refeq{LambExT} reads
\bea
\ln \frak a(x)
&=& -{\pi\beta\over\gamma}{1\over\cosh{\pi \over \gamma}x}
+{\pi\over\pi-\gamma}\ln\omega \nonumber\\
&&+\int_{-\infty}^\infty\left[k(x-y)\ln \frak A(y)
-k(x-y-i\gamma+i\epsilon)\ln\overline {\frak A}(y)\right]{\rm d}y,
\label{IntEqxxz}
\eea
and
\be
\L=-\beta e_0+{1\over 2\gamma}
\int_{-\infty}^\infty{\ln[\A\bA(x)]\over\cosh{\pi \over \gamma}x}\difx.
\label{eigenxxz}
\ee
where
\be
k(x)\to{1\over
2\pi}\int_{-\infty}^{\infty}{\sinh\left({\pi\over 2}-\gamma\right)k\cos(kx)
\over2\cosh{\gamma\over 2}k\sinh{\pi-\gamma\over 2}k}\difk,
\ee
and we have used
\be
c(x)\to {1\over 2\gamma\cosh{\pi\over\gamma}x}.
\ee
One of the subtleties in the derivation of \refeq{IntEqxxz} is the
'renormalization' of the additive constant due to the contribution of the
constant term in $k(x)$ \refeq{kernel} in the limit of $K\to\infty$.
This can be checked directly by considering the asymptotic behaviour of
the two sides of \refeq{IntEqxxz} for $x\to\infty$ ($\ln\a(\infty)=2\ln\omega$
etc.). It appears that the critical limit of \refeq{IntEqT} for $r=-1$ becomes
inconsistent. The missing eigenvalue is described in terms of (pure)
hole excitations
with $N/2-1$ Bethe ansatz numbers, $N$ being the Trotter number.

In general the excited states are described by equations identical to
\refeq{IntEqxxz}, however, with more complicated integration paths $\cal L$.
Proceeding as in section 3 or simply taking the limit of
\refeq{IntEq1}, \refeq{LambEx1} we find for 1-string excitations
\bea
\ln \frak a(x)
&=& -{\pi\beta\over\gamma}{1\over\cosh{\pi \over \gamma}x}
+\pi i+{\pi\over\pi-\gamma}\ln\omega \nonumber\\
&&-K(x-(\theta_1+i\lam))-K(x-(\theta_2+i\lam))
+\ln{\sinh{\pi\over\pi-\gamma}(x-(y_0+i\gamma/2))
\over\sinh{\pi\over\pi-\gamma}(x-(y_0-i\gamma/2))}\nonumber\\
&&+\int_{\cal L}\left[k(x-y)\ln \frak A(y)
-k(x-y-i\gamma+i\epsilon)\ln\overline {\frak A}(y)\right]{\rm d}y,
\eea
and
\be
\L=-\beta e_0+
\ln\left[\tanh{\pi\over 2\gamma}\theta_1\tanh{\pi\over 2\gamma}\theta_2\right]
+{1\over 2\gamma}
\int_{\cal L}{\ln[\A\bA(x)]\over\cosh{\pi \over \gamma}x}\difx,
\ee
where
\be
K(x)\to
i\int_{-\infty}^{\infty}{\sinh\left({\pi\over 2}-\gamma\right)k\sin(kx)\over
2k\cosh{\gamma\over 2}k\sinh{\pi-\gamma\over 2}k}\difk.
\ee
For 2-string excitations the equations are identical upon the replacement of
$y_0$ by $y_+$. Excitations corresponding to $N/2-1$ Bethe ansatz numbers
are described by similar equations where the $y_0$ dependent term is
replaced by the additive constant $i{\pi-2\gamma\over\pi-\gamma}\gamma$.
The equations for the largest eigenvalue for the critical $XXZ$ chain in
zero magnetic field have also been derived in \refto{Klum92} and
\refto{DestdeVeW92} using the non-linear integral equation approach to
the study of Bethe ansatz equations.

We do not dwell any longer on these integral equations as we aim now at
an analysis of {\it all} eigenvalues in the low-temperature limit using
an approach with reference to \refto{KlumWZ93}. There the {\it finite-size}
corrections to all eigenvalues of the six-vertex model transfer matrix were
calculated analytically. The integral equations in \refto{KlumWZ93} are
identical to \refeq{IntEqxxz} and those for the excitations provided we
exchange the 'inhomogeneities'
\be
-{\pi\beta\over\gamma}{1\over\cosh{\pi \over \gamma}x}
\longleftrightarrow
N\ln\tanh\left({\pi x\over 2\gamma}\right),\label{inhomog}
\ee
where now $N$ denotes the system size of the (homogeneous) six-vertex model
in \refto{KlumWZ93}.
The functions \refeq{inhomog} have the same asymptotic behaviour
if we identify
\be
v\beta=N.\label{corres}
\ee
In \refto{KlumWZ93} the $1/N$ corrections to the eigenvalues were found just
from the asymptotic behaviour of the function on the \rhs of \refeq{inhomog}.
This means that the $1/\beta (=T)$ contributions to the eigenvalues of the
quantum transfer matrix can be obtained from \refto{KlumWZ93} via the
correspondence \refeq{corres}. The result is
\be
\ln\L=-\beta e_0-{2\pi\over v} T (\bx-\bc/12)+o(T^2)+i P_0,\label{result}
\ee
with
\be
\bc=1+{3\beta^2h^2\over 2\pi(\pi-\gamma)},\label{effc}
\ee
and
\be
\bx={1-\gamma/\pi\over 2}S^2+{1\over 2(1-\gamma/\pi)}m^2+
i{m\over\pi-\gamma}{h\over 2T},
\ee
where the magnetic field entered via \refeq{omeg}, $S$, $m$ are integers,
and $P_0=(S-m)\pi$ is a lattice momentum.
A condition for
the validity of \refeq{result} is the finiteness of $\omega$, i.e.
$h=O(T)$.

Next we discuss the physical consequences of \refeq{result}.
Applying \refeq{freeEn} to the largest eigenvalue ($\bx=0$) we calculate
the free energy density at low temperatures and magnetic fields
\be
f(T,h)=e_0-{\pi\over 6v}\left(T^2+{3h^2\over 2\pi(\pi-\gamma)}\right),
\ee
which is consistent with the scaling prediction $f(T)=f(0)-(\pi c/6v)T^2$
derived from conformal invariance \refto{Affl86}.
The central charge $c$ (not to be confused
with the related quantity \refeq{effc}) of the underlying field theory is 1.
The low-temperature behaviour of the zero-field susceptibility is
\be
\chi={1\over 2v(\pi-\gamma)}.
\ee
The asymptotic behaviour of the correlation functions is determined by the
next-largest eigenvalues \refeq{length} with non-zero $\bx$
\be
C_R\sim \left({\L\over\L_\max}\right)^R\sim
\cos (P_0R) \e^{\hbox{$ -{2 \pi\over v}\bx TR$}}.\label{asympCorr}
\ee
The 2-point function for spin components in axial direction is described
by $S=0$, $m=\pm 1$. For planar spin components we have the selection rule
$S=1$, $m=0$.
For non-zero magnetic field the quantity $\bx$ is complex if the quantum number
$m\neq 0$. The dominant behaviour of physical correlation functions is
determined by a pair $\pm m$, such that
\be
C_R\sim
\cos \left(m{\pi\over \pi-\gamma}{h\over v}R\right) \e^{\hbox{$
-{2 \pi\over v}\Re(\bx)TR$}},
\ee
where we have ignored for the moment the alternating sign for a lattice
moment $P_0=\pi$. The exponential decay of the correlation functions
can be obtained from conformal invariance \refto{Card84a,BogK89}
\be
C_R\sim\left({\Lambda\over\Lambda_{\rm max}}\right)^R\sim
\e^{\hbox{$ -{2\pi\over v}x T R$}},
\label{corr}
\ee
but not the prefactor whose oscillations in general
are incommensurate with the lattice. Obviously, the scaling dimensions $x$ of
the underlying field theory are given by $x=\Re(\bx)$ in accordance with
finite-size calculations, see for instance \refto{KlumWZ93} and references
therein.

%%%%%%%%%%%%%%%%%%%%%%%%%%%%%%%%%%%%%
%biquadratisches Modell und Numerik
%%%%%%%%%%%%%%%%%%%%%%%%%%%%%%%%%%%%%
\section{The non-critical $XXZ$ chain}

\setcounter{equation}{0}

Lastly we study the modulus $k\to 0$ limit of the $XYZ$ chain
where $K\to\pi/2$, $K'\to\infty$, and \refeq{InterCoeff} reads
\be
J_X={1\over 2\sinh\lam},\qquad
{{J_Y}} =  {1\over 2\sinh\lam},\qquad
{{J_Z}} = - {\cosh\lam\over 2\sinh\lam},
\label{Inter0}
\ee
where $\lam$ is any positive real number. In this case again two interaction
coefficients are equal. The third one is larger in absolute value.
This limit of the $XYZ$ chain is the non-critical antiferromagnetic
$XXZ$ chain. The ferromagnetic $XXZ$ chain can also be obtained in the
limit $k\to 0$, however with fixed $K'-\lam$, cf. \refeq{IntRel}.

The thermodynamics of this model are given by \refeq{IntEqT}, \refeq{LambExT}
for the antiferromagnetic regime, and by \refeq{Eqferro}, \refeq{Lamferro}
for the ferromagnetic case
where the limit $k\to 0$ can be performed in a straightforward way. There are
two largest eigenvalues of the quantum transfer matrix which are exponentially
close in the low-temperature limit. From these eigenvalues the free energy
and the leading correlation length $\xi$ can be calculated. Upon approaching
zero temperature the length $\xi$ diverges heralding the onset of
long-range order. All other lengths remain finite.
In the limit of the one-dimensional classical Ising model ($\lam\to\infty$)
the integral equations turn into simple algebraic equations from which the
well-known results are recovered easily.

In Fig. 4a the specific heat and $\xi$ are shown for anisotropy
$\cosh\lam=3/2$, with a convenient normalization such that
$J_X=J_Y=1$ and $J_Z=-3/2$. In Fig. 4b these curves are given for the
same Hamiltonian with ferromagnetic sign. Note the divergence of the
curves at $T=0$ with an essential singularity $\exp(\pm\Delta/T)$ where
$\Delta={1\over 2}\cdot 0.173...$ and $\Delta=1$ in Fig. 4a and b,
respectively.
The quantities $0.173...$ and $1$ are the energy gaps of the excitations for
the model with antiferromagnetic and ferromagnetic sign.

The antiferromagnetic $XXZ$ chain of the previous paragraph is related
to the socalled spin-1 biquadratic chain
\be
H_{\hbox{biq}}= -\sum_{j=1}^N (\vec S_j \vec S_{j +1})^2,\label{HamBiq}
\ee
where $\vec S_j$ denotes the vector of spin-1 operators. The relation was
first observed in \refto{ParkBiq}. The most systematic way to establish this
employs the Temperley-Lieb algebra \refto{BarbB89}
which is satisfied by the biquadratic
chain and the previous $XXZ$ chain. The ground state energies and gaps
coincide. For a calculation of the singlet-singlet
correlation length at $T=0$ see \refto{KlumBiq}.

Here we want to describe how the thermodynamics of the biquadratic chain
are tackled. The problem of the socalled Temperley-Lieb equivalence is the
unknown degeneracy of each energy level of the biquadratic chain, otherwise
its spectrum is identical to that of the $XXZ$ chain
(for certain boundary conditions). We therefore
choose a different approach using the classical two-dimensional counterpart
to the biquadratic chain \refto{KlumBiq}. This model is an exactly
solvable $3$-state vertex model. We repeat the reasoning of section 2,
derive the corresponding quantum transfer matrix which is an inhomogeneous
$3$-state vertex transfer matrix. At this point we make use of the
Temperley-Lieb correspondence. The two largest eigenvalues of the quantum
transfer matrix of the biquadratic chain are equal to the two largest
eigenvalues of the transfer matrix of the $XXZ$ chain
\be
\L_{\rm biq}(T)=\L_{XXZ}(T,\omega),
\ee
provided that an external magnetic field is applied to the $XXZ$ chain such
that $\omega+1/\omega=3$. (More generally, the quantum transfer matrix
eigenvalues for the biquadratic chain in a field $h_b$ and the $XXZ$
chain in a field $h$ coincide if $1+2\cosh{h_b/ T}=2\cosh{h/ 2T}$.)
The proof of this exact relation will be published elsewhere. In Fig. 5
the specific heat and the leading correlation length is shown. Note that
the divergence at low temperature is described by the same essential
singularity as in the a.f. $XXZ$ case. The values of the specific heat
in Fig. 5 are generally larger than those in Fig. 4a. This is due to
the larger high-temperature entropy of the biquadratic chain. In this
limit the solution to \refeq{IntEqT} is $\a=\omega^2$ and $\ba=1/\omega^2$.
The integral in \refeq{ResultFree} can be done, yielding
\be
f\sim -T \ln(\omega+1/\omega)=-T\ln 3,
\ee
with a high-temperature entropy $\ln 3$ due to the three states per site
for the biquadratic model. The curve for the specific heat shown in Fig. 5
compares quite well with the results of \refto{PaczP90}. A discrepancy appears
at low temperatures since the biquadratic model was treated by numerical
diagonalization of finite chains in \refto{PaczP90}.

The thermodynamics for the biquadratic chain with opposite sign can also
be treated, just by introducing a negative temperature for \refeq{HamBiq}.
Then equations \refeq{Eqferro}, \refeq{Lamferro} have to be solved. We do
not show numerical results for this case, because the computations at low
temperature are difficult as it appears that
the integration paths have to be deformed.
However, the low-temperature limit can be studied analytically. The result
for the two largest eigenvalues is quite interesting
\be
\L_\max=-\beta e_0+\ln\omega,
\qquad\L=-\beta e_0-\ln\omega,
\ee
which implies a residual entropy $S_0=\ln\omega=0.962...$ and a
finite correlation length $\xi=1/(2\ln\omega)=0.519...$,
with the explicit value $\omega=(3+\sqrt{5})/2$.
The value of the non-zero $S_0$ was derived earlier \refto{Muell}. The
result for the leading correlation length is novel, its finiteness
implies absence of long-range order at $T=0$.

\section*{Acknowledgments}

I am grateful to A. Schadschneider and T. Wehner for useful discussions.

%%%%%%%%%%%%%%%%%%%%%%%%%%%%%%%%%%%%%
%Appendix A
%%%%%%%%%%%%%%%%%%%%%%%%%%%%%%%%%%%%%

\renewcommand{\theequation}{A.\arabic{equation}}

\section*{Appendix~A}

\setcounter{equation}{0}

In this appendix we list the definitions of the Jacobian elliptic functions
and related quantities. Elliptic functions are complex, meromorphic and
doubly periodic functions. Here we denote the argument by $u$. Another
parameter which determines the periods is the nome $q$ with $0<q<1$. The
quarter-periods $K$, $K'$, the modulus k and the conjugate modulus $k'$ are
defined by
\bea
K&=&{1 \over 2}\pi \prod_{n=1}^\infty {\biggl({1+q^{2n-1} \over 1-q^{2n-1}}
\cdot
{1-q^{2n} \over 1+q^{2n}}\biggr)^2},\cr
K'&=&\pi^{-1} K\, \ln(q^{-1}),\cr
k&=&4 q^{1 \over 2} \prod_{n=1}^\infty {\biggl({1+q^{2n} \over
1+q^{2n-1}}\biggr)^4},\cr
k'&=& \prod_{n=1}^\infty {\biggl({1-q^{2n-1} \over 1+q^{2n-1}}\biggr)^4},
\eea
such that $k^2+k'^2=1$ and $q=\exp(-\pi K'/K)$.

The Jacobian elliptic functions $\sn$, $\cn$, $\dn$ with periods $4K,4iK'$,
and the related $\snh$, $\cnh$, $\dnh$ functions are
\be
\vcenter{\openup1\jot
\halign {$\hfil\displaystyle #$ &$ {\displaystyle #} \hfil $ &$ \quad
{\displaystyle #} \hfil$ &$ {\displaystyle #} \hfil$\cr
\sn u &= k^{-{1 \over 2}}{H(u) \over \Theta (u)}, &\snh u &= -i \,\sn iu ,\cr
\cn u &= \left({k' \over k}\right)^{{1 \over 2}}{H_1(u) \over \Theta (u)},
&\cnh u &= \cn iu ,\cr
\dn u &= k'^{{1 \over 2}}{\Theta_1 (u) \over \Theta (u)}, &\dnh u &= \dn iu
,\cr}}
\ee
where the theta functions $H(u)$, $\Theta(u)$ are defined by
\bea
H(u) &=& 2 q^{1 \over 4} \sin {\pi u \over 2K} \prod_{n=1}^\infty
{\left(1-2 q^{2n} \cos {\pi u \over K}+q^{4n}\right)(1-q^{2n})},\cr
\Theta(u) &=&  \prod_{n=1}^\infty
{\left(1-2 q^{2n-1} \cos {\pi u \over K}+q^{4n-2}\right)(1-q^{2n})},\cr
H_1(u) &=& H(u+K),\cr
\Theta_1(u) &=& \Theta(u+K).
\eea
These functions satisfy a bewildering variety of identities of which
we cite only
\bea
H(u+iK')&=&iq^{-1/4}\exp\left(-{\pi i u\over 2K}\right)\Theta(u),\cr
\Theta(u+iK')&=&iq^{-1/4}\exp\left(-{\pi i u\over 2K}\right)H(u),
\eea
and
\be
\sn(u+iK')={1\over k\, \sn u}.\label{funcsn}
\ee
A more comprehensive
summary can be found in \refto{AbramS64} and \refto{Baxt82b}.

In the main body of the paper a function $h(v)$ is used which is defined by
\be
h(v)=-i\Theta(0)H(iv)\Theta(iv),
\ee
enjoying the properties
\be
h(v+i2K)=-h(v),\qquad h(v+K')=-q^{-1/2}\exp(\pi v/K)h(v).
\ee

\renewcommand{\theequation}{B.\arabic{equation}}

\section*{Appendix~B}

\setcounter{equation}{0}

Here we list the Fourier transforms of some functions appearing
in section 3. The main strategy of the solution procedure in
section 3 is to formulate functional equations for certain functions
and to solve them by Fourier transforms. All functions which we have
to deal with are $4K$ periodic. We therefore introduce the Fourier transform
and its inverse
\bea
f(x)&=&\ksum f_k\, \e^{i{\pi\over 2K}kx},\cr
f_k&=&{1\over 4K}\dint f(x)\, \e^{-i{\pi\over 2K}kx}\difx.\label{Fourier}
\eea
As we have to study the logarithms of periodic functions $g(x)$, it may happen
that $\ln g(x)$ is not periodic, but acquires a phase shift
$\ln g(x+4K)=\ln g(x)+$ const. In that case we apply \refeq{Fourier} to
the derivative of $\ln g(x)$ or introduce a linear term in \refeq{Fourier}.
For the functions $h$ and $\sn$ we have the explicit formulae
\be
\ln h\left({i\over 2}x\right)=\ln\gamma+{\pi i\over 4K}x
+\ksumx {\e^{i{\pi\over 2K}kx}\over k(1-q^{-k})},
\ee
with
\be
\gamma=q^{1/4}\Theta(0)\jprod(1-q^{2j})^2,
\ee
and
\be
\ln(ik^{1/2}\sn \,x)={1\over 4}\ln q+i{\pi x\over 2K}
+\ksumx{\e^{i{\pi\over K}kx}\over k(1+q^{-k})}.
\ee

In section 3 we take the Fourier transforms of equations relating different
functions at different values of the argument. One has to be aware that
a Fourier representation is convergent only in certain strips of the
complex plane. For the largest eigenvalues for instance,
i.e. real Bethe ansatz numbers,
$\ln q(x)$ is analytic in each strip $n2K'<\Im x<(n+1)2K'$, but singular
on the boundaries. We take
the strip $-2K'<\Im x<0$ as the 'fundamental' regime where we define
the Fourier transform. All expressions of
$\ln q(x)$ with arguments $x$ in other strips are related to the 'fundamental'
regime by (quasi) periodicity. In this way we find the appropriate versions
of \refeq{diffRep} and the appropriate expression for $\a(x)$
\bea
\L(x-i\lam)&=&{rs\over\omega}q^{N/2}\exp\left({N\pi\over 4K}(3\lam-x_0)\right)
\P_1(x-i2\lam)\P_2(x+i(2K'-2\lam))\cr
&&{q(x+i(\lam-2K'))\over q(x-i\lam)}\A(x),\cr
\L(x+i\lam)&=&{s\omega}\exp\left({N\pi\over 4K}(\lam-x_0)\right)
\P_1(x+i(2\lam-2K'))\P_2(x+i2\lam)\cr
&&{q(x-i\lam)\over q(x+i(\lam-2K'))}\bA(x),\cr
\a(x)&=&r\omega^2 q^{-N/2}\exp\left({N\pi\over 4K}(x_0-3\lam)\right)
{\P_1(x)\over\P_1(x-i2\lam)}
{\P_2(x)\over\P_2(x+i(2K'-2\lam))}\cr
&&{q(x-i3\lam)\over q(x+i(\lam-2K'))}.
\eea

The kernel \refeq{kernel} satisfies two important functional equations
\be
k(x)+k(x-i2\lam)={i\over 2\pi}
\left[\ln{h\left(i{K_2\over 2K}x,k_2\right)\over
       h\left(i{K_2\over 2K}(x-i2\lam),k_2\right)}\right]',\label{funcEqk1}
\ee
and
\be
k(x)-k(x-i2(K'-\lam))={i\over 2\pi}
\left\{\ln\left[\sn\left({K_1\over 2K}x,k_1\right)
\sn\left({K_1\over 2K}(x-i2(K'-\lam)),k_1\right)\right]\right\}',
\ee
which may serve for an analytic continuation of $k(x)$ outside the strip
of convergence of \refeq{kernel}.

Lastly we give two expressions of the function $c(x)$ \refeq{Defc} in terms
of standard elliptic functions
\bea
c(x)&=&-i{K_1\over 4\pi K}[\ln\sn]'\left({K_1\over 2K}x-i{K_1'\over 2}\right),
\label{Repc1}\\
c(x)&=&{K_3\over 4\pi K}\dn\left({K_3\over 2K}x,k_3\right),\qquad
{K_3'\over K_3}={\lam\over 2K},\label{Repc2}
\eea
where the sn function carries the modulus $k_1$ \refeq{k1} and
the dn function a new modulus $k_3$.

%%%%%%%%%%%%%%%%%%%%%%%%%%%%%%%%%%%%%
%Alternative fuer Attraktion
%%%%%%%%%%%%%%%%%%%%%%%%%%%%%%%%%%%%%
\renewcommand{\theequation}{C.\arabic{equation}}

\section*{Appendix~C}

\setcounter{equation}{0}

In section 3 we have derived a set of integral equations which determine
the thermodynamics of the $XYZ$ chain. However, for the model with attractive
interaction ($K'/2<\lam<K'$) the analysis of equations \refeq{IntEqT},
\refeq{LambExT} becomes difficult, particularly in numerical applications
as the integration paths have to be shifted into the upper or lower half
plane. We can avoid this by an alternative approach to the case
$K'/2<\lam<K'$. From \refeq{InterCoeff} we find
\be
J_X(\lam)=+J_X(K'-\lam),\qquad
J_Y(\lam)=-J_Y(K'-\lam),\qquad
J_Z(\lam)=+J_Z(K'-\lam).\label{IntRel}
\ee
Recalling that due to a unitary transformation
we are free to change any two signs of the interaction
coefficients \refto{Baxt82b} we see that the interchange of $\lam$ with
$K'-\lam$ amounts to the replacement $H\to-H$. Instead of treating the
thermodynamics of the chain with crossing parameter $\lam$ at temperature
$T$ we may study the chain with crossing parameter $K'-\lam$ at temperature
$-T$. This is actually possible.

{}From now on we consider the model for a crossing parameter $\lam$ with
$0<\lam<K'/2$ at negative temperatures $-T$ ($T>0$).
In this case the auxiliary functions $\a$ and $\ba$ in
\refeq{SubCond} are no longer small for large values of $N$ and $\beta$.
Instead we find that we have to take the reciprocals
\bea
\a(x)={p(x-i\lam)},\cr
\ba(x)={1\over p(x+i\lam)}.
\eea
In terms of these functions we find two representations of the eigenvalue
function $\L$
\bea
\L(x-i\lam)&=&{s\omega}\P(x){q(x-i3\lam)\over q(x-i\lam)}\A(x),\cr
\L(x+i\lam)&=&{s\over\omega}\P(x){q(x+i3\lam)\over q(x+i\lam)}\bA(x).
\eea
Proceeding similarly to section 3 we derive the integral equation
\bea
\ln\a(x)&=&
{N\over 2}\ln\left[\sn{K_2\over 2K}(x+i(x_0-2\lam))\over
\sn{K_2\over 2K}(x-ix_0)\right]\cr
&&+\dint k(x-y)\ln\A(y)\dify-\dint k(x-y-i2\lam+i\epsilon)\ln\bA(y)\dify\cr
&&+D,
\eea
where the $\sn$ functions carry the modulus $k_2$ \refeq{k2} and $D$ is
a constant
\be
D=-\ln\omega +{\pi i\over 2K}\jsum y_j.
\ee
The kernel $k(x)$ is defined by
\be
k(x)={1\over 4K}\left[{1\over 2}-
\sum_{k=1}^\infty{\sh(k(K'-2\lam))\ch(i k x)\over
\ch(k(K'-\lam))\sh(k\lam)}\right],
\ee
and is identical to \refeq{kernel} after a substitution of $\lam$ by
$K'-\lam$. Unfortunately, in this case the constant $D$ cannot be expressed
in terms of the auxiliary functions. Instead of this missing
relation we find the subsidiary condition
\be
{1\over 4K}\dint\ln\bA(x)\difx-{1\over 4K}\dint\ln\A(x)\difx=2\ln\omega,
\ee
which completes the integral equation. For $\ba$ a similar integral
equation is valid.
The eigenvalue is given by
\bea
\ln\L(x)&=&\ln r+{N\pi\over 4K}(x_0-\lam)\cr
&&+{N}\ksumo{\ch(k(K'-\lam))-\ch(ik x)\ch(k (K'-x_0))\over
k\ch(k (K'-\lam))\sh(k K')}\ch(k(K'-2\lam))\cr
&&+\dint [c(x-y-i(K'-2\lam))\ln\A(y)
+c(x-y+i(K'-2\lam))\ln\bA(y)]\dify,\cr
&&
\eea
where $c(x)$ is defined by
\be
c(x)={1\over 4K}\left[{1\over 2}+\ksumo{\ch(ik x)\over \ch(k\(K'-\lam))}
\right],
\ee
which is identical to \refeq{Defc} after a substitution of $\lam$ by
$K'-\lam$. In the limit of infinite Trotter number $N$ we find
\bea
\ln\a(x)&=&
-{2\pi\beta}c(x+i(K'-2\lam))+D\cr
&&+\dint k(x-y)\ln\A(y)\dify-\dint k(x-y-i2\lam+i\epsilon)\ln\bA(y)\dify,\cr
&&\label{Eqferro}
\eea
and
\bea
\ln\L&=&\ln r+{\pi\beta\over 4K}\left[1+
2\ksumo{\sh(k (K'-\lam))\ch(k(K'-2\lam))\over
\ch(k (K'-\lam))\sh(k K')}\right]\cr
&&+\dint [c(x+i(K'-2\lam))\ln\A(x)+c(x-i(K'-2\lam))\ln\bA(x)]\difx.\cr
&&\label{Lamferro}
\eea
These equations determine the free energy density and the leading correlation
length of the quantum chain in the attractive regime.
Here we do not present a study of
the next-leading eigenvalues due to 1- and 2-strings.

\newpage

\def\and{and\ }

\def\eds{eds.\ }

\def\edi{ed.\ }
\references

\def\mtb{M. T. Batchelor}
\def\rjb{R. J. Baxter}
\def\dk{D. Kim}
\def\pap{P. A. Pearce}
\def\nyr{N. Yu. Reshetikhin}
\def\ak{A. Kl\"umper}

\refis{AbramS64} M. Abramowitz, I. A. Stegun, ``Handbook of
Mathematical Functions", Washington, U.S. National Bureau of Standards 1964;
New York, Dover 1965.

\refis{Affl86} I. Affleck, \prl 56, 746, 1986

\refis{AKLT87} I. Affleck, T. Kennedy, E. H. Lieb \and H. Tasaki, \prl 59,
799, 1987

\refis{AKLT88} I. Affleck, T. Kennedy, E. H. Lieb \and H. Tasaki, \cmp 115,
477, 1988

\refis{AfflGSZ89} I. Affleck, D. Gepner, H. J. Schulz \and T. Ziman,
\jpa 22, 511, 1989

\refis{AkutDW89} Y. Akutsu, T. Deguchi \and M. Wadati, in Braid Group, Knot
Theory and Statistical
Mechanics, \eds C. N. Yang \and M. L. Ge, World Scientific, Singapore, 1989

\refis{AkutKW86a} Y. Akutsu, A. Kuniba \and M. Wadati,\jpj 55, 1466, 1986

\refis{AkutKW86b} Y. Akutsu, A. Kuniba \and M. Wadati,\jpj 55, 2907, 1986

\refis{AlcaBB87} F. C. Alcaraz, M. N. Barber \and \mtb,\prl 58, 771, 1987

\refis{AlcaBB88} F. C. Alcaraz, M. N. Barber \and \mtb,\annp 182, 280, 1988

\refis{AlcaBGR88} F. C. Alcaraz, M. Baake, U. Grimm \and V. Rittenberg,
\jpa 21, L117, 1988

\refis{AlcaM89}  F. C. Alcaraz \and M. J. Martins, \jpa 22, 1829, 1989

\refis{AlcaM90}  F. C. Alcaraz \and M. J. Martins, \jpa 23, 1439-51, 1990

\refis{Alex75} S. Alexander, \pla 54, 353-4, 1975

\refis{AndrBF84} G. E. Andrews, \rjb\ \and P. J. Forrester, \jsp 35, 193,
1984

\refis{BarbBP87} \mtb, M.N. Barber \and \pap,\jsp 49, 1117, 1987

\refis{BarbB89} M.N. Barber \and \mtb, \prb 40, 4621, 1989

\refis{Barb91} M.N. Barber, \physica A 170, 221, 1991

\refis{Bariev82} R. Z. Bariev, \tmp 49, 1021, 1982

\refis{BarievKSZ93} R. Z. Bariev, A. Kl\"{u}mper, A. Schadschneider
\and J. Zittartz, \jpa 26, 1249, 1993

\refis{BarouchM71} E. Barouch \and B. M. McCoy, \pra 3, 786, 1971

\refis{Baxt70} \rjb,\jmp 11, 3116, 1970

\refis{Baxt71b} \rjb,\prl 26, 834, 1971

\refis{Baxt72} \rjb,\annp 70, 193, 1972

\refis{Baxt73} \rjb,\jsp 8, 25, 1973

\refis{Baxt80} \rjb,\jpa 13, L61--70, 1980

\refis{Baxt81a} \rjb,\physica 106A, 18--27, 1981

\refis{Baxt81b} \rjb,\jsp 26, 427--52, 1981

\refis{Baxt82a} \rjb,\jsp 28, 1, 1982

\refis{Baxt82b} \rjb, ``Exactly Solved Models in Statistical Mechanics",
Academic Press, London, 1982.

\refis{BaxtP82} \rjb\space \and \pap,\jpa 15, 897, 1982

\refis{BaxtP83} \rjb\space \and \pap,\jpa 16, 2239, 1983

\refis{BazhR89} V.V. Bazhanov \and \nyr,\ijmpa 4, 115--42, 1989

\refis{BazhB93} V.V. Bazhanov \and \rjb,\physica A 194, 390--396, 1993

\refis{BednorzM86} J. G. Bednorz \and K. A. M"uller, \zpb 64, 189, 1986

\refis{BelaPZ84} A. A. Belavin, A. M. Polyakov \and A. B. Zamolodchikov,
\npb 241, 333, 1984

\refis{Bethe31} H. A. Bethe,\zp 71, 205, 1931

\refis{BlotCN86} H. W. J. Bl\"ote, J. L. Cardy \and M. P. Nightingale, \prl
56, 742,
1986

\refis{BogK89} N. M. Bogoliubov \and V. E. Korepin, \ijmpb 3, 427-439, 1989

\refis{BretzD71} M. Bretz \and J. G. Dash, \prl 27, 647, 1971

\refis{Bretz77} M. Bretz, \prl 38, 501, 1977

\refis{Buy86} W. J. L. Buyers, R. M. Morra, R. L. Armstrong, P. Gerlach
\and K. Hirakawa, \prl 56, 371, 1986

\refis{Morra88} R. M. Morra, W. J. L. Buyers, R. L. Armstrong \and K. Hirakawa,
\prb 38, 543, 1988

\refis{Stei87} M. Steiner, K. Kakurai, J. K. Kjems, D. Petitgrand \and R. Pynn,
\jappp 61, 3953, 1987

\refis{Tun90} Z. Tun, W. J. L. Buyers, R. L. Armstrong, K. Hirakawa \and
B. Briat, \prb 42, 4677, 1990

\refis{Tun91} Z. Tun, W. J. L. Buyers, A. Harrison \and J. A. Rayne, \prb 43,
13331, 1991

\refis{Ren87} J. P. Renard, M. Verdaguer, L. P. Regnault, W. A. C. Erkelens,
J. Rossa-Mignod \and W. G. Stirling, \eurolett 3, 945, 1987

\refis{Ren88} J. P. Renard, M. Verdaguer, L. P. Regnault, W. A. C. Erkelens,
J. Rossa-Mignod, J. Ribas, W. G. Stirling \and C. Vettier, \jappp 63, 3538,
1988

\refis{Reg89} L. P. Regnault, J. Rossa-Mignod, J. P. Renard, M. Verdaguer
\and C. Vettier, \physica B 156 \& 157, 247, 1989

\refis{Colom87} P. Colombet, S. Lee, G. Ouvrard \and R. Brec, \jcr, 134, 1987

\refis{deGroot82} H. J. M. de Groot, L. J . de Jongh, R. D. Willet \and
J. Reeyk, \jappp 53, 8038, 1982

\refis{Capp88} A. Cappelli, Recent Results in Two-Dimensional Conformal
Field
Theory, in Proceedings of the XXIV International Conference on High Energy
Physics,
\eds R. Kotthaus \and J. K\"uhn, Springer, Berlin, 1988

\refis{CappIZ87a} A. Cappelli, C. Itzykson \and J.-B. Zuber, \npb {280
[FS18]},
445--65, 1987

\refis{CappIZ87b} A. Cappelli, C. Itzykson \and J.-B. Zuber, \cmp 113,
1--26, 1987

\refis{Card84a} J. L. Cardy, \jpa 17, L385, 1984

\refis{Card86a} J. L. Cardy, \npb {270 [FS16]}, 186, 1986

\refis{Card86b} J. L. Cardy, \npb {275 [FS17]}, 200, 1986

\refis{Card88} J. L. Cardy, ``Phase Transitions and Critical Phenomena,
Vol.11",
\eds C. Domb \and J.L. Lebowitz, Academic Press, London 1988

\refis{Card89} J. L. Cardy, Conformal Invariance and Statistical Mechanics,
in Les
Houches, Session XLIV, Fields, Strings and Critical Phenomena, \eds E.
Br\'ezin \and
J. Zinn-Justin, 1989

\refis{ChoiKK90} J.-Y. Choi, K. Kwon \and D. Kim, \eurolett xx, to appear,
1990

\refis{ChoiKP89} J.-Y. Choi, D. Kim \and \pap, \jpa 22, 1661--71, 1989

\refis{CvetDS80} D. M. Cvetkovic, M. Doob \and H. Sachs, ``Spectra of
Graphs", Academic Press, London 1980

\refis{DateJKMO87} E. Date, M. Jimbo, A. Kuniba, T. Miwa, \and M. Okado,
\npb
B290, 231--273, 1987
%%%fusionRSOS paper I

\refis{DateJKMO88} E. Date, M. Jimbo, A. Kuniba, T. Miwa, \and M. Okado,
\aspm 16,
17, 1988
%%%fusion RSOS paper II

\refis{DateJMO86} E. Date, M. Jimbo, T. Miwa \and M. Okado,\lmp 12, 209,
1986

\refis{DateJMO87} E. Date, M. Jimbo, T. Miwa \and M. Okado,\prb 35, 2105--7,
1987

\refis{DaviP90} B. Davies \and \pap, \ijmpb {}, this issue, 1990

\refis{deVeK87} H. J. de Vega \and M. Karowski, \npb {285 [FS19]}, 619, 1987

\refis{deVeW85} H. J. de Vega \and F. Woynarovich,\npb 251, 439, 1985

\refis{deVeW90} H. J. de Vega \and F. Woynarovich,\jpa 23, 1613, 1990

\refis{DestdeVeW92} C. Destri \and H. J. de Vega,\prl 69, 2313, 1992

\refis{diFrSZ87} P. di Francesco, H. Saleur \and J.-B. Zuber, \jsp 49,
57--79, 1987

\refis{diFrZ89} P. di Francesco \and J.-B. Zuber, $SU(N)$ Lattice Models
Associated
with Graphs, Saclay preprint SPhT/89-92, 1989

\refis{DijkVV88} R. Dijkgraaf, E. Verlinde \and H. Verlinde, in Proceedings
of the
1987 Copenhagen Conference, World Scientific, 1988

\refis{DijkVVV89} R. Dijkgraaf, C. Vafa, E. Verlinde \and H. Verlinde,\cmp
123, 485, 1989

\refis{DombG76} ``Phase Transitions and Critical Phenomena, Vol.6",
Academic Press, London 1976

\refis{FateZ85} V. A. Fateev \and A. B. Zamolodchikov, \jetp 62, 215, 1985

\refis{FendG89} P. Fendley \and P. Ginsparg, \npb 324, 549--80, 1989

\refis{FodaN89} O. Foda \and B. Nienhuis, \npb {},{},1989

\refis{FrieQS84} D. Friedan, Z. Qiu \and S. Shenker, \prl 52, 1575, 1984; in
``Vertex Operators in Mathematics and Physics", \eds J. Lepowsky, S.
Mandelstam \and
I.M. Singer, Springer, 1984

\refis{FrahmK90} H. Frahm \and V. E. Korepin, \prb 42, 10553, 1990

\refis{FrahmYF90} H. Frahm, N.-C. Yu \and M. Fowler, \npb 336, 396, 1990

\refis{GepnQ87} D. Gepner \and Z. Qiu, \npb 285, 423--53, 1987

\refis{Gins88} P. Ginsparg,\npb {295 [FS21]}, 153--70, 1988

\refis{Gins89a} P. Ginsparg, Applied Conformal Field Theory, in Les
Houches,
Session XLIV, Fields, Strings and Critical Phenomena, \eds E. Br\'ezin \and
J.
Zinn-Justin, 1989

\refis{Gins89b} P. Ginsparg, Some Statistical Mechanical Models and
Conformal Field
Theories, Trieste Spring School Lectures, HUTP-89/A027

\refis{GradR80} I.S. Gradshteyn \and I.M. Ryzhik, ``Tables of Integrals,
Series and Products", Academic
Press, New York, 1980.

\refis{Grif72} R. B. Griffiths, ``Phase Transitions and Critical Phenomena,
Vol.1",\eds C. Domb \and M. S. Green, Academic Press, London 1972

\refis{Hald83a} F. D. M. Haldane, \prl 50, 1153, 1983

\refis{Hald83b} F. D. M. Haldane, \pla 93, 464, 1983

\refis{Hame85} C. J. Hamer,\jpa 18, L1133, 1985

\refis{Hame86} C. J. Hamer,\jpa 19, 3335, 1986

\refis{Hirsch89a} J. E. Hirsch, \pla 134, 451, 1989

\refis{Hirsch89b} J. E. Hirsch, \physica  C 158, 326, 1989

\refis{Huse82} D. A. Huse,\prl 49, 1121--4, 1982

\refis{Huse84} D. A. Huse, \prb 30, 3908, 1984

\refis{ItzySZ88} C. Itzykson, H. Saleur \and J-B. Zuber, ``Conformal
Invariance and Applications to
Statistical Mechanics", World Scientific, Singapore, 1988

\refis{JimbM84} M. Jimbo \and T. Miwa, \aspm 4, 97--119, 1984

\refis{JimbMO87} M. Jimbo, T. Miwa \and M. Okado, \lmp 14, 123--31, 1987

\refis{JimbMO88} M. Jimbo, T. Miwa \and M. Okado, \cmp 116, 507--25, 1988

\refis{JimbMT89} M. Jimbo, T. Miwa \and A. Tsuchiya,``Integrable Systems in
Quantum Field Theory and
Statistical Mechanics", \aspm 19, ,1989

\refis{JohnKM73} J.D. Johnson, S. Krinsky, \and B.M. McCoy,\pra 8, 2526,
1973

\refis{Kac79} V. G. Kac, \lnp 94, 441--445, 1979

\refis{KadaB79} L. P. Kadanoff \and A. C. Brown, \annp 121, 318--42, 1979

\refis{Karo88} M. Karowski, \npb {300 [FS22]}, 473, 1988

\refis{Kato87} A. Kato, \mpla 2, 585, 1987

\refis{KimP87} \dk\space \and \pap,\jpa 20, L451--6, 1987

\refis{KimP89}  \dk\space \and \pap,\jpa 22, 1439--50, 1989

\refis{Kiri89} E. B. Kiritsis, \plb  217, 427, 1989

\refis{KiriR86} A. N. Kirillov \and N. Yu. Reshetikhin,\jpa 19, 565, 1986

\refis{KiriR87} A. N. Kirillov \and N. Yu. Reshetikhin,\jpa 20, 1565, 1987

\refis{KlassM90} T. R. Klassen \and E. Melzer, \npb 338, 485, 1990

\refis{KlassM91} T. R. Klassen \and E. Melzer, \npb 350, 635, 1991

\refis{KlumBiq} A. Kl\"{u}mper, \eurolett 9, 815, 1989; \jpa 23, 809, 1990

\refis{KlumB90} A. Kl\"{u}mper \and \mtb,\jpa 23, L189, 1990

\refis{KlumBP91} A. Kl\"{u}mper, \mtb \ \and \pap, \jpa 24, 3111--3133, 1991

\refis{KlumP91} A. Kl\"{u}mper \and \pap, \jsp 64, 13--76, 1991

\refis{Klum92c} A. Kl\"{u}mper, unver"offentlichte Rechnungen, 1992

\refis{KlumZ88} A. Kl\"{u}mper \and J. Zittartz,\zpb 71, 495, 1988

\refis{KlumZ89} A. Kl\"{u}mper \and J. Zittartz,\zpb 75, 371, 1989

\refis{KlumZ8VM} A. Kl\"{u}mper \and J. Zittartz,\zpb 71, 495, 1988;
\zpb 75, 371, 1989

\refis{KlumSZ89} A. Kl\"{u}mper, A. Schadschneider \and J. Zittartz,
\zpb 76, 247, 1989

\refis{KlumSZMPG} A. Kl\"{u}mper, A. Schadschneider \and J. Zittartz,
\jpa 24, L955-L959, 1991; \zpb 87, 281-287, 1992

\refis{Klum89} \ak, \eurolett 9, 815, 1989

\refis{KlumP92} \ak\  \and \pap, \physica 183A, 304-350, 1992

\refis{Klum92} \ak , Ann. Physik 1, 540, (1992)

\refis{Klum92b} \ak, in preparation

\refis{KlumWZ93} \ak, T. Wehner \and J. Zittartz, submitted to {\it J. Phys.
A.}

\refis{Knabe88} S. Knabe, \jsp 52, 627, 1988

\refis{Koma} T. Koma, \ptp 78, 1213, 1987; \bf 81, \rm 783, (1989)

\refis{KorepinS90} V. E. Korepin \and N. A. Slavnov, \npb 340, 759, 1990

\refis{ItsIK92} A. R. Its, A. G. Izergin \and V. E. Korepin, \physica D 54,
351, 1992

\refis{ItsIKS93} A. R. Its, A. G. Izergin, V. E. Korepin \and N. A. Slavnov,
\prl 70, 1704, 1993

\refis{KuniY88} A. Kuniba \and T. Yajima,\jsp 52, 829, 1988

\refis{KuliRS81} P. P. Kulish, N. Yu. Reshetikhin \and E. K. Sklyanin, \lmp
5, 393, 1981

\refis{Kuniba92} A. Kuniba, ``Thermodynamics of the $U_q(X_r^{(1)})$ Bethe
Ansatz System with $q$ a Root of Unity", ANU preprint (1991)

\refis{Lewi58} L. Lewin, Dilogarithms and Associated Functions, MacDonald,
London, 1958

\refis{LiebWu68} E. H. Lieb \and F. Y. Wu, \prl 20, 1445, 1968

\refis{LutherP74} A. Luther \and I. Peschel, \prb 9, 2911, 1974

\refis{Martins91} M. J. Martins, \prl 22, 419, 1991 and private communication
(1991)
%\refis{Martins91} M. J. Martins, ``private communication"  1991

\refis{Muell} E. M\"uller-Hartmann, unpublished results, (1989)

\refis{Muell89} E. M\"uller-Hartmann, unver"offentlichte Ergebnisse, (1989)

\refis{Mura89} J. Murakami, \aspm 19, 399--415, 1989

\refis{Nien87} B. Nienhuis, in Phase Transitions and Critical Phenomena,
Vol.11,
\eds C. Domb \and J.L. Lebowitz, Academic Press, 1987

\refis{NahmRT92} W. Nahm, A. Recknagel \and M. Terhoven, Preprint ``Dilogarithm
identities in conformal field theory'', 1992

\refis{NighB86} M. P. Nightingale \and H. W. J. Bl"ote, \prb 33, 659, 1986

\refis{OwczB87} A. L. Owczarek \and \rjb,\jsp 49, 1093, 1987

\refis{ParkBiq} J. B. Parkinson,\jpc 20, L1029, 1987; \jpc 21, 3793, 1988;
\jphc 8, 1413, 1988

\refis{ParkB85} J. B. Parkinson \and J. C. Bonner, \prb 32, 4703, 1985

\refis{PaczP90} I. D. Paczek \and J. B. Parkinson,\jpcon 2, 5373, 1990

\refis{Pasq87a} V. Pasquier,\npb {285 [FS19]}, 162, 1987

\refis{Pasq87b} V. Pasquier,\jpa 20, {L217, L221}, 1987

\refis{Pasq87c} V. Pasquier,\jpa 20, {L1229, 5707}, 1987

\refis{Pasq88} V. Pasquier,\npb {B295 [FS21]}, 491--510, 1988

\refis{Pear85} \pap,\jpa 18, 3217--26, 1985

\refis{Pear87prl} \pap,\prl 58, 1502--4, 1987

\refis{Pear87jpa} \pap,\jpa 20, 6463--9, 1987

\refis{Pear90ijmpb} \pap,\ijmpb 4, 715--34, 1990

\refis{PearB90} \pap\space \and \mtb, \jsp 60, 77--135, 1990

\refis{PearK87} \pap \and \dk, \jpa, 20, 6471-85, 1987

\refis{PearS88} \pap\space \and K. A. Seaton,\prl 60, 1347, 1988

\refis{PearS89} \pap\space \and K. A. Seaton,\annp 193, 326, 1989

\refis{PearS90} \pap\space \and K. A. Seaton,\jpa 23, 1191--1206, 1990

\refis{Pear91} \pap, Row Transfer Matrix Functional Equations for
$A$--$D$--$E$ Lattice Models,
 to be published, 1991

\refis{PearK91} \pap\space \and A. Kl\"umper, \prl 66, 974, 1991

\refis{Pear92} \pap, \ijmpa 7, Suppl.1B, 791, 1992

\refis{Resh83jetp} \nyr,\jetp 57, 691, 1983
%  analytic ansatz

\refis{Resh83lmp} \nyr, \lmp 7, 205--13, 1983
%  inversion identities for vertex models

\refis{Sale88} H. Saleur, Lattice Models and Conformal Field Theories, in
Carg\`ese
School on Common Trends in Condensed Matter and Particle Physics, 1988

\refis{SaleB89} H. Saleur \and M. Bauer, \npb 320, 591--624, 1989

\refis{SaleD87} H. Saleur \and P. di Francesco, Two Dimensional Critical
Models on a
Torus, in Brasov Summer School on Conformal Invariance and String Theory,
1987

\refis{Samuel73} E. J. Samuelson, \prl 31, 936, 1973

\refis{SeatP89} K. A. Seaton \and \pap\space, \jpa 22, 2567--76, 1989

\refis{Strog79} Yu. G. Stroganov, \pla 74, 116, 1979

\refis{Suth70} B. Sutherland, \jmp 11, 3183, 1970

\refis{Suzuki85} M. Suzuki, \prb 31, 2957, 1985

\refis{SuzukiI87} M. Suzuki \and M. Inoue, \ptp 78, 787, 1987

\refis{Suzuki87} M. Suzuki, in ``Quantum Monte Carlo Methods in
Equilibrium and Nonequilibrium Systems",
\edi M. Suzuki, Springer Verlag, 1987

\refis{SuzukiAW90} J. Suzuki, Y. Akutsu \and M. Wadati, \jpj 59, 2667-2680,
1990

\refis{SuzukiNW92} J. Suzuki, T. Nagao \and M. Wadati, \ijmpb 6, 1119, 1992

\refis{Tak71} M. Takahashi, \ptp 46, 401, 1971

\refis{TakTBA} M. Takahashi, \ptp 46, 401, 1971; \ptp 50, 1519, 1973

\refis{Tak91} M. Takahashi, \prb 43, 5788, 1991; \prb 44, 12382, 1991

\refis{Tak91a} M. Takahashi, \prb 43, 5788, 1991

\refis{Tak91b} M. Takahashi, \prb 44, 12382, 1991

\refis{TempL71} H. N. V. Temperley \and E. H. Lieb, \prs 322, 251, 1971

\refis{Tetel82} M. G. Tetel'man, \jetp 55, 306, 1982

\refis{TruS83} T. T. Truong \and K. D. Schotte, \npb 220, 77, 1983

\refis{Tsun91} H. Tsunetsugu, \jpj 60, 1460, 1991

\refis{vonGR87} G. von Gehlen \and V. Rittenberg, \jpa 20, 227, 1987

\refis{WadaDA89} M. Wadati, T. Deguchi \and Y. Akutsu, \prep 180, 247--332,
1989

\refis{Woyn87} F. Woynarovich, \prl 59, 259, 1987

\refis{WoynE87} F. Woynarovich \and H.-P. Eckle, \jpa 20, L97, 1987

\refis{Yang69} C. N. Yang \and C. P. Yang, \jmp 10, 1115, 1969

\refis{Yang66} C. N. Yang \and C. P. Yang, \pr 147, 303, 1966; 150, 321

\refis{Yang62} C. N. Yang, \rmp 34, 691, 1962

\refis{Yang67} C. N. Yang, \prl 19, 1312, 1967

\refis{YangG89} C. N. Yang \and M. L. Ge (Editors), Braid Group, Knot Theory
and Statistical
Mechanics, World Scientific, Singapore, 1989

\refis{ZamoF80} A. B. Zamolodchikov \and V. Fateev, \sjnp 32, 198, 1980

\refis{Zamo80} A. B. Zamolodchikov, \jetp 52, 325, 1980; \cmp 79, 489, 1981

\refis{Zamo91} Al. B. Zamolodchikov, \plb 253, 391--4, 1991; \npb 358,
497--523, 1991

\refis{Zamo91a} Al. B. Zamolodchikov, \plb 253, 391--4, 1991

\refis{Zamo91b} Al. B. Zamolodchikov, \npb 358, 497--523, 1991

\endreferences

\newpage

\section*{Captions}

\begin{itemize}

\item[Fig. 1]
The eight arrow configurations allowed at a vertex and the corresponding
Boltzmann weights.

\item[Fig. 2]
Graphical representation of the 'crossing' symmetry expressing the
invariance of the Boltzmann weights of a vertex under a combined rotation
by $90^o$ and the substitution of $v$ by $-v$.

\item[Fig. 3]
Representation of a $N\times L$ square lattice with a distribution of
spectral parameters $\pm v_0=\pm(\lam-\beta/N)$ homogeneous in each row
and alternating in columns. $N$ denotes the 'Trotter' number and $L$
the length of the quantum chain.
The column-to-column transfer matrix is identical
to the transfer matrix with row inhomogeneity depicted on the right.

The study of the quantum chain in an external magnetic field leads to
the introduction of a horizontal seam into the lattice model and
modified boundary conditions for the column-to-column transfer matrix.

\item[Fig. 4]
(a) Depiction of the specific heat and the leading correlation length
for the anisotropic Heisenberg chain with $J_X=J_Y=1$ and $J_Z=-3/2$.
(b) Specific heat and leading correlation length
for the ferromagnetic case $J_X=J_Y=1$ and $J_Z=3/2$.
Note that the curves for the correlation length have been scaled by factors
of 1/100 and 1/1000, respectively.

\item[Fig. 5]
Depiction of the specific heat and the leading correlation length
for the spin-1 biquadratic chain. The values for $\xi$ have been scaled by
a factor of 1/100.

\end{itemize}

\end{document}

%%%%%%%%%%%%%%%%%%%%%%% Fig 1&2 %%%%%%%%%%%%%%%%%%%%%%%%%%%%%%%%%%%
%!
%%Title: /tmp/xfig-export004082
%%Creator: fig2dev
%%CreationDate: Tue Jun  8 16:45:05 1993
%%For: kluemper@sun12 (Andreas Kluemper)
%%Pages: 0
%%BoundingBox: 0 0 351 567
%%EndComments
/$F2psDict 32 dict def
$F2psDict begin
	$F2psDict /mtrx matrix put

	end
	/$F2psBegin {$F2psDict begin /$F2psEnteredState save def} def
	/$F2psEnd {$F2psEnteredState restore end} def

$F2psBegin
1 setlinecap 1 setlinejoin
-94 40 translate
0.000000 567.000000 translate 0.900 -0.900 scale
612 0 translate 90 rotate
1.000 setlinewidth
% Polyline
newpath 44 139 moveto 114 139 lineto stroke
gsave  0.000 setgray fill grestore stroke
% Polyline
newpath 79 104 moveto 79 174 lineto stroke
gsave  0.000 setgray fill grestore stroke
% Polyline
newpath 79 118 moveto 76 125 lineto 83 125 lineto 79 118 lineto closepath gsave
 0.000 setgray fill grestore stroke
% Polyline
newpath 76 160 moveto 79 153 lineto 83 160 lineto 76 160 lineto closepath gsave
 0.000 setgray fill grestore stroke
% Polyline
newpath 58 143 moveto 65 139 lineto 58 136 lineto 58 143 lineto closepath gsave
 0.000 setgray fill grestore stroke
% Polyline
newpath 93 136 moveto 93 143 lineto 100 139 lineto 93 136 lineto closepath
gsave  0.000 setgray fill grestore stroke
% Polyline
newpath 194 139 moveto 124 139 lineto stroke
gsave  0.000 setgray fill grestore stroke
% Polyline
newpath 159 174 moveto 159 104 lineto stroke
gsave  0.000 setgray fill grestore stroke
% Polyline
newpath 159 160 moveto 162 153 lineto 155 153 lineto 159 160 lineto closepath
gsave  0.000 setgray fill grestore stroke
% Polyline
newpath 162 118 moveto 159 125 lineto 155 118 lineto 162 118 lineto closepath
gsave  0.000 setgray fill grestore stroke
% Polyline
newpath 180 135 moveto 173 139 lineto 180 142 lineto 180 135 lineto closepath
gsave  0.000 setgray fill grestore stroke
% Polyline
newpath 145 142 moveto 145 135 lineto 138 139 lineto 145 142 lineto closepath
gsave  0.000 setgray fill grestore stroke
% Polyline
newpath 239 104 moveto 239 174 lineto stroke
gsave  0.000 setgray fill grestore stroke
% Polyline
newpath 274 139 moveto 204 139 lineto stroke
gsave  0.000 setgray fill grestore stroke
% Polyline
newpath 260 139 moveto 253 136 lineto 253 143 lineto 260 139 lineto closepath
gsave  0.000 setgray fill grestore stroke
% Polyline
newpath 218 136 moveto 225 139 lineto 218 143 lineto 218 136 lineto closepath
gsave  0.000 setgray fill grestore stroke
% Polyline
newpath 235 118 moveto 239 125 lineto 242 118 lineto 235 118 lineto closepath
gsave  0.000 setgray fill grestore stroke
% Polyline
newpath 242 153 moveto 235 153 lineto 239 160 lineto 242 153 lineto closepath
gsave  0.000 setgray fill grestore stroke
% Polyline
newpath 319 174 moveto 319 104 lineto stroke
gsave  0.000 setgray fill grestore stroke
% Polyline
newpath 284 139 moveto 354 139 lineto stroke
gsave  0.000 setgray fill grestore stroke
% Polyline
newpath 298 139 moveto 305 142 lineto 305 135 lineto 298 139 lineto closepath
gsave  0.000 setgray fill grestore stroke
% Polyline
newpath 340 142 moveto 333 139 lineto 340 135 lineto 340 142 lineto closepath
gsave  0.000 setgray fill grestore stroke
% Polyline
newpath 323 160 moveto 319 153 lineto 316 160 lineto 323 160 lineto closepath
gsave  0.000 setgray fill grestore stroke
% Polyline
newpath 316 125 moveto 323 125 lineto 319 118 lineto 316 125 lineto closepath
gsave  0.000 setgray fill grestore stroke
% Polyline
newpath 364 139 moveto 434 139 lineto stroke
gsave  0.000 setgray fill grestore stroke
% Polyline
newpath 399 104 moveto 399 174 lineto stroke
gsave  0.000 setgray fill grestore stroke
% Polyline
newpath 399 118 moveto 396 125 lineto 402 125 lineto 399 118 lineto closepath
gsave  0.000 setgray fill grestore stroke
% Polyline
newpath 378 142 moveto 385 139 lineto 378 136 lineto 378 142 lineto closepath
gsave  0.000 setgray fill grestore stroke
% Polyline
newpath 396 153 moveto 399 160 lineto 402 153 lineto 396 153 lineto closepath
gsave  0.000 setgray fill grestore stroke
% Polyline
newpath 413 139 moveto 420 136 lineto 420 142 lineto 413 139 lineto closepath
gsave  0.000 setgray fill grestore stroke
% Polyline
newpath 479 174 moveto 479 104 lineto stroke
gsave  0.000 setgray fill grestore stroke
% Polyline
newpath 444 139 moveto 514 139 lineto stroke
gsave  0.000 setgray fill grestore stroke
% Polyline
newpath 458 139 moveto 465 142 lineto 465 136 lineto 458 139 lineto closepath
gsave  0.000 setgray fill grestore stroke
% Polyline
newpath 482 160 moveto 479 153 lineto 476 160 lineto 482 160 lineto closepath
gsave  0.000 setgray fill grestore stroke
% Polyline
newpath 493 142 moveto 500 139 lineto 493 136 lineto 493 142 lineto closepath
gsave  0.000 setgray fill grestore stroke
% Polyline
newpath 479 125 moveto 476 118 lineto 482 118 lineto 479 125 lineto closepath
gsave  0.000 setgray fill grestore stroke
% Polyline
newpath 559 174 moveto 559 104 lineto stroke
gsave  0.000 setgray fill grestore stroke
% Polyline
newpath 562 160 moveto 559 153 lineto 556 160 lineto 562 160 lineto closepath
gsave  0.000 setgray fill grestore stroke
% Polyline
newpath 559 125 moveto 556 118 lineto 562 118 lineto 559 125 lineto closepath
gsave  0.000 setgray fill grestore stroke
% Polyline
newpath 594 139 moveto 524 139 lineto stroke
gsave  0.000 setgray fill grestore stroke
% Polyline
newpath 580 136 moveto 573 139 lineto 580 142 lineto 580 136 lineto closepath
gsave  0.000 setgray fill grestore stroke
% Polyline
newpath 545 139 moveto 538 142 lineto 538 136 lineto 545 139 lineto closepath
gsave  0.000 setgray fill grestore stroke
% Polyline
newpath 604 139 moveto 674 139 lineto stroke
gsave  0.000 setgray fill grestore stroke
% Polyline
newpath 618 139 moveto 625 142 lineto 625 136 lineto 618 139 lineto closepath
gsave  0.000 setgray fill grestore stroke
% Polyline
newpath 653 142 moveto 660 139 lineto 653 136 lineto 653 142 lineto closepath
gsave  0.000 setgray fill grestore stroke
% Polyline
newpath 639 174 moveto 639 104 lineto stroke
gsave  0.000 setgray fill grestore stroke
% Polyline
newpath 639 160 moveto 642 153 lineto 636 153 lineto 639 160 lineto closepath
gsave  0.000 setgray fill grestore stroke
% Polyline
newpath 642 125 moveto 639 118 lineto 636 125 lineto 642 125 lineto closepath
gsave  0.000 setgray fill grestore stroke
/Times-Italic findfont 18.000 scalefont setfont
74 209 moveto
1 -1 scale
(1) gsave  0.000 rotate show grestore 1 -1 scale
/Times-Italic findfont 18.000 scalefont setfont
154 209 moveto
1 -1 scale
(2) gsave  0.000 rotate show grestore 1 -1 scale
/Times-Italic findfont 18.000 scalefont setfont
234 209 moveto
1 -1 scale
(3) gsave  0.000 rotate show grestore 1 -1 scale
/Times-Italic findfont 18.000 scalefont setfont
314 209 moveto
1 -1 scale
(4) gsave  0.000 rotate show grestore 1 -1 scale
/Times-Italic findfont 18.000 scalefont setfont
394 209 moveto
1 -1 scale
(5) gsave  0.000 rotate show grestore 1 -1 scale
/Times-Italic findfont 18.000 scalefont setfont
474 209 moveto
1 -1 scale
(6) gsave  0.000 rotate show grestore 1 -1 scale
/Times-Italic findfont 18.000 scalefont setfont
554 209 moveto
1 -1 scale
(7) gsave  0.000 rotate show grestore 1 -1 scale
/Times-Italic findfont 18.000 scalefont setfont
634 209 moveto
1 -1 scale
(8) gsave  0.000 rotate show grestore 1 -1 scale
/Times-Italic findfont 18.000 scalefont setfont
114 269 moveto
1 -1 scale
(a) gsave  0.000 rotate show grestore 1 -1 scale
/Times-Italic findfont 18.000 scalefont setfont
274 269 moveto
1 -1 scale
(b) gsave  0.000 rotate show grestore 1 -1 scale
/Times-Italic findfont 18.000 scalefont setfont
434 269 moveto
1 -1 scale
(c) gsave  0.000 rotate show grestore 1 -1 scale
/Times-Italic findfont 18.000 scalefont setfont
594 269 moveto
1 -1 scale
(d) gsave  0.000 rotate show grestore 1 -1 scale
% Polyline
newpath 239 424 moveto 239 494 lineto stroke
gsave  0.000 setgray fill grestore stroke
% Polyline
newpath 204 459 moveto 274 459 lineto stroke
gsave  0.000 setgray fill grestore stroke
/Times-Italic findfont 18.000 scalefont setfont
244 454 moveto
1 -1 scale
(v) gsave  0.000 rotate show grestore 1 -1 scale
% Polyline
newpath 359 494 moveto 359 424 lineto stroke
gsave  0.000 setgray fill grestore stroke
% Polyline
newpath 324 459 moveto 394 459 lineto stroke
gsave  0.000 setgray fill grestore stroke
/Times-Italic findfont 18.000 scalefont setfont
354 454 moveto
1 -1 scale
(-v) gsave 90.012 rotate show grestore 1 -1 scale
/Times-Italic findfont 18.000 scalefont setfont
294 464 moveto
1 -1 scale
(=) gsave  0.000 rotate show grestore 1 -1 scale
$F2psEnd

%%%%%%%%%%%%%%%%%%%%%%% Fig 3 %%%%%%%%%%%%%%%%%%%%%%%%%%%%%%%%%%%
%!
%%Title: /tmp/xfig-export004082
%%Creator: fig2dev
%%CreationDate: Tue Jun  8 16:42:01 1993
%%For: kluemper@sun12 (Andreas Kluemper)
%%Pages: 0
%%BoundingBox: 0 0 441 621
%%EndComments
/$F2psDict 32 dict def
$F2psDict begin
	$F2psDict /mtrx matrix put

	end
	/$F2psBegin {$F2psDict begin /$F2psEnteredState save def} def
	/$F2psEnd {$F2psEnteredState restore end} def

$F2psBegin
1 setlinecap 1 setlinejoin
-4 9 translate
0.000000 621.000000 translate 0.900 -0.900 scale
612 0 translate 90 rotate
1.000 setlinewidth
% Polyline
newpath 119 24 moveto 119 4 lineto stroke
stroke
% Polyline
newpath 114 19 moveto 124 29 lineto stroke
stroke
% Polyline
newpath 114 29 moveto 124 19 lineto stroke
stroke
% Polyline
newpath 519 24 moveto 519 4 lineto stroke
stroke
% Polyline
newpath 514 19 moveto 524 29 lineto stroke
stroke
% Polyline
newpath 514 29 moveto 524 19 lineto stroke
stroke
% Polyline
newpath 439 24 moveto 439 4 lineto stroke
stroke
% Polyline
newpath 434 19 moveto 444 29 lineto stroke
stroke
% Polyline
newpath 434 29 moveto 444 19 lineto stroke
stroke
% Polyline
newpath 279 24 moveto 279 4 lineto stroke
stroke
% Polyline
newpath 274 19 moveto 284 29 lineto stroke
stroke
% Polyline
newpath 274 29 moveto 284 19 lineto stroke
stroke
% Polyline
newpath 199 24 moveto 199 4 lineto stroke
stroke
% Polyline
newpath 194 19 moveto 204 29 lineto stroke
stroke
% Polyline
newpath 194 29 moveto 204 19 lineto stroke
stroke
/Times-Italic findfont 30.000 scalefont setfont
139 369 moveto
1 -1 scale
(v) gsave  0.000 rotate show grestore 1 -1 scale
/Times-Italic findfont 20.000 scalefont setfont
149 379 moveto
1 -1 scale
(o) gsave  0.000 rotate show grestore 1 -1 scale
/Times-Italic findfont 20.000 scalefont setfont
124 369 moveto
1 -1 scale
(+) gsave  0.000 rotate show grestore 1 -1 scale
/Times-Italic findfont 30.000 scalefont setfont
219 369 moveto
1 -1 scale
(v) gsave  0.000 rotate show grestore 1 -1 scale
/Times-Italic findfont 20.000 scalefont setfont
229 379 moveto
1 -1 scale
(o) gsave  0.000 rotate show grestore 1 -1 scale
/Times-Italic findfont 20.000 scalefont setfont
204 369 moveto
1 -1 scale
(+) gsave  0.000 rotate show grestore 1 -1 scale
/Times-Italic findfont 30.000 scalefont setfont
299 369 moveto
1 -1 scale
(v) gsave  0.000 rotate show grestore 1 -1 scale
/Times-Italic findfont 20.000 scalefont setfont
309 379 moveto
1 -1 scale
(o) gsave  0.000 rotate show grestore 1 -1 scale
/Times-Italic findfont 20.000 scalefont setfont
284 369 moveto
1 -1 scale
(+) gsave  0.000 rotate show grestore 1 -1 scale
/Times-Italic findfont 30.000 scalefont setfont
379 369 moveto
1 -1 scale
(v) gsave  0.000 rotate show grestore 1 -1 scale
/Times-Italic findfont 20.000 scalefont setfont
389 379 moveto
1 -1 scale
(o) gsave  0.000 rotate show grestore 1 -1 scale
/Times-Italic findfont 20.000 scalefont setfont
364 369 moveto
1 -1 scale
(+) gsave  0.000 rotate show grestore 1 -1 scale
/Times-Italic findfont 30.000 scalefont setfont
459 369 moveto
1 -1 scale
(v) gsave  0.000 rotate show grestore 1 -1 scale
/Times-Italic findfont 20.000 scalefont setfont
469 379 moveto
1 -1 scale
(o) gsave  0.000 rotate show grestore 1 -1 scale
/Times-Italic findfont 20.000 scalefont setfont
444 369 moveto
1 -1 scale
(+) gsave  0.000 rotate show grestore 1 -1 scale
/Times-Italic findfont 30.000 scalefont setfont
139 209 moveto
1 -1 scale
(v) gsave  0.000 rotate show grestore 1 -1 scale
/Times-Italic findfont 20.000 scalefont setfont
149 219 moveto
1 -1 scale
(o) gsave  0.000 rotate show grestore 1 -1 scale
/Times-Italic findfont 20.000 scalefont setfont
124 209 moveto
1 -1 scale
(+) gsave  0.000 rotate show grestore 1 -1 scale
/Times-Italic findfont 30.000 scalefont setfont
219 209 moveto
1 -1 scale
(v) gsave  0.000 rotate show grestore 1 -1 scale
/Times-Italic findfont 20.000 scalefont setfont
229 219 moveto
1 -1 scale
(o) gsave  0.000 rotate show grestore 1 -1 scale
/Times-Italic findfont 20.000 scalefont setfont
204 209 moveto
1 -1 scale
(+) gsave  0.000 rotate show grestore 1 -1 scale
/Times-Italic findfont 30.000 scalefont setfont
299 209 moveto
1 -1 scale
(v) gsave  0.000 rotate show grestore 1 -1 scale
/Times-Italic findfont 20.000 scalefont setfont
309 219 moveto
1 -1 scale
(o) gsave  0.000 rotate show grestore 1 -1 scale
/Times-Italic findfont 20.000 scalefont setfont
284 209 moveto
1 -1 scale
(+) gsave  0.000 rotate show grestore 1 -1 scale
/Times-Italic findfont 30.000 scalefont setfont
379 209 moveto
1 -1 scale
(v) gsave  0.000 rotate show grestore 1 -1 scale
/Times-Italic findfont 20.000 scalefont setfont
389 219 moveto
1 -1 scale
(o) gsave  0.000 rotate show grestore 1 -1 scale
/Times-Italic findfont 20.000 scalefont setfont
364 209 moveto
1 -1 scale
(+) gsave  0.000 rotate show grestore 1 -1 scale
/Times-Italic findfont 30.000 scalefont setfont
459 209 moveto
1 -1 scale
(v) gsave  0.000 rotate show grestore 1 -1 scale
/Times-Italic findfont 20.000 scalefont setfont
469 219 moveto
1 -1 scale
(o) gsave  0.000 rotate show grestore 1 -1 scale
/Times-Italic findfont 20.000 scalefont setfont
444 209 moveto
1 -1 scale
(+) gsave  0.000 rotate show grestore 1 -1 scale
/Times-Italic findfont 30.000 scalefont setfont
138 125 moveto
1 -1 scale
(v) gsave  0.000 rotate show grestore 1 -1 scale
/Times-Italic findfont 20.000 scalefont setfont
148 135 moveto
1 -1 scale
(o) gsave  0.000 rotate show grestore 1 -1 scale
% Polyline
newpath 124 119 moveto 134 119 lineto stroke
gsave  0.000 setgray fill grestore stroke
/Times-Italic findfont 30.000 scalefont setfont
218 125 moveto
1 -1 scale
(v) gsave  0.000 rotate show grestore 1 -1 scale
/Times-Italic findfont 20.000 scalefont setfont
228 135 moveto
1 -1 scale
(o) gsave  0.000 rotate show grestore 1 -1 scale
% Polyline
newpath 204 119 moveto 214 119 lineto stroke
gsave  0.000 setgray fill grestore stroke
/Times-Italic findfont 30.000 scalefont setfont
298 125 moveto
1 -1 scale
(v) gsave  0.000 rotate show grestore 1 -1 scale
/Times-Italic findfont 20.000 scalefont setfont
308 135 moveto
1 -1 scale
(o) gsave  0.000 rotate show grestore 1 -1 scale
% Polyline
newpath 284 119 moveto 294 119 lineto stroke
gsave  0.000 setgray fill grestore stroke
/Times-Italic findfont 30.000 scalefont setfont
378 125 moveto
1 -1 scale
(v) gsave  0.000 rotate show grestore 1 -1 scale
/Times-Italic findfont 20.000 scalefont setfont
388 135 moveto
1 -1 scale
(o) gsave  0.000 rotate show grestore 1 -1 scale
% Polyline
newpath 364 119 moveto 374 119 lineto stroke
gsave  0.000 setgray fill grestore stroke
/Times-Italic findfont 30.000 scalefont setfont
458 125 moveto
1 -1 scale
(v) gsave  0.000 rotate show grestore 1 -1 scale
/Times-Italic findfont 20.000 scalefont setfont
468 135 moveto
1 -1 scale
(o) gsave  0.000 rotate show grestore 1 -1 scale
% Polyline
newpath 444 119 moveto 454 119 lineto stroke
gsave  0.000 setgray fill grestore stroke
/Times-Italic findfont 30.000 scalefont setfont
138 285 moveto
1 -1 scale
(v) gsave  0.000 rotate show grestore 1 -1 scale
/Times-Italic findfont 20.000 scalefont setfont
148 295 moveto
1 -1 scale
(o) gsave  0.000 rotate show grestore 1 -1 scale
% Polyline
newpath 124 279 moveto 134 279 lineto stroke
gsave  0.000 setgray fill grestore stroke
/Times-Italic findfont 30.000 scalefont setfont
218 285 moveto
1 -1 scale
(v) gsave  0.000 rotate show grestore 1 -1 scale
/Times-Italic findfont 20.000 scalefont setfont
228 295 moveto
1 -1 scale
(o) gsave  0.000 rotate show grestore 1 -1 scale
% Polyline
newpath 204 279 moveto 214 279 lineto stroke
gsave  0.000 setgray fill grestore stroke
/Times-Italic findfont 30.000 scalefont setfont
298 285 moveto
1 -1 scale
(v) gsave  0.000 rotate show grestore 1 -1 scale
/Times-Italic findfont 20.000 scalefont setfont
308 295 moveto
1 -1 scale
(o) gsave  0.000 rotate show grestore 1 -1 scale
% Polyline
newpath 284 279 moveto 294 279 lineto stroke
gsave  0.000 setgray fill grestore stroke
/Times-Italic findfont 30.000 scalefont setfont
378 285 moveto
1 -1 scale
(v) gsave  0.000 rotate show grestore 1 -1 scale
/Times-Italic findfont 20.000 scalefont setfont
388 295 moveto
1 -1 scale
(o) gsave  0.000 rotate show grestore 1 -1 scale
% Polyline
newpath 364 279 moveto 374 279 lineto stroke
gsave  0.000 setgray fill grestore stroke
/Times-Italic findfont 30.000 scalefont setfont
458 285 moveto
1 -1 scale
(v) gsave  0.000 rotate show grestore 1 -1 scale
/Times-Italic findfont 20.000 scalefont setfont
468 295 moveto
1 -1 scale
(o) gsave  0.000 rotate show grestore 1 -1 scale
% Polyline
newpath 444 279 moveto 454 279 lineto stroke
gsave  0.000 setgray fill grestore stroke
/Times-Italic findfont 30.000 scalefont setfont
645 125 moveto
1 -1 scale
(v) gsave 90.012 rotate show grestore 1 -1 scale
/Times-Italic findfont 20.000 scalefont setfont
655 115 moveto
1 -1 scale
(o) gsave 90.012 rotate show grestore 1 -1 scale
/Times-Italic findfont 20.000 scalefont setfont
645 140 moveto
1 -1 scale
(+) gsave 90.012 rotate show grestore 1 -1 scale
/Times-Italic findfont 30.000 scalefont setfont
645 205 moveto
1 -1 scale
(v) gsave 90.012 rotate show grestore 1 -1 scale
/Times-Italic findfont 20.000 scalefont setfont
655 195 moveto
1 -1 scale
(o) gsave 90.012 rotate show grestore 1 -1 scale
% Polyline
newpath 639 219 moveto 639 209 lineto stroke
gsave  0.000 setgray fill grestore stroke
/Times-Italic findfont 30.000 scalefont setfont
645 365 moveto
1 -1 scale
(v) gsave 90.012 rotate show grestore 1 -1 scale
/Times-Italic findfont 20.000 scalefont setfont
655 355 moveto
1 -1 scale
(o) gsave 90.012 rotate show grestore 1 -1 scale
% Polyline
newpath 639 379 moveto 639 369 lineto stroke
gsave  0.000 setgray fill grestore stroke
/Times-Italic findfont 30.000 scalefont setfont
645 285 moveto
1 -1 scale
(v) gsave 90.012 rotate show grestore 1 -1 scale
/Times-Italic findfont 20.000 scalefont setfont
655 275 moveto
1 -1 scale
(o) gsave 90.012 rotate show grestore 1 -1 scale
/Times-Italic findfont 20.000 scalefont setfont
645 300 moveto
1 -1 scale
(+) gsave 90.012 rotate show grestore 1 -1 scale
% Polyline
newpath 79 144 moveto 559 144 lineto stroke
gsave  0.000 setgray fill grestore stroke
% Polyline
newpath 79 64 moveto 559 64 lineto stroke
gsave  0.000 setgray fill grestore stroke
% Polyline
newpath 79 224 moveto 559 224 lineto stroke
gsave  0.000 setgray fill grestore stroke
% Polyline
newpath 79 304 moveto 559 304 lineto stroke
gsave  0.000 setgray fill grestore stroke
% Polyline
newpath 79 384 moveto 559 384 lineto stroke
gsave  0.000 setgray fill grestore stroke
% Polyline
newpath 119 24 moveto 119 424 lineto stroke
gsave  0.000 setgray fill grestore stroke
% Polyline
newpath 199 24 moveto 199 424 lineto stroke
gsave  0.000 setgray fill grestore stroke
% Polyline
newpath 279 24 moveto 279 424 lineto stroke
gsave  0.000 setgray fill grestore stroke
% Polyline
newpath 439 24 moveto 439 424 lineto stroke
gsave  0.000 setgray fill grestore stroke
% Polyline
newpath 519 24 moveto 519 424 lineto stroke
gsave  0.000 setgray fill grestore stroke
% Polyline
newpath 19 254 moveto 19 424 lineto stroke
gsave  0.000 setgray fill grestore stroke
newpath 21.000 416.000 moveto 19.000 424.000 lineto 17.000 416.000 lineto
stroke
% Polyline
newpath 349 484 moveto 559 484 lineto stroke
gsave  0.000 setgray fill grestore stroke
newpath 551.000 482.000 moveto 559.000 484.000 lineto 551.000 486.000 lineto
stroke
% Polyline
newpath 289 484 moveto 79 484 lineto stroke
gsave  0.000 setgray fill grestore stroke
newpath 87.000 486.000 moveto 79.000 484.000 lineto 87.000 482.000 lineto
stroke
% Polyline
newpath 19 204 moveto 19 4 lineto stroke
stroke
newpath 17.000 12.000 moveto 19.000 4.000 lineto 21.000 12.000 lineto stroke
% Polyline
newpath 620 65 moveto 700 65 lineto stroke
gsave  0.000 setgray fill grestore stroke
% Polyline
newpath 620 145 moveto 700 145 lineto stroke
gsave  0.000 setgray fill grestore stroke
% Polyline
newpath 620 225 moveto 700 225 lineto stroke
gsave  0.000 setgray fill grestore stroke
% Polyline
newpath 620 305 moveto 700 305 lineto stroke
gsave  0.000 setgray fill grestore stroke
% Polyline
newpath 620 385 moveto 700 385 lineto stroke
gsave  0.000 setgray fill grestore stroke
3.000 setlinewidth
% Polyline
newpath 354 19 moveto 364 29 lineto stroke
stroke
% Polyline
newpath 354 29 moveto 364 19 lineto stroke
stroke
% Polyline
newpath 654 19 moveto 664 29 lineto stroke
stroke
% Polyline
newpath 654 29 moveto 664 19 lineto stroke
stroke
% Polyline
newpath 359 4 moveto 359 424 lineto stroke
stroke
% Polyline
newpath 659 4 moveto 659 424 lineto stroke
stroke
/Times-Italic findfont 24.000 scalefont setfont
314 494 moveto
1 -1 scale
(L) gsave  0.000 rotate show grestore 1 -1 scale
/Times-Italic findfont 24.000 scalefont setfont
9 234 moveto
1 -1 scale
(N) gsave  0.000 rotate show grestore 1 -1 scale
$F2psEnd

%%%%%%%%%%%%%%%%%%%%%%% Fig 4a %%%%%%%%%%%%%%%%%%%%%%%%%%%%%%%%%%%

%!
save
/w {setlinewidth} def
/b {lineto} def
/n {newpath} def
.25 .25 scale
1 setlinejoin
n
 stroke
 0.0039 setgray
 stroke
00004 w
stroke
n
 1410 0811 moveto
  1410 0732 b
 stroke
 0.0039 setgray
 stroke
00004 w
stroke
n
 1413 0811 moveto
  1413 0732 b
 stroke
 0.0039 setgray
 stroke
00004 w
stroke
n
 1387 0811 moveto
  1383 0789 b
  1383 0811 b
  1440 0811 b
  1440 0789 b
  1436 0811 b
 stroke
 0.0039 setgray
 stroke
00004 w
stroke
n
 1399 0732 moveto
  1425 0732 b
 stroke
 0.0039 setgray
 stroke
00004 w
stroke
n
 0585 0985 moveto
  0577 0983 b
  0573 0976 b
  0570 0964 b
  0570 0956 b
  0573 0944 b
  0577 0937 b
  0585 0935 b
  0589 0935 b
  0597 0937 b
  0601 0944 b
  0604 0956 b
  0604 0964 b
  0601 0976 b
  0597 0983 b
  0589 0985 b
  0585 0985 b
 stroke
 0.0039 setgray
 stroke
00004 w
stroke
n
 0585 0985 moveto
  0580 0983 b
  0577 0980 b
  0575 0976 b
  0573 0964 b
  0573 0956 b
  0575 0944 b
  0577 0940 b
  0580 0937 b
  0585 0935 b
 stroke
 0.0039 setgray
 stroke
00004 w
stroke
n
 0589 0935 moveto
  0594 0937 b
  0597 0940 b
  0599 0944 b
  0601 0956 b
  0601 0964 b
  0599 0976 b
  0597 0980 b
  0594 0983 b
  0589 0985 b
 stroke
 0.0039 setgray
 stroke
00004 w
stroke
n
 0637 0956 moveto
  0684 0956 b
 stroke
 0.0039 setgray
 stroke
00004 w
stroke
n
 0668 1150 moveto
  0637 1150 b
 stroke
 0.0039 setgray
 stroke
00004 w
stroke
n
 0537 1327 moveto
  0534 1324 b
  0537 1322 b
  0539 1324 b
  0537 1327 b
 stroke
 0.0039 setgray
 stroke
00004 w
stroke
n
 0577 1363 moveto
  0582 1365 b
  0589 1372 b
  0589 1322 b
 stroke
 0.0039 setgray
 stroke
00004 w
stroke
n
 0587 1370 moveto
  0587 1322 b
 stroke
 0.0039 setgray
 stroke
00004 w
stroke
n
 0577 1322 moveto
  0599 1322 b
 stroke
 0.0039 setgray
 stroke
00004 w
stroke
n
 0637 1344 moveto
  0684 1344 b
 stroke
 0.0039 setgray
 stroke
00004 w
stroke
n
 0668 1537 moveto
  0637 1537 b
 stroke
 0.0039 setgray
 stroke
00004 w
stroke
n
 0537 1714 moveto
  0534 1711 b
  0537 1709 b
  0539 1711 b
  0537 1714 b
 stroke
 0.0039 setgray
 stroke
00004 w
stroke
n
 0573 1750 moveto
  0575 1747 b
  0573 1745 b
  0570 1747 b
  0570 1750 b
  0573 1755 b
  0575 1757 b
  0582 1759 b
  0592 1759 b
  0599 1757 b
  0601 1755 b
  0604 1750 b
  0604 1745 b
  0601 1740 b
  0594 1735 b
  0582 1731 b
  0577 1728 b
  0573 1723 b
  0570 1716 b
  0570 1709 b
 stroke
 0.0039 setgray
 stroke
00004 w
stroke
n
 0592 1759 moveto
  0597 1757 b
  0599 1755 b
  0601 1750 b
  0601 1745 b
  0599 1740 b
  0592 1735 b
  0582 1731 b
 stroke
 0.0039 setgray
 stroke
00004 w
stroke
n
 0570 1714 moveto
  0573 1716 b
  0577 1716 b
  0589 1711 b
  0597 1711 b
  0601 1714 b
  0604 1716 b
 stroke
 0.0039 setgray
 stroke
00004 w
stroke
n
 0577 1716 moveto
  0589 1709 b
  0599 1709 b
  0601 1711 b
  0604 1716 b
  0604 1721 b
 stroke
 0.0039 setgray
 stroke
00004 w
stroke
n
 0637 1731 moveto
  0684 1731 b
 stroke
 0.0039 setgray
 stroke
00004 w
stroke
n
 0668 1924 moveto
  0637 1924 b
 stroke
 0.0039 setgray
 stroke
00004 w
stroke
n
 0537 2101 moveto
  0534 2099 b
  0537 2096 b
  0539 2099 b
  0537 2101 b
 stroke
 0.0039 setgray
 stroke
00004 w
stroke
n
 0573 2139 moveto
  0575 2137 b
  0573 2135 b
  0570 2137 b
  0570 2139 b
  0575 2144 b
  0582 2147 b
  0592 2147 b
  0599 2144 b
  0601 2139 b
  0601 2132 b
  0599 2127 b
  0592 2125 b
  0585 2125 b
 stroke
 0.0039 setgray
 stroke
00004 w
stroke
n
 0592 2147 moveto
  0597 2144 b
  0599 2139 b
  0599 2132 b
  0597 2127 b
  0592 2125 b
 stroke
 0.0039 setgray
 stroke
00004 w
stroke
n
 0592 2125 moveto
  0597 2123 b
  0601 2118 b
  0604 2113 b
  0604 2106 b
  0601 2101 b
  0599 2099 b
  0592 2096 b
  0582 2096 b
  0575 2099 b
  0573 2101 b
  0570 2106 b
  0570 2108 b
  0573 2111 b
  0575 2108 b
  0573 2106 b
 stroke
 0.0039 setgray
 stroke
00004 w
stroke
n
 0599 2120 moveto
  0601 2113 b
  0601 2106 b
  0599 2101 b
  0597 2099 b
  0592 2096 b
 stroke
 0.0039 setgray
 stroke
00004 w
stroke
n
 0637 2118 moveto
  0684 2118 b
 stroke
 0.0039 setgray
 stroke
00004 w
stroke
n
 0668 2311 moveto
  0637 2311 b
 stroke
 0.0039 setgray
 stroke
00004 w
stroke
n
 0537 2488 moveto
  0534 2486 b
  0537 2483 b
  0539 2486 b
  0537 2488 b
 stroke
 0.0039 setgray
 stroke
00004 w
stroke
n
 0592 2529 moveto
  0592 2483 b
 stroke
 0.0039 setgray
 stroke
00004 w
stroke
n
 0594 2534 moveto
  0594 2483 b
 stroke
 0.0039 setgray
 stroke
00004 w
stroke
n
 0594 2534 moveto
  0568 2498 b
  0606 2498 b
 stroke
 0.0039 setgray
 stroke
00004 w
stroke
n
 0585 2483 moveto
  0601 2483 b
 stroke
 0.0039 setgray
 stroke
00004 w
stroke
n
 0637 2505 moveto
  0684 2505 b
 stroke
 0.0039 setgray
 stroke
00004 w
stroke
n
 0637 0956 moveto
  0637 2505 b
 stroke
 0.0039 setgray
 stroke
00004 w
stroke
n
 2186 0956 moveto
  2139 0956 b
 stroke
 0.0039 setgray
 stroke
00004 w
stroke
n
 2155 1150 moveto
  2186 1150 b
 stroke
 0.0039 setgray
 stroke
00004 w
stroke
n
 2186 1344 moveto
  2139 1344 b
 stroke
 0.0039 setgray
 stroke
00004 w
stroke
n
 2155 1537 moveto
  2186 1537 b
 stroke
 0.0039 setgray
 stroke
00004 w
stroke
n
 2186 1731 moveto
  2139 1731 b
 stroke
 0.0039 setgray
 stroke
00004 w
stroke
n
 2155 1924 moveto
  2186 1924 b
 stroke
 0.0039 setgray
 stroke
00004 w
stroke
n
 2186 2118 moveto
  2139 2118 b
 stroke
 0.0039 setgray
 stroke
00004 w
stroke
n
 2155 2311 moveto
  2186 2311 b
 stroke
 0.0039 setgray
 stroke
00004 w
stroke
n
 2186 2505 moveto
  2139 2505 b
 stroke
 0.0039 setgray
 stroke
00004 w
stroke
n
 2186 2505 moveto
  2186 0956 b
 stroke
 0.0039 setgray
 stroke
00004 w
stroke
n
 0635 0929 moveto
  0628 0926 b
  0623 0919 b
  0621 0907 b
  0621 0900 b
  0623 0888 b
  0628 0881 b
  0635 0878 b
  0640 0878 b
  0647 0881 b
  0652 0888 b
  0654 0900 b
  0654 0907 b
  0652 0919 b
  0647 0926 b
  0640 0929 b
  0635 0929 b
 stroke
 0.0039 setgray
 stroke
00004 w
stroke
n
 0635 0929 moveto
  0630 0926 b
  0628 0924 b
  0625 0919 b
  0623 0907 b
  0623 0900 b
  0625 0888 b
  0628 0883 b
  0630 0881 b
 stroke
 0.0039 setgray
 stroke
00004 w
stroke
n
 0630 0881 moveto
  0635 0878 b
 stroke
 0.0039 setgray
 stroke
00004 w
stroke
n
 0640 0878 moveto
  0645 0881 b
  0647 0883 b
  0649 0888 b
  0652 0900 b
  0652 0907 b
  0649 0919 b
  0647 0924 b
  0645 0926 b
  0640 0929 b
 stroke
 0.0039 setgray
 stroke
00004 w
stroke
n
 0637 0956 moveto
  0637 1003 b
 stroke
 0.0039 setgray
 stroke
00004 w
stroke
n
 0682 0987 moveto
  0682 0956 b
 stroke
 0.0039 setgray
 stroke
00004 w
stroke
n
 0726 0956 moveto
  0726 0987 b
 stroke
 0.0039 setgray
 stroke
00004 w
stroke
n
 0770 0987 moveto
  0770 0956 b
 stroke
 0.0039 setgray
 stroke
00004 w
stroke
n
 0814 0956 moveto
  0814 0987 b
 stroke
 0.0039 setgray
 stroke
00004 w
stroke
n
 0859 1003 moveto
  0859 0956 b
 stroke
 0.0039 setgray
 stroke
00004 w
stroke
n
 0833 0883 moveto
  0831 0881 b
  0833 0878 b
  0836 0881 b
  0833 0883 b
 stroke
 0.0039 setgray
 stroke
00004 w
stroke
n
 0872 0929 moveto
  0867 0905 b
 stroke
 0.0039 setgray
 stroke
00004 w
stroke
n
 0867 0905 moveto
  0872 0909 b
  0879 0912 b
  0886 0912 b
  0893 0909 b
  0898 0905 b
  0901 0897 b
  0901 0893 b
  0898 0885 b
  0893 0881 b
  0886 0878 b
  0879 0878 b
  0872 0881 b
  0869 0883 b
  0867 0888 b
  0867 0890 b
  0869 0893 b
  0872 0890 b
  0869 0888 b
 stroke
 0.0039 setgray
 stroke
00004 w
stroke
n
 0886 0912 moveto
  0891 0909 b
  0896 0905 b
  0898 0897 b
  0898 0893 b
  0896 0885 b
  0891 0881 b
  0886 0878 b
 stroke
 0.0039 setgray
 stroke
00004 w
stroke
n
 0872 0929 moveto
  0896 0929 b
 stroke
 0.0039 setgray
 stroke
00004 w
stroke
n
 0872 0926 moveto
  0884 0926 b
  0896 0929 b
 stroke
 0.0039 setgray
 stroke
00004 w
stroke
n
 0903 0956 moveto
  0903 0987 b
 stroke
 0.0039 setgray
 stroke
00004 w
stroke
n
 0947 0987 moveto
  0947 0956 b
 stroke
 0.0039 setgray
 stroke
00004 w
stroke
n
 0991 0956 moveto
  0991 0987 b
 stroke
 0.0039 setgray
 stroke
00004 w
stroke
n
 1036 0987 moveto
  1036 0956 b
 stroke
 0.0039 setgray
 stroke
00004 w
stroke
n
 1020 0919 moveto
  1025 0921 b
  1032 0929 b
  1032 0878 b
 stroke
 0.0039 setgray
 stroke
00004 w
stroke
n
 1029 0926 moveto
  1029 0878 b
 stroke
 0.0039 setgray
 stroke
00004 w
stroke
n
 1020 0878 moveto
  1041 0878 b
 stroke
 0.0039 setgray
 stroke
00004 w
stroke
n
 1080 0883 moveto
  1077 0881 b
  1080 0878 b
  1082 0881 b
  1080 0883 b
 stroke
 0.0039 setgray
 stroke
00004 w
stroke
n
 1128 0929 moveto
  1121 0926 b
  1116 0919 b
  1113 0907 b
  1113 0900 b
  1116 0888 b
  1121 0881 b
  1128 0878 b
  1133 0878 b
  1140 0881 b
  1145 0888 b
  1147 0900 b
  1147 0907 b
  1145 0919 b
  1140 0926 b
  1133 0929 b
  1128 0929 b
 stroke
 0.0039 setgray
 stroke
00004 w
stroke
n
 1128 0929 moveto
  1123 0926 b
  1121 0924 b
  1118 0919 b
  1116 0907 b
  1116 0900 b
  1118 0888 b
  1121 0883 b
  1123 0881 b
  1128 0878 b
 stroke
 0.0039 setgray
 stroke
00004 w
stroke
n
 1133 0878 moveto
  1137 0881 b
  1140 0883 b
  1142 0888 b
  1145 0900 b
  1145 0907 b
  1142 0919 b
  1140 0924 b
  1137 0926 b
  1133 0929 b
 stroke
 0.0039 setgray
 stroke
00004 w
stroke
n
 1080 0956 moveto
  1080 1003 b
 stroke
 0.0039 setgray
 stroke
00004 w
stroke
n
 1124 0987 moveto
  1124 0956 b
 stroke
 0.0039 setgray
 stroke
00004 w
stroke
n
 1168 0956 moveto
  1168 0987 b
 stroke
 0.0039 setgray
 stroke
00004 w
stroke
n
 1213 0987 moveto
  1213 0956 b
 stroke
 0.0039 setgray
 stroke
00004 w
stroke
n
 1257 0956 moveto
  1257 0987 b
 stroke
 0.0039 setgray
 stroke
00004 w
stroke
n
 1301 1003 moveto
  1301 0956 b
 stroke
 0.0039 setgray
 stroke
00004 w
stroke
n
 1241 0919 moveto
  1246 0921 b
  1253 0929 b
  1253 0878 b
 stroke
 0.0039 setgray
 stroke
00004 w
stroke
n
 1251 0926 moveto
  1251 0878 b
 stroke
 0.0039 setgray
 stroke
00004 w
stroke
n
 1241 0878 moveto
  1263 0878 b
 stroke
 0.0039 setgray
 stroke
00004 w
stroke
n
 1301 0883 moveto
  1299 0881 b
  1301 0878 b
  1303 0881 b
  1301 0883 b
 stroke
 0.0039 setgray
 stroke
00004 w
stroke
n
 1339 0929 moveto
  1335 0905 b
 stroke
 0.0039 setgray
 stroke
00004 w
stroke
n
 1335 0905 moveto
  1339 0909 b
  1347 0912 b
  1354 0912 b
  1361 0909 b
  1366 0905 b
  1368 0897 b
  1368 0893 b
  1366 0885 b
  1361 0881 b
  1354 0878 b
  1347 0878 b
  1339 0881 b
  1337 0883 b
  1335 0888 b
  1335 0890 b
  1337 0893 b
  1339 0890 b
  1337 0888 b
 stroke
 0.0039 setgray
 stroke
00004 w
stroke
n
 1354 0912 moveto
  1359 0909 b
  1363 0905 b
  1366 0897 b
  1366 0893 b
  1363 0885 b
  1359 0881 b
  1354 0878 b
 stroke
 0.0039 setgray
 stroke
00004 w
stroke
n
 1339 0929 moveto
  1363 0929 b
 stroke
 0.0039 setgray
 stroke
00004 w
stroke
n
 1339 0926 moveto
  1351 0926 b
  1363 0929 b
 stroke
 0.0039 setgray
 stroke
00004 w
stroke
n
 1345 0956 moveto
  1345 0987 b
 stroke
 0.0039 setgray
 stroke
00004 w
stroke
n
 1390 0987 moveto
  1390 0956 b
 stroke
 0.0039 setgray
 stroke
00004 w
stroke
n
 1434 0956 moveto
  1434 0987 b
 stroke
 0.0039 setgray
 stroke
00004 w
stroke
n
 1478 0987 moveto
  1478 0956 b
 stroke
 0.0039 setgray
 stroke
00004 w
stroke
n
 1457 0919 moveto
  1460 0917 b
  1457 0914 b
  1455 0917 b
  1455 0919 b
  1457 0924 b
  1460 0926 b
  1467 0929 b
  1477 0929 b
  1484 0926 b
  1486 0924 b
  1489 0919 b
  1489 0914 b
  1486 0909 b
  1479 0905 b
  1467 0900 b
  1462 0897 b
  1457 0893 b
  1455 0885 b
  1455 0878 b
 stroke
 0.0039 setgray
 stroke
00004 w
stroke
n
 1477 0929 moveto
  1481 0926 b
  1484 0924 b
  1486 0919 b
  1486 0914 b
  1484 0909 b
  1477 0905 b
  1467 0900 b
 stroke
 0.0039 setgray
 stroke
00004 w
stroke
n
 1455 0883 moveto
  1457 0885 b
  1462 0885 b
  1474 0881 b
  1481 0881 b
  1486 0883 b
  1489 0885 b
 stroke
 0.0039 setgray
 stroke
00004 w
stroke
n
 1462 0885 moveto
  1474 0878 b
  1484 0878 b
  1486 0881 b
  1489 0885 b
  1489 0890 b
 stroke
 0.0039 setgray
 stroke
00004 w
stroke
n
 1522 0883 moveto
  1520 0881 b
  1522 0878 b
  1525 0881 b
  1522 0883 b
 stroke
 0.0039 setgray
 stroke
00004 w
stroke
n
 1570 0929 moveto
  1563 0926 b
  1558 0919 b
  1556 0907 b
  1556 0900 b
  1558 0888 b
  1563 0881 b
  1570 0878 b
  1575 0878 b
  1582 0881 b
  1587 0888 b
  1589 0900 b
  1589 0907 b
  1587 0919 b
  1582 0926 b
  1575 0929 b
  1570 0929 b
 stroke
 0.0039 setgray
 stroke
00004 w
stroke
n
 1570 0929 moveto
  1566 0926 b
  1563 0924 b
  1561 0919 b
  1558 0907 b
  1558 0900 b
  1561 0888 b
  1563 0883 b
  1566 0881 b
  1570 0878 b
 stroke
 0.0039 setgray
 stroke
00004 w
stroke
n
 1575 0878 moveto
  1580 0881 b
  1582 0883 b
  1585 0888 b
  1587 0900 b
  1587 0907 b
  1585 0919 b
  1582 0924 b
  1580 0926 b
  1575 0929 b
 stroke
 0.0039 setgray
 stroke
00004 w
stroke
n
 1522 0956 moveto
  1522 1003 b
 stroke
 0.0039 setgray
 stroke
00004 w
stroke
n
 1567 0987 moveto
  1567 0956 b
 stroke
 0.0039 setgray
 stroke
00004 w
stroke
n
 1611 0956 moveto
  1611 0987 b
 stroke
 0.0039 setgray
 stroke
00004 w
stroke
n
 1655 0987 moveto
  1655 0956 b
 stroke
 0.0039 setgray
 stroke
00004 w
stroke
n
 1699 0956 moveto
  1699 0987 b
 stroke
 0.0039 setgray
 stroke
00004 w
stroke
n
 1743 1003 moveto
  1743 0956 b
 stroke
 0.0039 setgray
 stroke
00004 w
stroke
n
 1679 0919 moveto
  1681 0917 b
  1679 0914 b
  1676 0917 b
  1676 0919 b
  1679 0924 b
  1681 0926 b
  1688 0929 b
  1698 0929 b
  1705 0926 b
  1707 0924 b
  1710 0919 b
  1710 0914 b
  1707 0909 b
  1700 0905 b
  1688 0900 b
  1683 0897 b
  1679 0893 b
  1676 0885 b
  1676 0878 b
 stroke
 0.0039 setgray
 stroke
00004 w
stroke
n
 1698 0929 moveto
  1703 0926 b
  1705 0924 b
  1707 0919 b
  1707 0914 b
  1705 0909 b
  1698 0905 b
  1688 0900 b
 stroke
 0.0039 setgray
 stroke
00004 w
stroke
n
 1676 0883 moveto
  1679 0885 b
  1683 0885 b
  1695 0881 b
  1703 0881 b
  1707 0883 b
  1710 0885 b
 stroke
 0.0039 setgray
 stroke
00004 w
stroke
n
 1683 0885 moveto
  1695 0878 b
  1705 0878 b
 stroke
 0.0039 setgray
 stroke
00004 w
stroke
n
 1705 0878 moveto
  1707 0881 b
  1710 0885 b
  1710 0890 b
 stroke
 0.0039 setgray
 stroke
00004 w
stroke
n
 1743 0883 moveto
  1741 0881 b
  1743 0878 b
  1746 0881 b
  1743 0883 b
 stroke
 0.0039 setgray
 stroke
00004 w
stroke
n
 1782 0929 moveto
  1777 0905 b
 stroke
 0.0039 setgray
 stroke
00004 w
stroke
n
 1777 0905 moveto
  1782 0909 b
  1789 0912 b
  1796 0912 b
  1803 0909 b
  1808 0905 b
  1811 0897 b
  1811 0893 b
  1808 0885 b
  1803 0881 b
  1796 0878 b
  1789 0878 b
  1782 0881 b
  1780 0883 b
  1777 0888 b
  1777 0890 b
  1780 0893 b
  1782 0890 b
  1780 0888 b
 stroke
 0.0039 setgray
 stroke
00004 w
stroke
n
 1796 0912 moveto
  1801 0909 b
  1806 0905 b
  1808 0897 b
  1808 0893 b
  1806 0885 b
  1801 0881 b
  1796 0878 b
 stroke
 0.0039 setgray
 stroke
00004 w
stroke
n
 1782 0929 moveto
  1806 0929 b
 stroke
 0.0039 setgray
 stroke
00004 w
stroke
n
 1782 0926 moveto
  1794 0926 b
  1806 0929 b
 stroke
 0.0039 setgray
 stroke
00004 w
stroke
n
 1788 0956 moveto
  1788 0987 b
 stroke
 0.0039 setgray
 stroke
00004 w
stroke
n
 1832 0987 moveto
  1832 0956 b
 stroke
 0.0039 setgray
 stroke
00004 w
stroke
n
 1876 0956 moveto
  1876 0987 b
 stroke
 0.0039 setgray
 stroke
00004 w
stroke
n
 1920 0987 moveto
  1920 0956 b
 stroke
 0.0039 setgray
 stroke
00004 w
stroke
n
 1900 0921 moveto
  1902 0919 b
  1900 0917 b
  1897 0919 b
  1897 0921 b
  1902 0926 b
  1909 0929 b
  1919 0929 b
  1926 0926 b
  1929 0921 b
  1929 0914 b
  1926 0909 b
  1919 0907 b
  1912 0907 b
 stroke
 0.0039 setgray
 stroke
00004 w
stroke
n
 1919 0929 moveto
  1924 0926 b
  1926 0921 b
  1926 0914 b
  1924 0909 b
  1919 0907 b
 stroke
 0.0039 setgray
 stroke
00004 w
stroke
n
 1919 0907 moveto
  1924 0905 b
  1929 0900 b
  1931 0895 b
  1931 0888 b
  1929 0883 b
  1926 0881 b
  1919 0878 b
  1909 0878 b
  1902 0881 b
  1900 0883 b
  1897 0888 b
  1897 0890 b
  1900 0893 b
  1902 0890 b
  1900 0888 b
 stroke
 0.0039 setgray
 stroke
00004 w
stroke
n
 1926 0902 moveto
  1929 0895 b
  1929 0888 b
  1926 0883 b
  1924 0881 b
  1919 0878 b
 stroke
 0.0039 setgray
 stroke
00004 w
stroke
n
 1965 0883 moveto
  1962 0881 b
  1965 0878 b
  1967 0881 b
  1965 0883 b
 stroke
 0.0039 setgray
 stroke
00004 w
stroke
n
 2013 0929 moveto
  2005 0926 b
  2001 0919 b
  1998 0907 b
  1998 0900 b
  2001 0888 b
  2005 0881 b
  2013 0878 b
  2017 0878 b
  2025 0881 b
  2030 0888 b
  2032 0900 b
  2032 0907 b
  2030 0919 b
  2025 0926 b
  2017 0929 b
  2013 0929 b
 stroke
 0.0039 setgray
 stroke
00004 w
stroke
n
 2013 0929 moveto
  2008 0926 b
  2005 0924 b
  2003 0919 b
  2001 0907 b
  2001 0900 b
  2003 0888 b
  2005 0883 b
  2008 0881 b
  2013 0878 b
 stroke
 0.0039 setgray
 stroke
00004 w
stroke
n
 2017 0878 moveto
  2022 0881 b
  2025 0883 b
  2027 0888 b
  2030 0900 b
  2030 0907 b
  2027 0919 b
  2025 0924 b
  2022 0926 b
  2017 0929 b
 stroke
 0.0039 setgray
 stroke
00004 w
stroke
n
 1965 0956 moveto
  1965 1003 b
 stroke
 0.0039 setgray
 stroke
00004 w
stroke
n
 2009 0987 moveto
  2009 0956 b
 stroke
 0.0039 setgray
 stroke
00004 w
stroke
n
 2053 0956 moveto
  2053 0987 b
 stroke
 0.0039 setgray
 stroke
00004 w
stroke
n
 2097 0987 moveto
  2097 0956 b
 stroke
 0.0039 setgray
 stroke
00004 w
stroke
n
 2142 0956 moveto
  2142 0987 b
 stroke
 0.0039 setgray
 stroke
00004 w
stroke
n
 2186 1003 moveto
  2186 0956 b
 stroke
 0.0039 setgray
 stroke
00004 w
stroke
n
 2121 0921 moveto
  2123 0919 b
  2121 0917 b
  2119 0919 b
  2119 0921 b
  2123 0926 b
  2131 0929 b
  2140 0929 b
  2147 0926 b
  2150 0921 b
  2150 0914 b
  2147 0909 b
  2140 0907 b
  2133 0907 b
 stroke
 0.0039 setgray
 stroke
00004 w
stroke
n
 2140 0929 moveto
  2145 0926 b
  2147 0921 b
  2147 0914 b
  2145 0909 b
  2140 0907 b
 stroke
 0.0039 setgray
 stroke
00004 w
stroke
n
 2140 0907 moveto
  2145 0905 b
  2150 0900 b
  2152 0895 b
  2152 0888 b
  2150 0883 b
  2147 0881 b
  2140 0878 b
  2131 0878 b
  2123 0881 b
  2121 0883 b
  2119 0888 b
  2119 0890 b
  2121 0893 b
  2123 0890 b
  2121 0888 b
 stroke
 0.0039 setgray
 stroke
00004 w
stroke
n
 2147 0902 moveto
  2150 0895 b
 stroke
 0.0039 setgray
 stroke
00004 w
stroke
n
 2150 0895 moveto
  2150 0888 b
  2147 0883 b
  2145 0881 b
  2140 0878 b
 stroke
 0.0039 setgray
 stroke
00004 w
stroke
n
 2186 0883 moveto
  2184 0881 b
  2186 0878 b
  2188 0881 b
  2186 0883 b
 stroke
 0.0039 setgray
 stroke
00004 w
stroke
n
 2224 0929 moveto
  2219 0905 b
 stroke
 0.0039 setgray
 stroke
00004 w
stroke
n
 2219 0905 moveto
  2224 0909 b
  2231 0912 b
  2239 0912 b
  2246 0909 b
  2251 0905 b
  2253 0897 b
  2253 0893 b
  2251 0885 b
  2246 0881 b
  2239 0878 b
  2231 0878 b
  2224 0881 b
  2222 0883 b
  2219 0888 b
  2219 0890 b
  2222 0893 b
  2224 0890 b
  2222 0888 b
 stroke
 0.0039 setgray
 stroke
00004 w
stroke
n
 2239 0912 moveto
  2243 0909 b
  2248 0905 b
  2251 0897 b
  2251 0893 b
  2248 0885 b
  2243 0881 b
  2239 0878 b
 stroke
 0.0039 setgray
 stroke
00004 w
stroke
n
 2224 0929 moveto
  2248 0929 b
 stroke
 0.0039 setgray
 stroke
00004 w
stroke
n
 2224 0926 moveto
  2236 0926 b
  2248 0929 b
 stroke
 0.0039 setgray
 stroke
00004 w
stroke
n
 2186 0956 moveto
  0637 0956 b
 stroke
 0.0039 setgray
 stroke
00004 w
stroke
n
 0637 2505 moveto
  0637 2458 b
 stroke
 0.0039 setgray
 stroke
00004 w
stroke
n
 0682 2474 moveto
  0682 2505 b
 stroke
 0.0039 setgray
 stroke
00004 w
stroke
n
 0726 2505 moveto
  0726 2474 b
 stroke
 0.0039 setgray
 stroke
00004 w
stroke
n
 0770 2474 moveto
  0770 2505 b
 stroke
 0.0039 setgray
 stroke
00004 w
stroke
n
 0814 2505 moveto
  0814 2474 b
 stroke
 0.0039 setgray
 stroke
00004 w
stroke
n
 0859 2458 moveto
  0859 2505 b
 stroke
 0.0039 setgray
 stroke
00004 w
stroke
n
 0903 2505 moveto
  0903 2474 b
 stroke
 0.0039 setgray
 stroke
00004 w
stroke
n
 0947 2474 moveto
  0947 2505 b
 stroke
 0.0039 setgray
 stroke
00004 w
stroke
n
 0991 2505 moveto
  0991 2474 b
 stroke
 0.0039 setgray
 stroke
00004 w
stroke
n
 1036 2474 moveto
  1036 2505 b
 stroke
 0.0039 setgray
 stroke
00004 w
stroke
n
 1080 2505 moveto
  1080 2458 b
 stroke
 0.0039 setgray
 stroke
00004 w
stroke
n
 1124 2474 moveto
  1124 2505 b
 stroke
 0.0039 setgray
 stroke
00004 w
stroke
n
 1168 2505 moveto
  1168 2474 b
 stroke
 0.0039 setgray
 stroke
00004 w
stroke
n
 1213 2474 moveto
  1213 2505 b
 stroke
 0.0039 setgray
 stroke
00004 w
stroke
n
 1257 2505 moveto
  1257 2474 b
 stroke
 0.0039 setgray
 stroke
00004 w
stroke
n
 1301 2458 moveto
  1301 2505 b
 stroke
 0.0039 setgray
 stroke
00004 w
stroke
n
 1345 2505 moveto
  1345 2474 b
 stroke
 0.0039 setgray
 stroke
00004 w
stroke
n
 1390 2474 moveto
  1390 2505 b
 stroke
 0.0039 setgray
 stroke
00004 w
stroke
n
 1434 2505 moveto
  1434 2474 b
 stroke
 0.0039 setgray
 stroke
00004 w
stroke
n
 1478 2474 moveto
  1478 2505 b
 stroke
 0.0039 setgray
 stroke
00004 w
stroke
n
 1522 2505 moveto
  1522 2458 b
 stroke
 0.0039 setgray
 stroke
00004 w
stroke
n
 1567 2474 moveto
  1567 2505 b
 stroke
 0.0039 setgray
 stroke
00004 w
stroke
n
 1611 2505 moveto
  1611 2474 b
 stroke
 0.0039 setgray
 stroke
00004 w
stroke
n
 1655 2474 moveto
  1655 2505 b
 stroke
 0.0039 setgray
 stroke
00004 w
stroke
n
 1699 2505 moveto
  1699 2474 b
 stroke
 0.0039 setgray
 stroke
00004 w
stroke
n
 1743 2458 moveto
  1743 2505 b
 stroke
 0.0039 setgray
 stroke
00004 w
stroke
n
 1788 2505 moveto
  1788 2474 b
 stroke
 0.0039 setgray
 stroke
00004 w
stroke
n
 1832 2474 moveto
  1832 2505 b
 stroke
 0.0039 setgray
 stroke
00004 w
stroke
n
 1876 2505 moveto
  1876 2474 b
 stroke
 0.0039 setgray
 stroke
00004 w
stroke
n
 1920 2474 moveto
  1920 2505 b
 stroke
 0.0039 setgray
 stroke
00004 w
stroke
n
 1965 2505 moveto
  1965 2458 b
 stroke
 0.0039 setgray
 stroke
00004 w
stroke
n
 2009 2474 moveto
  2009 2505 b
 stroke
 0.0039 setgray
 stroke
00004 w
stroke
n
 2053 2505 moveto
  2053 2474 b
 stroke
 0.0039 setgray
 stroke
00004 w
stroke
n
 2097 2474 moveto
  2097 2505 b
 stroke
 0.0039 setgray
 stroke
00004 w
stroke
n
 2142 2505 moveto
  2142 2474 b
 stroke
 0.0039 setgray
 stroke
00004 w
stroke
n
 2186 2458 moveto
  2186 2505 b
 stroke
 0.0039 setgray
 stroke
00004 w
stroke
n
 0637 2505 moveto
  2186 2505 b
 stroke
 0.0039 setgray
 stroke
00002 w
stroke
n
 1016 1196 moveto
  1012 1194 b
  1009 1192 b
  1009 1190 b
  1012 1187 b
  1018 1185 b
  1025 1185 b
 stroke
 0.0039 setgray
 stroke
00002 w
stroke
n
 1018 1185 moveto
  1009 1183 b
  1005 1181 b
  1003 1176 b
  1003 1172 b
  1007 1167 b
  1014 1165 b
  1021 1165 b
 stroke
 0.0039 setgray
 stroke
00002 w
stroke
n
 1018 1185 moveto
  1012 1183 b
  1007 1181 b
  1005 1176 b
  1005 1172 b
  1009 1167 b
  1014 1165 b
 stroke
 0.0039 setgray
 stroke
00002 w
stroke
n
 1014 1165 moveto
  1005 1163 b
  1000 1160 b
  0998 1156 b
  0998 1151 b
  1003 1147 b
  1014 1142 b
  1016 1140 b
  1016 1136 b
  1012 1133 b
  1007 1133 b
 stroke
 0.0039 setgray
 stroke
00002 w
stroke
n
 1014 1165 moveto
  1007 1163 b
  1003 1160 b
  1000 1156 b
  1000 1151 b
  1005 1147 b
  1014 1142 b
 stroke
 0.0039 setgray
 stroke
00002 w
stroke
n
 1468 1851 moveto
  1465 1849 b
  1468 1847 b
  1470 1849 b
  1470 1851 b
  1465 1856 b
  1461 1858 b
  1454 1858 b
  1447 1856 b
  1443 1851 b
  1441 1845 b
  1441 1840 b
  1443 1833 b
  1447 1829 b
  1454 1827 b
  1458 1827 b
  1465 1829 b
  1470 1833 b
 stroke
 0.0039 setgray
 stroke
00002 w
stroke
n
 1454 1858 moveto
  1449 1856 b
  1445 1851 b
  1443 1845 b
  1443 1840 b
  1445 1833 b
  1449 1829 b
  1454 1827 b
 stroke
 0.0039 setgray
 stroke
00004 w
stroke
n
 0641 0957 moveto
  0641 0956 b
  0642 0957 b
  0642 0957 b
  0642 0957 b
  0642 0957 b
  0643 0957 b
  0643 0958 b
  0643 0958 b
  0643 0959 b
  0644 0960 b
  0644 0960 b
  0644 0961 b
  0645 0963 b
  0645 0964 b
  0645 0965 b
  0646 0967 b
  0646 0969 b
  0647 0971 b
  0647 0973 b
  0648 0975 b
  0648 0978 b
  0649 0980 b
  0650 0983 b
  0650 0986 b
  0651 0989 b
  0652 0992 b
  0652 0995 b
  0653 0997 b
  0654 1000 b
  0655 1004 b
  0656 1006 b
  0657 1010 b
  0658 1013 b
  0659 1016 b
  0660 1019 b
  0661 1022 b
  0662 1026 b
  0664 1029 b
  0665 1033 b
  0666 1036 b
  0668 1040 b
  0670 1044 b
  0671 1048 b
  0673 1053 b
  0675 1057 b
  0677 1062 b
  0679 1067 b
  0681 1073 b
  0683 1079 b
 stroke
 0.0039 setgray
 stroke
00004 w
stroke
n
 0683 1079 moveto
  0686 1085 b
  0688 1091 b
  0691 1098 b
  0694 1105 b
  0697 1112 b
  0700 1120 b
  0703 1129 b
  0707 1138 b
  0710 1148 b
  0714 1158 b
  0718 1168 b
  0723 1180 b
  0727 1192 b
  0732 1205 b
  0737 1219 b
  0742 1234 b
  0747 1249 b
  0753 1266 b
  0759 1285 b
  0766 1304 b
  0773 1325 b
  0780 1348 b
  0787 1373 b
  0795 1400 b
  0803 1430 b
  0812 1462 b
  0821 1496 b
  0831 1534 b
  0841 1575 b
  0852 1619 b
  0863 1665 b
  0875 1715 b
  0887 1767 b
  0901 1821 b
  0914 1877 b
  0929 1933 b
  0944 1989 b
  0960 2043 b
  0977 2096 b
  0995 2145 b
  1014 2189 b
  1034 2228 b
  1055 2260 b
  1077 2285 b
  1100 2302 b
  1124 2309 b
  1150 2309 b
  1177 2299 b
  1205 2280 b
 stroke
 0.0039 setgray
 stroke
00004 w
stroke
n
 1205 2280 moveto
  1235 2253 b
  1267 2219 b
  1300 2178 b
  1335 2131 b
  1371 2080 b
  1410 2025 b
  1451 1968 b
  1493 1909 b
  1538 1849 b
  1586 1790 b
  1636 1731 b
  1688 1674 b
  1744 1619 b
  1802 1566 b
  1863 1516 b
  1928 1468 b
  1996 1425 b
  2067 1383 b
  2142 1345 b
  2186 1325 b
 stroke
 0.0039 setgray
 stroke
00004 w
stroke
n
 0682 2505 moveto
  0683 2420 b
  0686 2300 b
  0688 2193 b
  0691 2095 b
  0694 2007 b
  0697 1926 b
  0700 1853 b
  0703 1787 b
  0707 1726 b
  0710 1670 b
  0714 1620 b
  0718 1573 b
  0723 1530 b
  0727 1491 b
  0732 1454 b
  0737 1421 b
  0742 1390 b
  0747 1361 b
  0753 1335 b
  0759 1310 b
  0766 1289 b
  0773 1279 b
  0780 1264 b
  0787 1249 b
  0795 1235 b
  0803 1222 b
  0812 1209 b
  0821 1197 b
  0831 1185 b
  0841 1173 b
  0852 1163 b
  0863 1152 b
  0875 1143 b
  0887 1133 b
  0901 1124 b
  0914 1116 b
  0929 1107 b
  0944 1100 b
  0960 1092 b
  0977 1085 b
  0995 1078 b
  1014 1072 b
  1034 1066 b
  1055 1060 b
  1077 1055 b
  1100 1050 b
  1124 1045 b
  1150 1041 b
  1177 1036 b
 stroke
 0.0039 setgray
 stroke
00004 w
stroke
n
 1177 1036 moveto
  1205 1032 b
  1235 1029 b
  1267 1025 b
  1300 1022 b
  1335 1019 b
  1371 1016 b
  1410 1014 b
  1451 1011 b
  1493 1009 b
  1538 1007 b
  1586 1005 b
  1636 1003 b
  1688 1001 b
  1744 1000 b
  1802 0998 b
  1863 0997 b
  1928 0996 b
  1996 0994 b
  2067 0993 b
  2142 0992 b
  2186 0992 b
 stroke
 showpage
n
.25 .25 scale
1 setlinejoin
 restore

%%%%%%%%%%%%%%%%%%%%%%% Fig 4b %%%%%%%%%%%%%%%%%%%%%%%%%%%%%%%%%%%

%!
save
/w {setlinewidth} def
/b {lineto} def
/n {newpath} def
.25 .25 scale
1 setlinejoin
n
 stroke
 0.0039 setgray
 stroke
00004 w
stroke
n
 1410 0811 moveto
  1410 0732 b
 stroke
 0.0039 setgray
 stroke
00004 w
stroke
n
 1413 0811 moveto
  1413 0732 b
 stroke
 0.0039 setgray
 stroke
00004 w
stroke
n
 1387 0811 moveto
  1383 0789 b
  1383 0811 b
  1440 0811 b
  1440 0789 b
  1436 0811 b
 stroke
 0.0039 setgray
 stroke
00004 w
stroke
n
 1399 0732 moveto
  1425 0732 b
 stroke
 0.0039 setgray
 stroke
00004 w
stroke
n
 0585 0985 moveto
  0577 0983 b
  0573 0976 b
  0570 0964 b
  0570 0956 b
  0573 0944 b
  0577 0937 b
  0585 0935 b
  0589 0935 b
  0597 0937 b
  0601 0944 b
  0604 0956 b
  0604 0964 b
  0601 0976 b
  0597 0983 b
  0589 0985 b
  0585 0985 b
 stroke
 0.0039 setgray
 stroke
00004 w
stroke
n
 0585 0985 moveto
  0580 0983 b
  0577 0980 b
  0575 0976 b
  0573 0964 b
  0573 0956 b
  0575 0944 b
  0577 0940 b
  0580 0937 b
  0585 0935 b
 stroke
 0.0039 setgray
 stroke
00004 w
stroke
n
 0589 0935 moveto
  0594 0937 b
  0597 0940 b
  0599 0944 b
  0601 0956 b
  0601 0964 b
  0599 0976 b
  0597 0980 b
  0594 0983 b
  0589 0985 b
 stroke
 0.0039 setgray
 stroke
00004 w
stroke
n
 0637 0956 moveto
  0684 0956 b
 stroke
 0.0039 setgray
 stroke
00004 w
stroke
n
 0668 1129 moveto
  0637 1129 b
 stroke
 0.0039 setgray
 stroke
00004 w
stroke
n
 0537 1284 moveto
  0534 1281 b
  0537 1279 b
  0539 1281 b
  0537 1284 b
 stroke
 0.0039 setgray
 stroke
00004 w
stroke
n
 0577 1320 moveto
  0582 1322 b
  0589 1329 b
  0589 1279 b
 stroke
 0.0039 setgray
 stroke
00004 w
stroke
n
 0587 1327 moveto
  0587 1279 b
 stroke
 0.0039 setgray
 stroke
00004 w
stroke
n
 0577 1279 moveto
  0599 1279 b
 stroke
 0.0039 setgray
 stroke
00004 w
stroke
n
 0637 1301 moveto
  0684 1301 b
 stroke
 0.0039 setgray
 stroke
00004 w
stroke
n
 0668 1473 moveto
  0637 1473 b
 stroke
 0.0039 setgray
 stroke
00004 w
stroke
n
 0537 1628 moveto
  0534 1625 b
  0537 1623 b
  0539 1625 b
  0537 1628 b
 stroke
 0.0039 setgray
 stroke
00004 w
stroke
n
 0573 1664 moveto
  0575 1661 b
  0573 1659 b
  0570 1661 b
  0570 1664 b
  0573 1669 b
  0575 1671 b
  0582 1673 b
  0592 1673 b
  0599 1671 b
  0601 1669 b
  0604 1664 b
  0604 1659 b
  0601 1654 b
  0594 1649 b
  0582 1645 b
  0577 1642 b
  0573 1637 b
  0570 1630 b
  0570 1623 b
 stroke
 0.0039 setgray
 stroke
00004 w
stroke
n
 0592 1673 moveto
  0597 1671 b
  0599 1669 b
  0601 1664 b
  0601 1659 b
  0599 1654 b
  0592 1649 b
  0582 1645 b
 stroke
 0.0039 setgray
 stroke
00004 w
stroke
n
 0570 1628 moveto
  0573 1630 b
  0577 1630 b
  0589 1625 b
  0597 1625 b
  0601 1628 b
  0604 1630 b
 stroke
 0.0039 setgray
 stroke
00004 w
stroke
n
 0577 1630 moveto
  0589 1623 b
  0599 1623 b
  0601 1625 b
  0604 1630 b
  0604 1635 b
 stroke
 0.0039 setgray
 stroke
00004 w
stroke
n
 0637 1645 moveto
  0684 1645 b
 stroke
 0.0039 setgray
 stroke
00004 w
stroke
n
 0668 1817 moveto
  0637 1817 b
 stroke
 0.0039 setgray
 stroke
00004 w
stroke
n
 0537 1972 moveto
  0534 1970 b
  0537 1967 b
  0539 1970 b
  0537 1972 b
 stroke
 0.0039 setgray
 stroke
00004 w
stroke
n
 0573 2010 moveto
  0575 2008 b
  0573 2006 b
  0570 2008 b
  0570 2010 b
  0575 2015 b
  0582 2018 b
  0592 2018 b
  0599 2015 b
  0601 2010 b
  0601 2003 b
  0599 1998 b
  0592 1996 b
  0585 1996 b
 stroke
 0.0039 setgray
 stroke
00004 w
stroke
n
 0592 2018 moveto
  0597 2015 b
  0599 2010 b
  0599 2003 b
  0597 1998 b
  0592 1996 b
 stroke
 0.0039 setgray
 stroke
00004 w
stroke
n
 0592 1996 moveto
  0597 1994 b
  0601 1989 b
  0604 1984 b
  0604 1977 b
  0601 1972 b
  0599 1970 b
  0592 1967 b
  0582 1967 b
  0575 1970 b
  0573 1972 b
  0570 1977 b
  0570 1979 b
  0573 1982 b
  0575 1979 b
  0573 1977 b
 stroke
 0.0039 setgray
 stroke
00004 w
stroke
n
 0599 1991 moveto
  0601 1984 b
  0601 1977 b
  0599 1972 b
  0597 1970 b
  0592 1967 b
 stroke
 0.0039 setgray
 stroke
00004 w
stroke
n
 0637 1989 moveto
  0684 1989 b
 stroke
 0.0039 setgray
 stroke
00004 w
stroke
n
 0668 2161 moveto
  0637 2161 b
 stroke
 0.0039 setgray
 stroke
00004 w
stroke
n
 0537 2316 moveto
  0534 2314 b
  0537 2311 b
  0539 2314 b
  0537 2316 b
 stroke
 0.0039 setgray
 stroke
00004 w
stroke
n
 0592 2357 moveto
  0592 2311 b
 stroke
 0.0039 setgray
 stroke
00004 w
stroke
n
 0594 2362 moveto
  0594 2311 b
 stroke
 0.0039 setgray
 stroke
00004 w
stroke
n
 0594 2362 moveto
  0568 2326 b
  0606 2326 b
 stroke
 0.0039 setgray
 stroke
00004 w
stroke
n
 0585 2311 moveto
  0601 2311 b
 stroke
 0.0039 setgray
 stroke
00004 w
stroke
n
 0637 2333 moveto
  0684 2333 b
 stroke
 0.0039 setgray
 stroke
00004 w
stroke
n
 0637 0956 moveto
  0637 2505 b
 stroke
 0.0039 setgray
 stroke
00004 w
stroke
n
 2186 0956 moveto
  2139 0956 b
 stroke
 0.0039 setgray
 stroke
00004 w
stroke
n
 2155 1129 moveto
  2186 1129 b
 stroke
 0.0039 setgray
 stroke
00004 w
stroke
n
 2186 1301 moveto
  2139 1301 b
 stroke
 0.0039 setgray
 stroke
00004 w
stroke
n
 2155 1473 moveto
  2186 1473 b
 stroke
 0.0039 setgray
 stroke
00004 w
stroke
n
 2186 1645 moveto
  2139 1645 b
 stroke
 0.0039 setgray
 stroke
00004 w
stroke
n
 2155 1817 moveto
  2186 1817 b
 stroke
 0.0039 setgray
 stroke
00004 w
stroke
n
 2186 1989 moveto
  2139 1989 b
 stroke
 0.0039 setgray
 stroke
00004 w
stroke
n
 2155 2161 moveto
  2186 2161 b
 stroke
 0.0039 setgray
 stroke
00004 w
stroke
n
 2186 2333 moveto
  2139 2333 b
 stroke
 0.0039 setgray
 stroke
00004 w
stroke
n
 2186 2505 moveto
  2186 0956 b
 stroke
 0.0039 setgray
 stroke
00004 w
stroke
n
 0635 0929 moveto
  0628 0926 b
  0623 0919 b
  0621 0907 b
  0621 0900 b
  0623 0888 b
  0628 0881 b
  0635 0878 b
  0640 0878 b
  0647 0881 b
  0652 0888 b
  0654 0900 b
  0654 0907 b
  0652 0919 b
  0647 0926 b
  0640 0929 b
  0635 0929 b
 stroke
 0.0039 setgray
 stroke
00004 w
stroke
n
 0635 0929 moveto
  0630 0926 b
  0628 0924 b
  0625 0919 b
  0623 0907 b
  0623 0900 b
  0625 0888 b
  0628 0883 b
  0630 0881 b
 stroke
 0.0039 setgray
 stroke
00004 w
stroke
n
 0630 0881 moveto
  0635 0878 b
 stroke
 0.0039 setgray
 stroke
00004 w
stroke
n
 0640 0878 moveto
  0645 0881 b
  0647 0883 b
  0649 0888 b
  0652 0900 b
  0652 0907 b
  0649 0919 b
  0647 0924 b
  0645 0926 b
  0640 0929 b
 stroke
 0.0039 setgray
 stroke
00004 w
stroke
n
 0637 0956 moveto
  0637 1003 b
 stroke
 0.0039 setgray
 stroke
00004 w
stroke
n
 0734 0987 moveto
  0734 0956 b
 stroke
 0.0039 setgray
 stroke
00004 w
stroke
n
 0806 0883 moveto
  0803 0881 b
  0806 0878 b
  0808 0881 b
  0806 0883 b
 stroke
 0.0039 setgray
 stroke
00004 w
stroke
n
 0842 0919 moveto
  0844 0917 b
  0842 0914 b
  0839 0917 b
  0839 0919 b
  0842 0924 b
  0844 0926 b
  0851 0929 b
  0861 0929 b
  0868 0926 b
  0871 0924 b
  0873 0919 b
  0873 0914 b
  0871 0909 b
  0863 0905 b
  0851 0900 b
  0847 0897 b
  0842 0893 b
  0839 0885 b
  0839 0878 b
 stroke
 0.0039 setgray
 stroke
00004 w
stroke
n
 0861 0929 moveto
  0866 0926 b
  0868 0924 b
  0871 0919 b
  0871 0914 b
  0868 0909 b
  0861 0905 b
  0851 0900 b
 stroke
 0.0039 setgray
 stroke
00004 w
stroke
n
 0839 0883 moveto
  0842 0885 b
  0847 0885 b
  0859 0881 b
  0866 0881 b
  0871 0883 b
  0873 0885 b
 stroke
 0.0039 setgray
 stroke
00004 w
stroke
n
 0847 0885 moveto
  0859 0878 b
  0868 0878 b
  0871 0881 b
  0873 0885 b
  0873 0890 b
 stroke
 0.0039 setgray
 stroke
00004 w
stroke
n
 0831 0956 moveto
  0831 1003 b
 stroke
 0.0039 setgray
 stroke
00004 w
stroke
n
 0928 0987 moveto
  0928 0956 b
 stroke
 0.0039 setgray
 stroke
00004 w
stroke
n
 0999 0883 moveto
  0997 0881 b
  0999 0878 b
  1002 0881 b
  0999 0883 b
 stroke
 0.0039 setgray
 stroke
00004 w
stroke
n
 1055 0924 moveto
  1055 0878 b
 stroke
 0.0039 setgray
 stroke
00004 w
stroke
n
 1057 0929 moveto
  1057 0878 b
 stroke
 0.0039 setgray
 stroke
00004 w
stroke
n
 1057 0929 moveto
  1031 0893 b
  1069 0893 b
 stroke
 0.0039 setgray
 stroke
00004 w
stroke
n
 1047 0878 moveto
  1064 0878 b
 stroke
 0.0039 setgray
 stroke
00004 w
stroke
n
 1025 0956 moveto
  1025 1003 b
 stroke
 0.0039 setgray
 stroke
00004 w
stroke
n
 1121 0987 moveto
  1121 0956 b
 stroke
 0.0039 setgray
 stroke
00004 w
stroke
n
 1193 0883 moveto
  1191 0881 b
  1193 0878 b
  1195 0881 b
  1193 0883 b
 stroke
 0.0039 setgray
 stroke
00004 w
stroke
n
 1255 0921 moveto
  1253 0919 b
  1255 0917 b
  1258 0919 b
  1258 0921 b
  1255 0926 b
  1250 0929 b
  1243 0929 b
  1236 0926 b
  1231 0921 b
  1229 0917 b
  1226 0907 b
  1226 0893 b
  1229 0885 b
  1234 0881 b
  1241 0878 b
  1246 0878 b
  1253 0881 b
  1258 0885 b
  1260 0893 b
  1260 0895 b
  1258 0902 b
  1253 0907 b
  1246 0909 b
  1243 0909 b
  1236 0907 b
  1231 0902 b
  1229 0895 b
 stroke
 0.0039 setgray
 stroke
00004 w
stroke
n
 1243 0929 moveto
  1238 0926 b
  1234 0921 b
  1231 0917 b
  1229 0907 b
  1229 0893 b
  1231 0885 b
  1236 0881 b
  1241 0878 b
 stroke
 0.0039 setgray
 stroke
00004 w
stroke
n
 1246 0878 moveto
  1250 0881 b
  1255 0885 b
  1258 0893 b
  1258 0895 b
  1255 0902 b
  1250 0907 b
  1246 0909 b
 stroke
 0.0039 setgray
 stroke
00004 w
stroke
n
 1218 0956 moveto
  1218 1003 b
 stroke
 0.0039 setgray
 stroke
00004 w
stroke
n
 1315 0987 moveto
  1315 0956 b
 stroke
 0.0039 setgray
 stroke
00004 w
stroke
n
 1386 0883 moveto
  1384 0881 b
  1386 0878 b
  1389 0881 b
  1386 0883 b
 stroke
 0.0039 setgray
 stroke
00004 w
stroke
n
 1432 0929 moveto
  1425 0926 b
  1423 0921 b
  1423 0914 b
  1425 0909 b
  1432 0907 b
  1442 0907 b
  1449 0909 b
  1451 0914 b
  1451 0921 b
  1449 0926 b
  1442 0929 b
  1432 0929 b
 stroke
 0.0039 setgray
 stroke
00004 w
stroke
n
 1432 0929 moveto
  1427 0926 b
  1425 0921 b
  1425 0914 b
  1427 0909 b
  1432 0907 b
 stroke
 0.0039 setgray
 stroke
00004 w
stroke
n
 1442 0907 moveto
  1446 0909 b
  1449 0914 b
  1449 0921 b
  1446 0926 b
  1442 0929 b
 stroke
 0.0039 setgray
 stroke
00004 w
stroke
n
 1432 0907 moveto
  1425 0905 b
  1423 0902 b
  1420 0897 b
  1420 0888 b
  1423 0883 b
  1425 0881 b
  1432 0878 b
  1442 0878 b
  1449 0881 b
  1451 0883 b
  1454 0888 b
  1454 0897 b
  1451 0902 b
  1449 0905 b
  1442 0907 b
 stroke
 0.0039 setgray
 stroke
00004 w
stroke
n
 1432 0907 moveto
  1427 0905 b
  1425 0902 b
  1423 0897 b
  1423 0888 b
  1425 0883 b
  1427 0881 b
  1432 0878 b
 stroke
 0.0039 setgray
 stroke
00004 w
stroke
n
 1442 0878 moveto
  1446 0881 b
  1449 0883 b
  1451 0888 b
  1451 0897 b
  1449 0902 b
  1446 0905 b
  1442 0907 b
 stroke
 0.0039 setgray
 stroke
00004 w
stroke
n
 1412 0956 moveto
  1412 1003 b
 stroke
 0.0039 setgray
 stroke
00004 w
stroke
n
 1508 0987 moveto
  1508 0956 b
 stroke
 0.0039 setgray
 stroke
00004 w
stroke
n
 1545 0919 moveto
  1550 0921 b
  1557 0929 b
  1557 0878 b
 stroke
 0.0039 setgray
 stroke
00004 w
stroke
n
 1555 0926 moveto
  1555 0878 b
 stroke
 0.0039 setgray
 stroke
00004 w
stroke
n
 1545 0878 moveto
  1567 0878 b
 stroke
 0.0039 setgray
 stroke
00004 w
stroke
n
 1605 0883 moveto
  1603 0881 b
  1605 0878 b
  1608 0881 b
  1605 0883 b
 stroke
 0.0039 setgray
 stroke
00004 w
stroke
n
 1653 0929 moveto
  1646 0926 b
  1641 0919 b
  1639 0907 b
  1639 0900 b
  1641 0888 b
  1646 0881 b
  1653 0878 b
  1658 0878 b
  1665 0881 b
  1670 0888 b
  1672 0900 b
  1672 0907 b
  1670 0919 b
  1665 0926 b
  1658 0929 b
  1653 0929 b
 stroke
 0.0039 setgray
 stroke
00004 w
stroke
n
 1653 0929 moveto
  1648 0926 b
  1646 0924 b
  1644 0919 b
  1641 0907 b
  1641 0900 b
  1644 0888 b
  1646 0883 b
  1648 0881 b
  1653 0878 b
 stroke
 0.0039 setgray
 stroke
00004 w
stroke
n
 1658 0878 moveto
  1663 0881 b
  1665 0883 b
  1668 0888 b
  1670 0900 b
  1670 0907 b
  1668 0919 b
  1665 0924 b
  1663 0926 b
  1658 0929 b
 stroke
 0.0039 setgray
 stroke
00004 w
stroke
n
 1605 0956 moveto
  1605 1003 b
 stroke
 0.0039 setgray
 stroke
00004 w
stroke
n
 1702 0987 moveto
  1702 0956 b
 stroke
 0.0039 setgray
 stroke
00004 w
stroke
n
 1799 0956 moveto
  1799 1003 b
 stroke
 0.0039 setgray
 stroke
00004 w
stroke
n
 1896 0987 moveto
  1896 0956 b
 stroke
 0.0039 setgray
 stroke
00004 w
stroke
n
 1932 0919 moveto
  1937 0921 b
  1944 0929 b
  1944 0878 b
 stroke
 0.0039 setgray
 stroke
00004 w
stroke
n
 1942 0926 moveto
  1942 0878 b
 stroke
 0.0039 setgray
 stroke
00004 w
stroke
n
 1932 0878 moveto
  1954 0878 b
 stroke
 0.0039 setgray
 stroke
00004 w
stroke
n
 1992 0883 moveto
  1990 0881 b
  1992 0878 b
  1995 0881 b
  1992 0883 b
 stroke
 0.0039 setgray
 stroke
00004 w
stroke
n
 2048 0924 moveto
  2048 0878 b
 stroke
 0.0039 setgray
 stroke
00004 w
stroke
n
 2050 0929 moveto
  2050 0878 b
 stroke
 0.0039 setgray
 stroke
00004 w
stroke
n
 2050 0929 moveto
  2024 0893 b
  2062 0893 b
 stroke
 0.0039 setgray
 stroke
00004 w
stroke
n
 2040 0878 moveto
  2057 0878 b
 stroke
 0.0039 setgray
 stroke
00004 w
stroke
n
 1992 0956 moveto
  1992 1003 b
 stroke
 0.0039 setgray
 stroke
00004 w
stroke
n
 2089 0987 moveto
  2089 0956 b
 stroke
 0.0039 setgray
 stroke
00004 w
stroke
n
 2186 0956 moveto
  2186 1003 b
 stroke
 0.0039 setgray
 stroke
00004 w
stroke
n
 2186 0956 moveto
  0637 0956 b
 stroke
 0.0039 setgray
 stroke
00004 w
stroke
n
 0637 2505 moveto
  0637 2458 b
 stroke
 0.0039 setgray
 stroke
00004 w
stroke
n
 0734 2474 moveto
  0734 2505 b
 stroke
 0.0039 setgray
 stroke
00004 w
stroke
n
 0831 2505 moveto
  0831 2458 b
 stroke
 0.0039 setgray
 stroke
00004 w
stroke
n
 0928 2474 moveto
  0928 2505 b
 stroke
 0.0039 setgray
 stroke
00004 w
stroke
n
 1025 2505 moveto
  1025 2458 b
 stroke
 0.0039 setgray
 stroke
00004 w
stroke
n
 1121 2474 moveto
  1121 2505 b
 stroke
 0.0039 setgray
 stroke
00004 w
stroke
n
 1218 2505 moveto
  1218 2458 b
 stroke
 0.0039 setgray
 stroke
00004 w
stroke
n
 1315 2474 moveto
  1315 2505 b
 stroke
 0.0039 setgray
 stroke
00004 w
stroke
n
 1412 2505 moveto
  1412 2458 b
 stroke
 0.0039 setgray
 stroke
00004 w
stroke
n
 1508 2474 moveto
  1508 2505 b
 stroke
 0.0039 setgray
 stroke
00004 w
stroke
n
 1605 2505 moveto
  1605 2458 b
 stroke
 0.0039 setgray
 stroke
00004 w
stroke
n
 1702 2474 moveto
  1702 2505 b
 stroke
 0.0039 setgray
 stroke
00004 w
stroke
n
 1799 2505 moveto
  1799 2458 b
 stroke
 0.0039 setgray
 stroke
00004 w
stroke
n
 1896 2474 moveto
  1896 2505 b
 stroke
 0.0039 setgray
 stroke
00004 w
stroke
n
 1992 2505 moveto
  1992 2458 b
 stroke
 0.0039 setgray
 stroke
00004 w
stroke
n
 2089 2474 moveto
  2089 2505 b
 stroke
 0.0039 setgray
 stroke
00004 w
stroke
n
 2186 2505 moveto
  2186 2458 b
 stroke
 0.0039 setgray
 stroke
00004 w
stroke
n
 0637 2505 moveto
  2186 2505 b
 stroke
 0.0039 setgray
 stroke
00002 w
stroke
n
 1243 1190 moveto
  1238 1188 b
  1236 1185 b
  1236 1183 b
  1238 1181 b
  1245 1179 b
  1252 1179 b
 stroke
 0.0039 setgray
 stroke
00002 w
stroke
n
 1245 1179 moveto
  1236 1176 b
  1232 1174 b
  1229 1170 b
  1229 1165 b
  1234 1161 b
  1241 1158 b
  1247 1158 b
 stroke
 0.0039 setgray
 stroke
00002 w
stroke
n
 1245 1179 moveto
  1238 1176 b
  1234 1174 b
  1232 1170 b
  1232 1165 b
  1236 1161 b
  1241 1158 b
 stroke
 0.0039 setgray
 stroke
00002 w
stroke
n
 1241 1158 moveto
  1232 1156 b
  1227 1154 b
  1225 1149 b
  1225 1145 b
  1229 1140 b
  1241 1136 b
  1243 1134 b
  1243 1129 b
  1238 1127 b
  1234 1127 b
 stroke
 0.0039 setgray
 stroke
00002 w
stroke
n
 1241 1158 moveto
  1234 1156 b
  1229 1154 b
  1227 1149 b
  1227 1145 b
  1232 1140 b
  1241 1136 b
 stroke
 0.0039 setgray
 stroke
00002 w
stroke
n
 1445 1649 moveto
  1443 1647 b
  1445 1645 b
  1448 1647 b
  1448 1649 b
  1443 1654 b
  1439 1656 b
  1432 1656 b
  1425 1654 b
  1421 1649 b
  1418 1642 b
  1418 1638 b
  1421 1631 b
  1425 1627 b
  1432 1624 b
  1436 1624 b
  1443 1627 b
  1448 1631 b
 stroke
 0.0039 setgray
 stroke
00002 w
stroke
n
 1432 1656 moveto
  1427 1654 b
  1423 1649 b
  1421 1642 b
  1421 1638 b
  1423 1631 b
  1427 1627 b
  1432 1624 b
 stroke
 0.0039 setgray
 stroke
00004 w
stroke
n
 0637 0956 moveto
  0703 0957 b
  0706 0957 b
  0709 0957 b
  0712 0957 b
  0715 0957 b
  0718 0958 b
  0721 0958 b
  0725 0959 b
  0728 0961 b
  0732 0963 b
  0736 0966 b
  0740 0970 b
  0745 0975 b
  0749 0982 b
  0754 0991 b
  0759 1002 b
  0764 1016 b
  0769 1034 b
  0774 1055 b
  0780 1081 b
  0786 1111 b
  0792 1148 b
  0799 1189 b
  0805 1236 b
  0812 1289 b
  0820 1348 b
  0827 1412 b
  0835 1481 b
  0843 1555 b
  0852 1631 b
  0861 1710 b
  0870 1790 b
  0880 1869 b
  0890 1945 b
  0901 2018 b
  0912 2086 b
  0923 2146 b
  0935 2199 b
  0947 2243 b
  0960 2277 b
  0974 2301 b
  0988 2316 b
  1002 2321 b
  1017 2317 b
  1033 2305 b
  1050 2287 b
  1067 2262 b
  1085 2232 b
  1104 2198 b
 stroke
 0.0039 setgray
 stroke
00004 w
stroke
n
 1104 2198 moveto
  1123 2162 b
  1143 2124 b
  1164 2085 b
  1186 2045 b
  1209 2006 b
  1233 1967 b
  1258 1929 b
  1284 1894 b
  1310 1859 b
  1338 1826 b
  1368 1795 b
  1398 1766 b
  1430 1739 b
  1463 1712 b
  1497 1688 b
  1533 1665 b
  1570 1644 b
  1609 1623 b
  1650 1603 b
  1692 1585 b
  1736 1566 b
  1782 1549 b
  1829 1531 b
  1879 1515 b
  1931 1499 b
  1985 1481 b
  2041 1466 b
  2099 1450 b
  2160 1433 b
  2186 1427 b
 stroke
 0.0039 setgray
 stroke
00004 w
stroke
n
 0879 2505 moveto
  0880 2464 b
  0890 2220 b
  0901 2024 b
  0912 1865 b
  0923 1735 b
  0935 1628 b
  0947 1539 b
  0960 1465 b
  0974 1403 b
  0988 1351 b
  1002 1307 b
  1017 1270 b
  1033 1238 b
  1050 1210 b
  1067 1186 b
  1085 1165 b
  1104 1147 b
  1123 1131 b
  1143 1117 b
  1164 1105 b
  1186 1094 b
  1209 1084 b
  1233 1075 b
  1258 1067 b
  1284 1060 b
  1310 1054 b
  1338 1048 b
  1368 1043 b
  1398 1038 b
  1430 1033 b
  1463 1029 b
  1497 1026 b
  1533 1022 b
  1570 1019 b
  1609 1016 b
  1650 1013 b
  1692 1011 b
  1736 1009 b
  1782 1006 b
  1829 1004 b
  1879 1002 b
  1931 1001 b
  1985 0999 b
  2041 0997 b
  2099 0996 b
  2160 0995 b
  2186 0994 b
 stroke
 showpage
n
.25 .25 scale
1 setlinejoin
 restore

%%%%%%%%%%%%%%%%%%%%%%% Fig 5 %%%%%%%%%%%%%%%%%%%%%%%%%%%%%%%%%%%

%!
save
/w {setlinewidth} def
/b {lineto} def
/n {newpath} def
.25 .25 scale
1 setlinejoin
n
 stroke
 0.0039 setgray
 stroke
00004 w
stroke
n
 1410 0811 moveto
  1410 0732 b
 stroke
 0.0039 setgray
 stroke
00004 w
stroke
n
 1413 0811 moveto
  1413 0732 b
 stroke
 0.0039 setgray
 stroke
00004 w
stroke
n
 1387 0811 moveto
  1383 0789 b
  1383 0811 b
  1440 0811 b
  1440 0789 b
  1436 0811 b
 stroke
 0.0039 setgray
 stroke
00004 w
stroke
n
 1399 0732 moveto
  1425 0732 b
 stroke
 0.0039 setgray
 stroke
00004 w
stroke
n
 0585 0985 moveto
  0577 0983 b
  0573 0976 b
  0570 0964 b
  0570 0956 b
  0573 0944 b
  0577 0937 b
  0585 0935 b
  0589 0935 b
  0597 0937 b
  0601 0944 b
  0604 0956 b
  0604 0964 b
  0601 0976 b
  0597 0983 b
  0589 0985 b
  0585 0985 b
 stroke
 0.0039 setgray
 stroke
00004 w
stroke
n
 0585 0985 moveto
  0580 0983 b
  0577 0980 b
  0575 0976 b
  0573 0964 b
  0573 0956 b
  0575 0944 b
  0577 0940 b
  0580 0937 b
  0585 0935 b
 stroke
 0.0039 setgray
 stroke
00004 w
stroke
n
 0589 0935 moveto
  0594 0937 b
  0597 0940 b
  0599 0944 b
  0601 0956 b
  0601 0964 b
  0599 0976 b
  0597 0980 b
  0594 0983 b
  0589 0985 b
 stroke
 0.0039 setgray
 stroke
00004 w
stroke
n
 0637 0956 moveto
  0684 0956 b
 stroke
 0.0039 setgray
 stroke
00004 w
stroke
n
 0668 1067 moveto
  0637 1067 b
 stroke
 0.0039 setgray
 stroke
00004 w
stroke
n
 0537 1161 moveto
  0534 1159 b
  0537 1156 b
  0539 1159 b
  0537 1161 b
 stroke
 0.0039 setgray
 stroke
00004 w
stroke
n
 0577 1197 moveto
  0582 1199 b
  0589 1207 b
  0589 1156 b
 stroke
 0.0039 setgray
 stroke
00004 w
stroke
n
 0587 1204 moveto
  0587 1156 b
 stroke
 0.0039 setgray
 stroke
00004 w
stroke
n
 0577 1156 moveto
  0599 1156 b
 stroke
 0.0039 setgray
 stroke
00004 w
stroke
n
 0637 1178 moveto
  0684 1178 b
 stroke
 0.0039 setgray
 stroke
00004 w
stroke
n
 0668 1288 moveto
  0637 1288 b
 stroke
 0.0039 setgray
 stroke
00004 w
stroke
n
 0537 1382 moveto
  0534 1380 b
  0537 1377 b
  0539 1380 b
  0537 1382 b
 stroke
 0.0039 setgray
 stroke
00004 w
stroke
n
 0573 1418 moveto
  0575 1416 b
  0573 1413 b
  0570 1416 b
  0570 1418 b
  0573 1423 b
  0575 1425 b
  0582 1428 b
  0592 1428 b
  0599 1425 b
  0601 1423 b
  0604 1418 b
  0604 1413 b
  0601 1409 b
  0594 1404 b
  0582 1399 b
  0577 1397 b
  0573 1392 b
  0570 1385 b
  0570 1377 b
 stroke
 0.0039 setgray
 stroke
00004 w
stroke
n
 0592 1428 moveto
  0597 1425 b
  0599 1423 b
  0601 1418 b
  0601 1413 b
  0599 1409 b
  0592 1404 b
  0582 1399 b
 stroke
 0.0039 setgray
 stroke
00004 w
stroke
n
 0570 1382 moveto
  0573 1385 b
  0577 1385 b
  0589 1380 b
  0597 1380 b
  0601 1382 b
  0604 1385 b
 stroke
 0.0039 setgray
 stroke
00004 w
stroke
n
 0577 1385 moveto
  0589 1377 b
  0599 1377 b
  0601 1380 b
  0604 1385 b
  0604 1389 b
 stroke
 0.0039 setgray
 stroke
00004 w
stroke
n
 0637 1399 moveto
  0684 1399 b
 stroke
 0.0039 setgray
 stroke
00004 w
stroke
n
 0668 1509 moveto
  0637 1509 b
 stroke
 0.0039 setgray
 stroke
00004 w
stroke
n
 0537 1603 moveto
  0534 1601 b
  0537 1598 b
  0539 1601 b
  0537 1603 b
 stroke
 0.0039 setgray
 stroke
00004 w
stroke
n
 0573 1642 moveto
  0575 1639 b
  0573 1637 b
  0570 1639 b
  0570 1642 b
  0575 1646 b
  0582 1649 b
  0592 1649 b
  0599 1646 b
  0601 1642 b
  0601 1634 b
  0599 1630 b
  0592 1627 b
  0585 1627 b
 stroke
 0.0039 setgray
 stroke
00004 w
stroke
n
 0592 1649 moveto
  0597 1646 b
  0599 1642 b
  0599 1634 b
  0597 1630 b
  0592 1627 b
 stroke
 0.0039 setgray
 stroke
00004 w
stroke
n
 0592 1627 moveto
  0597 1625 b
  0601 1620 b
  0604 1615 b
  0604 1608 b
  0601 1603 b
  0599 1601 b
  0592 1598 b
  0582 1598 b
  0575 1601 b
  0573 1603 b
  0570 1608 b
  0570 1610 b
  0573 1613 b
  0575 1610 b
  0573 1608 b
 stroke
 0.0039 setgray
 stroke
00004 w
stroke
n
 0599 1622 moveto
  0601 1615 b
  0601 1608 b
  0599 1603 b
  0597 1601 b
  0592 1598 b
 stroke
 0.0039 setgray
 stroke
00004 w
stroke
n
 0637 1620 moveto
  0684 1620 b
 stroke
 0.0039 setgray
 stroke
00004 w
stroke
n
 0668 1731 moveto
  0637 1731 b
 stroke
 0.0039 setgray
 stroke
00004 w
stroke
n
 0537 1824 moveto
  0534 1822 b
  0537 1820 b
  0539 1822 b
  0537 1824 b
 stroke
 0.0039 setgray
 stroke
00004 w
stroke
n
 0592 1865 moveto
  0592 1820 b
 stroke
 0.0039 setgray
 stroke
00004 w
stroke
n
 0594 1870 moveto
  0594 1820 b
 stroke
 0.0039 setgray
 stroke
00004 w
stroke
n
 0594 1870 moveto
  0568 1834 b
  0606 1834 b
 stroke
 0.0039 setgray
 stroke
00004 w
stroke
n
 0585 1820 moveto
  0601 1820 b
 stroke
 0.0039 setgray
 stroke
00004 w
stroke
n
 0637 1841 moveto
  0684 1841 b
 stroke
 0.0039 setgray
 stroke
00004 w
stroke
n
 0668 1952 moveto
  0637 1952 b
 stroke
 0.0039 setgray
 stroke
00004 w
stroke
n
 0537 2046 moveto
  0534 2043 b
  0537 2041 b
  0539 2043 b
  0537 2046 b
 stroke
 0.0039 setgray
 stroke
00004 w
stroke
n
 0575 2091 moveto
  0570 2067 b
 stroke
 0.0039 setgray
 stroke
00004 w
stroke
n
 0570 2067 moveto
  0575 2072 b
  0582 2074 b
  0589 2074 b
  0597 2072 b
  0601 2067 b
  0604 2060 b
  0604 2055 b
  0601 2048 b
  0597 2043 b
  0589 2041 b
  0582 2041 b
  0575 2043 b
  0573 2046 b
  0570 2050 b
  0570 2053 b
  0573 2055 b
  0575 2053 b
  0573 2050 b
 stroke
 0.0039 setgray
 stroke
00004 w
stroke
n
 0589 2074 moveto
  0594 2072 b
  0599 2067 b
  0601 2060 b
  0601 2055 b
  0599 2048 b
  0594 2043 b
  0589 2041 b
 stroke
 0.0039 setgray
 stroke
00004 w
stroke
n
 0575 2091 moveto
  0599 2091 b
 stroke
 0.0039 setgray
 stroke
00004 w
stroke
n
 0575 2089 moveto
  0587 2089 b
  0599 2091 b
 stroke
 0.0039 setgray
 stroke
00004 w
stroke
n
 0637 2062 moveto
  0684 2062 b
 stroke
 0.0039 setgray
 stroke
00004 w
stroke
n
 0668 2173 moveto
  0637 2173 b
 stroke
 0.0039 setgray
 stroke
00004 w
stroke
n
 0537 2267 moveto
  0534 2264 b
  0537 2262 b
  0539 2264 b
  0537 2267 b
 stroke
 0.0039 setgray
 stroke
00004 w
stroke
n
 0599 2305 moveto
  0597 2303 b
  0599 2301 b
  0601 2303 b
  0601 2305 b
  0599 2310 b
  0594 2313 b
  0587 2313 b
  0580 2310 b
  0575 2305 b
  0573 2301 b
  0570 2291 b
  0570 2277 b
  0573 2269 b
  0577 2264 b
  0585 2262 b
  0589 2262 b
  0597 2264 b
  0601 2269 b
  0604 2277 b
  0604 2279 b
  0601 2286 b
  0597 2291 b
  0589 2293 b
  0587 2293 b
  0580 2291 b
  0575 2286 b
  0573 2279 b
 stroke
 0.0039 setgray
 stroke
00004 w
stroke
n
 0587 2313 moveto
  0582 2310 b
  0577 2305 b
  0575 2301 b
  0573 2291 b
  0573 2277 b
  0575 2269 b
  0580 2264 b
  0585 2262 b
 stroke
 0.0039 setgray
 stroke
00004 w
stroke
n
 0589 2262 moveto
  0594 2264 b
  0599 2269 b
  0601 2277 b
  0601 2279 b
  0599 2286 b
  0594 2291 b
  0589 2293 b
 stroke
 0.0039 setgray
 stroke
00004 w
stroke
n
 0637 2284 moveto
  0684 2284 b
 stroke
 0.0039 setgray
 stroke
00004 w
stroke
n
 0668 2394 moveto
  0637 2394 b
 stroke
 0.0039 setgray
 stroke
00004 w
stroke
n
 0537 2488 moveto
  0534 2486 b
  0537 2483 b
  0539 2486 b
  0537 2488 b
 stroke
 0.0039 setgray
 stroke
00004 w
stroke
n
 0570 2534 moveto
  0570 2519 b
 stroke
 0.0039 setgray
 stroke
00004 w
stroke
n
 0570 2524 moveto
  0573 2529 b
  0577 2534 b
  0582 2534 b
  0594 2526 b
  0599 2526 b
  0601 2529 b
  0604 2534 b
 stroke
 0.0039 setgray
 stroke
00004 w
stroke
n
 0573 2529 moveto
  0577 2531 b
  0582 2531 b
  0594 2526 b
 stroke
 0.0039 setgray
 stroke
00004 w
stroke
n
 0604 2534 moveto
  0604 2526 b
  0601 2519 b
  0592 2507 b
  0589 2502 b
  0587 2495 b
  0587 2483 b
 stroke
 0.0039 setgray
 stroke
00004 w
stroke
n
 0601 2519 moveto
  0589 2507 b
  0587 2502 b
  0585 2495 b
  0585 2483 b
 stroke
 0.0039 setgray
 stroke
00004 w
stroke
n
 0637 2505 moveto
  0684 2505 b
 stroke
 0.0039 setgray
 stroke
00004 w
stroke
n
 0637 0956 moveto
  0637 2505 b
 stroke
 0.0039 setgray
 stroke
00004 w
stroke
n
 2186 0956 moveto
  2139 0956 b
 stroke
 0.0039 setgray
 stroke
00004 w
stroke
n
 2155 1067 moveto
  2186 1067 b
 stroke
 0.0039 setgray
 stroke
00004 w
stroke
n
 2186 1178 moveto
  2139 1178 b
 stroke
 0.0039 setgray
 stroke
00004 w
stroke
n
 2155 1288 moveto
  2186 1288 b
 stroke
 0.0039 setgray
 stroke
00004 w
stroke
n
 2186 1399 moveto
  2139 1399 b
 stroke
 0.0039 setgray
 stroke
00004 w
stroke
n
 2155 1509 moveto
  2186 1509 b
 stroke
 0.0039 setgray
 stroke
00004 w
stroke
n
 2186 1620 moveto
  2139 1620 b
 stroke
 0.0039 setgray
 stroke
00004 w
stroke
n
 2155 1731 moveto
  2186 1731 b
 stroke
 0.0039 setgray
 stroke
00004 w
stroke
n
 2186 1841 moveto
  2139 1841 b
 stroke
 0.0039 setgray
 stroke
00004 w
stroke
n
 2155 1952 moveto
  2186 1952 b
 stroke
 0.0039 setgray
 stroke
00004 w
stroke
n
 2186 2062 moveto
  2139 2062 b
 stroke
 0.0039 setgray
 stroke
00004 w
stroke
n
 2155 2173 moveto
  2186 2173 b
 stroke
 0.0039 setgray
 stroke
00004 w
stroke
n
 2186 2284 moveto
  2139 2284 b
 stroke
 0.0039 setgray
 stroke
00004 w
stroke
n
 2155 2394 moveto
  2186 2394 b
 stroke
 0.0039 setgray
 stroke
00004 w
stroke
n
 2186 2505 moveto
  2139 2505 b
 stroke
 0.0039 setgray
 stroke
00004 w
stroke
n
 2186 2505 moveto
  2186 0956 b
 stroke
 0.0039 setgray
 stroke
00004 w
stroke
n
 0635 0929 moveto
  0628 0926 b
  0623 0919 b
  0621 0907 b
  0621 0900 b
  0623 0888 b
  0628 0881 b
  0635 0878 b
  0640 0878 b
  0647 0881 b
  0652 0888 b
  0654 0900 b
  0654 0907 b
  0652 0919 b
 stroke
 0.0039 setgray
 stroke
00004 w
stroke
n
 0652 0919 moveto
  0647 0926 b
  0640 0929 b
  0635 0929 b
 stroke
 0.0039 setgray
 stroke
00004 w
stroke
n
 0635 0929 moveto
  0630 0926 b
  0628 0924 b
  0625 0919 b
  0623 0907 b
  0623 0900 b
  0625 0888 b
  0628 0883 b
  0630 0881 b
  0635 0878 b
 stroke
 0.0039 setgray
 stroke
00004 w
stroke
n
 0640 0878 moveto
  0645 0881 b
  0647 0883 b
  0649 0888 b
  0652 0900 b
  0652 0907 b
  0649 0919 b
  0647 0924 b
  0645 0926 b
  0640 0929 b
 stroke
 0.0039 setgray
 stroke
00004 w
stroke
n
 0637 0956 moveto
  0637 1003 b
 stroke
 0.0039 setgray
 stroke
00004 w
stroke
n
 0682 0987 moveto
  0682 0956 b
 stroke
 0.0039 setgray
 stroke
00004 w
stroke
n
 0726 0956 moveto
  0726 0987 b
 stroke
 0.0039 setgray
 stroke
00004 w
stroke
n
 0770 0987 moveto
  0770 0956 b
 stroke
 0.0039 setgray
 stroke
00004 w
stroke
n
 0814 0956 moveto
  0814 0987 b
 stroke
 0.0039 setgray
 stroke
00004 w
stroke
n
 0859 1003 moveto
  0859 0956 b
 stroke
 0.0039 setgray
 stroke
00004 w
stroke
n
 0833 0883 moveto
  0831 0881 b
  0833 0878 b
  0836 0881 b
  0833 0883 b
 stroke
 0.0039 setgray
 stroke
00004 w
stroke
n
 0872 0929 moveto
  0867 0905 b
 stroke
 0.0039 setgray
 stroke
00004 w
stroke
n
 0867 0905 moveto
  0872 0909 b
  0879 0912 b
  0886 0912 b
  0893 0909 b
  0898 0905 b
  0901 0897 b
  0901 0893 b
  0898 0885 b
  0893 0881 b
  0886 0878 b
  0879 0878 b
  0872 0881 b
  0869 0883 b
  0867 0888 b
  0867 0890 b
  0869 0893 b
  0872 0890 b
  0869 0888 b
 stroke
 0.0039 setgray
 stroke
00004 w
stroke
n
 0886 0912 moveto
  0891 0909 b
  0896 0905 b
  0898 0897 b
  0898 0893 b
  0896 0885 b
  0891 0881 b
  0886 0878 b
 stroke
 0.0039 setgray
 stroke
00004 w
stroke
n
 0872 0929 moveto
  0896 0929 b
 stroke
 0.0039 setgray
 stroke
00004 w
stroke
n
 0872 0926 moveto
  0884 0926 b
  0896 0929 b
 stroke
 0.0039 setgray
 stroke
00004 w
stroke
n
 0903 0956 moveto
  0903 0987 b
 stroke
 0.0039 setgray
 stroke
00004 w
stroke
n
 0947 0987 moveto
  0947 0956 b
 stroke
 0.0039 setgray
 stroke
00004 w
stroke
n
 0991 0956 moveto
  0991 0987 b
 stroke
 0.0039 setgray
 stroke
00004 w
stroke
n
 1036 0987 moveto
  1036 0956 b
 stroke
 0.0039 setgray
 stroke
00004 w
stroke
n
 1020 0919 moveto
  1025 0921 b
  1032 0929 b
  1032 0878 b
 stroke
 0.0039 setgray
 stroke
00004 w
stroke
n
 1029 0926 moveto
  1029 0878 b
 stroke
 0.0039 setgray
 stroke
00004 w
stroke
n
 1020 0878 moveto
  1041 0878 b
 stroke
 0.0039 setgray
 stroke
00004 w
stroke
n
 1080 0883 moveto
  1077 0881 b
  1080 0878 b
  1082 0881 b
  1080 0883 b
 stroke
 0.0039 setgray
 stroke
00004 w
stroke
n
 1128 0929 moveto
  1121 0926 b
  1116 0919 b
  1113 0907 b
  1113 0900 b
  1116 0888 b
  1121 0881 b
  1128 0878 b
  1133 0878 b
  1140 0881 b
  1145 0888 b
  1147 0900 b
  1147 0907 b
  1145 0919 b
  1140 0926 b
  1133 0929 b
  1128 0929 b
 stroke
 0.0039 setgray
 stroke
00004 w
stroke
n
 1128 0929 moveto
  1123 0926 b
  1121 0924 b
  1118 0919 b
  1116 0907 b
  1116 0900 b
  1118 0888 b
  1121 0883 b
  1123 0881 b
  1128 0878 b
 stroke
 0.0039 setgray
 stroke
00004 w
stroke
n
 1133 0878 moveto
  1137 0881 b
  1140 0883 b
  1142 0888 b
  1145 0900 b
  1145 0907 b
  1142 0919 b
  1140 0924 b
  1137 0926 b
  1133 0929 b
 stroke
 0.0039 setgray
 stroke
00004 w
stroke
n
 1080 0956 moveto
  1080 1003 b
 stroke
 0.0039 setgray
 stroke
00004 w
stroke
n
 1124 0987 moveto
  1124 0956 b
 stroke
 0.0039 setgray
 stroke
00004 w
stroke
n
 1168 0956 moveto
  1168 0987 b
 stroke
 0.0039 setgray
 stroke
00004 w
stroke
n
 1213 0987 moveto
  1213 0956 b
 stroke
 0.0039 setgray
 stroke
00004 w
stroke
n
 1257 0956 moveto
  1257 0987 b
 stroke
 0.0039 setgray
 stroke
00004 w
stroke
n
 1301 1003 moveto
  1301 0956 b
 stroke
 0.0039 setgray
 stroke
00004 w
stroke
n
 1241 0919 moveto
  1246 0921 b
  1253 0929 b
  1253 0878 b
 stroke
 0.0039 setgray
 stroke
00004 w
stroke
n
 1251 0926 moveto
  1251 0878 b
 stroke
 0.0039 setgray
 stroke
00004 w
stroke
n
 1241 0878 moveto
  1263 0878 b
 stroke
 0.0039 setgray
 stroke
00004 w
stroke
n
 1301 0883 moveto
  1299 0881 b
  1301 0878 b
  1303 0881 b
  1301 0883 b
 stroke
 0.0039 setgray
 stroke
00004 w
stroke
n
 1339 0929 moveto
  1335 0905 b
 stroke
 0.0039 setgray
 stroke
00004 w
stroke
n
 1335 0905 moveto
  1339 0909 b
  1347 0912 b
  1354 0912 b
  1361 0909 b
  1366 0905 b
  1368 0897 b
  1368 0893 b
  1366 0885 b
  1361 0881 b
  1354 0878 b
  1347 0878 b
  1339 0881 b
  1337 0883 b
  1335 0888 b
  1335 0890 b
  1337 0893 b
  1339 0890 b
  1337 0888 b
 stroke
 0.0039 setgray
 stroke
00004 w
stroke
n
 1354 0912 moveto
  1359 0909 b
  1363 0905 b
  1366 0897 b
  1366 0893 b
  1363 0885 b
  1359 0881 b
  1354 0878 b
 stroke
 0.0039 setgray
 stroke
00004 w
stroke
n
 1339 0929 moveto
  1363 0929 b
 stroke
 0.0039 setgray
 stroke
00004 w
stroke
n
 1339 0926 moveto
  1351 0926 b
  1363 0929 b
 stroke
 0.0039 setgray
 stroke
00004 w
stroke
n
 1345 0956 moveto
  1345 0987 b
 stroke
 0.0039 setgray
 stroke
00004 w
stroke
n
 1390 0987 moveto
  1390 0956 b
 stroke
 0.0039 setgray
 stroke
00004 w
stroke
n
 1434 0956 moveto
  1434 0987 b
 stroke
 0.0039 setgray
 stroke
00004 w
stroke
n
 1478 0987 moveto
  1478 0956 b
 stroke
 0.0039 setgray
 stroke
00004 w
stroke
n
 1457 0919 moveto
  1460 0917 b
  1457 0914 b
  1455 0917 b
  1455 0919 b
  1457 0924 b
  1460 0926 b
  1467 0929 b
  1477 0929 b
  1484 0926 b
  1486 0924 b
  1489 0919 b
  1489 0914 b
  1486 0909 b
  1479 0905 b
  1467 0900 b
  1462 0897 b
  1457 0893 b
  1455 0885 b
  1455 0878 b
 stroke
 0.0039 setgray
 stroke
00004 w
stroke
n
 1477 0929 moveto
  1481 0926 b
  1484 0924 b
  1486 0919 b
  1486 0914 b
  1484 0909 b
  1477 0905 b
  1467 0900 b
 stroke
 0.0039 setgray
 stroke
00004 w
stroke
n
 1455 0883 moveto
  1457 0885 b
  1462 0885 b
  1474 0881 b
  1481 0881 b
  1486 0883 b
  1489 0885 b
 stroke
 0.0039 setgray
 stroke
00004 w
stroke
n
 1462 0885 moveto
  1474 0878 b
  1484 0878 b
  1486 0881 b
  1489 0885 b
  1489 0890 b
 stroke
 0.0039 setgray
 stroke
00004 w
stroke
n
 1522 0883 moveto
  1520 0881 b
  1522 0878 b
  1525 0881 b
  1522 0883 b
 stroke
 0.0039 setgray
 stroke
00004 w
stroke
n
 1570 0929 moveto
  1563 0926 b
  1558 0919 b
  1556 0907 b
  1556 0900 b
  1558 0888 b
  1563 0881 b
  1570 0878 b
  1575 0878 b
  1582 0881 b
  1587 0888 b
  1589 0900 b
  1589 0907 b
  1587 0919 b
  1582 0926 b
  1575 0929 b
  1570 0929 b
 stroke
 0.0039 setgray
 stroke
00004 w
stroke
n
 1570 0929 moveto
  1566 0926 b
  1563 0924 b
  1561 0919 b
  1558 0907 b
  1558 0900 b
  1561 0888 b
  1563 0883 b
  1566 0881 b
  1570 0878 b
 stroke
 0.0039 setgray
 stroke
00004 w
stroke
n
 1575 0878 moveto
  1580 0881 b
  1582 0883 b
  1585 0888 b
  1587 0900 b
  1587 0907 b
  1585 0919 b
  1582 0924 b
  1580 0926 b
  1575 0929 b
 stroke
 0.0039 setgray
 stroke
00004 w
stroke
n
 1522 0956 moveto
  1522 1003 b
 stroke
 0.0039 setgray
 stroke
00004 w
stroke
n
 1567 0987 moveto
  1567 0956 b
 stroke
 0.0039 setgray
 stroke
00004 w
stroke
n
 1611 0956 moveto
  1611 0987 b
 stroke
 0.0039 setgray
 stroke
00004 w
stroke
n
 1655 0987 moveto
  1655 0956 b
 stroke
 0.0039 setgray
 stroke
00004 w
stroke
n
 1699 0956 moveto
  1699 0987 b
 stroke
 0.0039 setgray
 stroke
00004 w
stroke
n
 1743 1003 moveto
  1743 0956 b
 stroke
 0.0039 setgray
 stroke
00004 w
stroke
n
 1679 0919 moveto
  1681 0917 b
  1679 0914 b
  1676 0917 b
  1676 0919 b
  1679 0924 b
  1681 0926 b
  1688 0929 b
  1698 0929 b
  1705 0926 b
  1707 0924 b
  1710 0919 b
  1710 0914 b
  1707 0909 b
  1700 0905 b
  1688 0900 b
  1683 0897 b
  1679 0893 b
  1676 0885 b
  1676 0878 b
 stroke
 0.0039 setgray
 stroke
00004 w
stroke
n
 1698 0929 moveto
  1703 0926 b
  1705 0924 b
  1707 0919 b
  1707 0914 b
  1705 0909 b
  1698 0905 b
  1688 0900 b
 stroke
 0.0039 setgray
 stroke
00004 w
stroke
n
 1676 0883 moveto
  1679 0885 b
  1683 0885 b
  1695 0881 b
  1703 0881 b
  1707 0883 b
  1710 0885 b
 stroke
 0.0039 setgray
 stroke
00004 w
stroke
n
 1683 0885 moveto
  1695 0878 b
  1705 0878 b
 stroke
 0.0039 setgray
 stroke
00004 w
stroke
n
 1705 0878 moveto
  1707 0881 b
  1710 0885 b
  1710 0890 b
 stroke
 0.0039 setgray
 stroke
00004 w
stroke
n
 1743 0883 moveto
  1741 0881 b
  1743 0878 b
  1746 0881 b
  1743 0883 b
 stroke
 0.0039 setgray
 stroke
00004 w
stroke
n
 1782 0929 moveto
  1777 0905 b
 stroke
 0.0039 setgray
 stroke
00004 w
stroke
n
 1777 0905 moveto
  1782 0909 b
  1789 0912 b
  1796 0912 b
  1803 0909 b
  1808 0905 b
  1811 0897 b
  1811 0893 b
  1808 0885 b
  1803 0881 b
  1796 0878 b
  1789 0878 b
  1782 0881 b
  1780 0883 b
  1777 0888 b
  1777 0890 b
  1780 0893 b
  1782 0890 b
  1780 0888 b
 stroke
 0.0039 setgray
 stroke
00004 w
stroke
n
 1796 0912 moveto
  1801 0909 b
  1806 0905 b
  1808 0897 b
  1808 0893 b
  1806 0885 b
  1801 0881 b
  1796 0878 b
 stroke
 0.0039 setgray
 stroke
00004 w
stroke
n
 1782 0929 moveto
  1806 0929 b
 stroke
 0.0039 setgray
 stroke
00004 w
stroke
n
 1782 0926 moveto
  1794 0926 b
  1806 0929 b
 stroke
 0.0039 setgray
 stroke
00004 w
stroke
n
 1788 0956 moveto
  1788 0987 b
 stroke
 0.0039 setgray
 stroke
00004 w
stroke
n
 1832 0987 moveto
  1832 0956 b
 stroke
 0.0039 setgray
 stroke
00004 w
stroke
n
 1876 0956 moveto
  1876 0987 b
 stroke
 0.0039 setgray
 stroke
00004 w
stroke
n
 1920 0987 moveto
  1920 0956 b
 stroke
 0.0039 setgray
 stroke
00004 w
stroke
n
 1900 0921 moveto
  1902 0919 b
  1900 0917 b
  1897 0919 b
  1897 0921 b
  1902 0926 b
  1909 0929 b
  1919 0929 b
  1926 0926 b
  1929 0921 b
  1929 0914 b
  1926 0909 b
  1919 0907 b
  1912 0907 b
 stroke
 0.0039 setgray
 stroke
00004 w
stroke
n
 1919 0929 moveto
  1924 0926 b
  1926 0921 b
  1926 0914 b
  1924 0909 b
  1919 0907 b
 stroke
 0.0039 setgray
 stroke
00004 w
stroke
n
 1919 0907 moveto
  1924 0905 b
  1929 0900 b
  1931 0895 b
  1931 0888 b
  1929 0883 b
  1926 0881 b
  1919 0878 b
  1909 0878 b
  1902 0881 b
  1900 0883 b
  1897 0888 b
  1897 0890 b
  1900 0893 b
  1902 0890 b
  1900 0888 b
 stroke
 0.0039 setgray
 stroke
00004 w
stroke
n
 1926 0902 moveto
  1929 0895 b
  1929 0888 b
  1926 0883 b
  1924 0881 b
  1919 0878 b
 stroke
 0.0039 setgray
 stroke
00004 w
stroke
n
 1965 0883 moveto
  1962 0881 b
  1965 0878 b
  1967 0881 b
  1965 0883 b
 stroke
 0.0039 setgray
 stroke
00004 w
stroke
n
 2013 0929 moveto
  2005 0926 b
  2001 0919 b
  1998 0907 b
  1998 0900 b
  2001 0888 b
  2005 0881 b
  2013 0878 b
  2017 0878 b
  2025 0881 b
  2030 0888 b
  2032 0900 b
  2032 0907 b
  2030 0919 b
  2025 0926 b
  2017 0929 b
  2013 0929 b
 stroke
 0.0039 setgray
 stroke
00004 w
stroke
n
 2013 0929 moveto
  2008 0926 b
  2005 0924 b
  2003 0919 b
  2001 0907 b
  2001 0900 b
  2003 0888 b
  2005 0883 b
  2008 0881 b
  2013 0878 b
 stroke
 0.0039 setgray
 stroke
00004 w
stroke
n
 2017 0878 moveto
  2022 0881 b
  2025 0883 b
  2027 0888 b
  2030 0900 b
  2030 0907 b
  2027 0919 b
  2025 0924 b
  2022 0926 b
  2017 0929 b
 stroke
 0.0039 setgray
 stroke
00004 w
stroke
n
 1965 0956 moveto
  1965 1003 b
 stroke
 0.0039 setgray
 stroke
00004 w
stroke
n
 2009 0987 moveto
  2009 0956 b
 stroke
 0.0039 setgray
 stroke
00004 w
stroke
n
 2053 0956 moveto
  2053 0987 b
 stroke
 0.0039 setgray
 stroke
00004 w
stroke
n
 2097 0987 moveto
  2097 0956 b
 stroke
 0.0039 setgray
 stroke
00004 w
stroke
n
 2142 0956 moveto
  2142 0987 b
 stroke
 0.0039 setgray
 stroke
00004 w
stroke
n
 2186 1003 moveto
  2186 0956 b
 stroke
 0.0039 setgray
 stroke
00004 w
stroke
n
 2121 0921 moveto
  2123 0919 b
  2121 0917 b
  2119 0919 b
  2119 0921 b
  2123 0926 b
  2131 0929 b
  2140 0929 b
  2147 0926 b
  2150 0921 b
  2150 0914 b
  2147 0909 b
  2140 0907 b
  2133 0907 b
 stroke
 0.0039 setgray
 stroke
00004 w
stroke
n
 2140 0929 moveto
  2145 0926 b
  2147 0921 b
  2147 0914 b
  2145 0909 b
  2140 0907 b
 stroke
 0.0039 setgray
 stroke
00004 w
stroke
n
 2140 0907 moveto
  2145 0905 b
  2150 0900 b
  2152 0895 b
  2152 0888 b
  2150 0883 b
  2147 0881 b
  2140 0878 b
  2131 0878 b
  2123 0881 b
  2121 0883 b
  2119 0888 b
  2119 0890 b
  2121 0893 b
  2123 0890 b
  2121 0888 b
 stroke
 0.0039 setgray
 stroke
00004 w
stroke
n
 2147 0902 moveto
  2150 0895 b
 stroke
 0.0039 setgray
 stroke
00004 w
stroke
n
 2150 0895 moveto
  2150 0888 b
  2147 0883 b
  2145 0881 b
  2140 0878 b
 stroke
 0.0039 setgray
 stroke
00004 w
stroke
n
 2186 0883 moveto
  2184 0881 b
  2186 0878 b
  2188 0881 b
  2186 0883 b
 stroke
 0.0039 setgray
 stroke
00004 w
stroke
n
 2224 0929 moveto
  2219 0905 b
 stroke
 0.0039 setgray
 stroke
00004 w
stroke
n
 2219 0905 moveto
  2224 0909 b
  2231 0912 b
  2239 0912 b
  2246 0909 b
  2251 0905 b
  2253 0897 b
  2253 0893 b
  2251 0885 b
  2246 0881 b
  2239 0878 b
  2231 0878 b
  2224 0881 b
  2222 0883 b
  2219 0888 b
  2219 0890 b
  2222 0893 b
  2224 0890 b
  2222 0888 b
 stroke
 0.0039 setgray
 stroke
00004 w
stroke
n
 2239 0912 moveto
  2243 0909 b
  2248 0905 b
  2251 0897 b
  2251 0893 b
  2248 0885 b
  2243 0881 b
  2239 0878 b
 stroke
 0.0039 setgray
 stroke
00004 w
stroke
n
 2224 0929 moveto
  2248 0929 b
 stroke
 0.0039 setgray
 stroke
00004 w
stroke
n
 2224 0926 moveto
  2236 0926 b
  2248 0929 b
 stroke
 0.0039 setgray
 stroke
00004 w
stroke
n
 2186 0956 moveto
  0637 0956 b
 stroke
 0.0039 setgray
 stroke
00004 w
stroke
n
 0637 2505 moveto
  0637 2458 b
 stroke
 0.0039 setgray
 stroke
00004 w
stroke
n
 0682 2474 moveto
  0682 2505 b
 stroke
 0.0039 setgray
 stroke
00004 w
stroke
n
 0726 2505 moveto
  0726 2474 b
 stroke
 0.0039 setgray
 stroke
00004 w
stroke
n
 0770 2474 moveto
  0770 2505 b
 stroke
 0.0039 setgray
 stroke
00004 w
stroke
n
 0814 2505 moveto
  0814 2474 b
 stroke
 0.0039 setgray
 stroke
00004 w
stroke
n
 0859 2458 moveto
  0859 2505 b
 stroke
 0.0039 setgray
 stroke
00004 w
stroke
n
 0903 2505 moveto
  0903 2474 b
 stroke
 0.0039 setgray
 stroke
00004 w
stroke
n
 0947 2474 moveto
  0947 2505 b
 stroke
 0.0039 setgray
 stroke
00004 w
stroke
n
 0991 2505 moveto
  0991 2474 b
 stroke
 0.0039 setgray
 stroke
00004 w
stroke
n
 1036 2474 moveto
  1036 2505 b
 stroke
 0.0039 setgray
 stroke
00004 w
stroke
n
 1080 2505 moveto
  1080 2458 b
 stroke
 0.0039 setgray
 stroke
00004 w
stroke
n
 1124 2474 moveto
  1124 2505 b
 stroke
 0.0039 setgray
 stroke
00004 w
stroke
n
 1168 2505 moveto
  1168 2474 b
 stroke
 0.0039 setgray
 stroke
00004 w
stroke
n
 1213 2474 moveto
  1213 2505 b
 stroke
 0.0039 setgray
 stroke
00004 w
stroke
n
 1257 2505 moveto
  1257 2474 b
 stroke
 0.0039 setgray
 stroke
00004 w
stroke
n
 1301 2458 moveto
  1301 2505 b
 stroke
 0.0039 setgray
 stroke
00004 w
stroke
n
 1345 2505 moveto
  1345 2474 b
 stroke
 0.0039 setgray
 stroke
00004 w
stroke
n
 1390 2474 moveto
  1390 2505 b
 stroke
 0.0039 setgray
 stroke
00004 w
stroke
n
 1434 2505 moveto
  1434 2474 b
 stroke
 0.0039 setgray
 stroke
00004 w
stroke
n
 1478 2474 moveto
  1478 2505 b
 stroke
 0.0039 setgray
 stroke
00004 w
stroke
n
 1522 2505 moveto
  1522 2458 b
 stroke
 0.0039 setgray
 stroke
00004 w
stroke
n
 1567 2474 moveto
  1567 2505 b
 stroke
 0.0039 setgray
 stroke
00004 w
stroke
n
 1611 2505 moveto
  1611 2474 b
 stroke
 0.0039 setgray
 stroke
00004 w
stroke
n
 1655 2474 moveto
  1655 2505 b
 stroke
 0.0039 setgray
 stroke
00004 w
stroke
n
 1699 2505 moveto
  1699 2474 b
 stroke
 0.0039 setgray
 stroke
00004 w
stroke
n
 1743 2458 moveto
  1743 2505 b
 stroke
 0.0039 setgray
 stroke
00004 w
stroke
n
 1788 2505 moveto
  1788 2474 b
 stroke
 0.0039 setgray
 stroke
00004 w
stroke
n
 1832 2474 moveto
  1832 2505 b
 stroke
 0.0039 setgray
 stroke
00004 w
stroke
n
 1876 2505 moveto
  1876 2474 b
 stroke
 0.0039 setgray
 stroke
00004 w
stroke
n
 1920 2474 moveto
  1920 2505 b
 stroke
 0.0039 setgray
 stroke
00004 w
stroke
n
 1965 2505 moveto
  1965 2458 b
 stroke
 0.0039 setgray
 stroke
00004 w
stroke
n
 2009 2474 moveto
  2009 2505 b
 stroke
 0.0039 setgray
 stroke
00004 w
stroke
n
 2053 2505 moveto
  2053 2474 b
 stroke
 0.0039 setgray
 stroke
00004 w
stroke
n
 2097 2474 moveto
  2097 2505 b
 stroke
 0.0039 setgray
 stroke
00004 w
stroke
n
 2142 2505 moveto
  2142 2474 b
 stroke
 0.0039 setgray
 stroke
00004 w
stroke
n
 2186 2458 moveto
  2186 2505 b
 stroke
 0.0039 setgray
 stroke
00004 w
stroke
n
 0637 2505 moveto
  2186 2505 b
 stroke
 0.0039 setgray
 stroke
00002 w
stroke
n
 1016 1116 moveto
  1012 1114 b
  1009 1112 b
  1009 1110 b
  1012 1107 b
  1018 1105 b
  1025 1105 b
 stroke
 0.0039 setgray
 stroke
00002 w
stroke
n
 1018 1105 moveto
  1009 1103 b
  1005 1100 b
  1003 1096 b
  1003 1091 b
  1007 1087 b
  1014 1085 b
  1021 1085 b
 stroke
 0.0039 setgray
 stroke
00002 w
stroke
n
 1018 1105 moveto
  1012 1103 b
  1007 1100 b
  1005 1096 b
  1005 1091 b
  1009 1087 b
  1014 1085 b
 stroke
 0.0039 setgray
 stroke
00002 w
stroke
n
 1014 1085 moveto
  1005 1082 b
  1000 1080 b
  0998 1076 b
  0998 1071 b
  1003 1067 b
  1014 1062 b
  1016 1060 b
  1016 1055 b
  1012 1053 b
  1007 1053 b
 stroke
 0.0039 setgray
 stroke
00002 w
stroke
n
 1014 1085 moveto
  1007 1082 b
  1003 1080 b
  1000 1076 b
  1000 1071 b
  1005 1067 b
  1014 1062 b
 stroke
 0.0039 setgray
 stroke
00002 w
stroke
n
 1468 1735 moveto
  1465 1733 b
  1468 1731 b
  1470 1733 b
  1470 1735 b
  1465 1740 b
  1461 1742 b
  1454 1742 b
  1447 1740 b
  1443 1735 b
  1441 1728 b
  1441 1724 b
  1443 1717 b
  1447 1713 b
  1454 1710 b
  1458 1710 b
  1465 1713 b
  1470 1717 b
 stroke
 0.0039 setgray
 stroke
00002 w
stroke
n
 1454 1742 moveto
  1449 1740 b
  1445 1735 b
  1443 1728 b
  1443 1724 b
  1445 1717 b
  1449 1713 b
  1454 1710 b
 stroke
 0.0039 setgray
 stroke
00004 w
stroke
n
 0641 0957 moveto
  0641 0957 b
  0642 0957 b
  0642 0957 b
  0642 0957 b
  0642 0957 b
  0643 0957 b
  0643 0958 b
  0643 0958 b
  0643 0959 b
  0644 0959 b
  0644 0960 b
  0644 0961 b
  0645 0962 b
  0645 0963 b
  0645 0964 b
  0646 0965 b
  0646 0967 b
  0647 0969 b
  0647 0970 b
  0648 0972 b
  0648 0974 b
  0649 0977 b
  0650 0979 b
  0650 0981 b
  0651 0983 b
  0652 0986 b
  0652 0988 b
  0653 0990 b
  0654 0993 b
  0655 0995 b
  0656 0998 b
  0657 1000 b
  0658 1003 b
  0659 1005 b
  0660 1008 b
  0661 1011 b
  0662 1014 b
  0664 1016 b
  0665 1020 b
  0666 1023 b
  0668 1026 b
  0670 1030 b
  0671 1033 b
  0673 1037 b
  0675 1041 b
  0677 1046 b
  0679 1050 b
  0681 1055 b
  0683 1060 b
 stroke
 0.0039 setgray
 stroke
00004 w
stroke
n
 0683 1060 moveto
  0686 1066 b
  0688 1072 b
  0691 1078 b
  0694 1084 b
  0697 1091 b
  0700 1098 b
  0703 1106 b
  0707 1115 b
  0710 1123 b
  0714 1133 b
  0718 1143 b
  0723 1154 b
  0727 1165 b
  0732 1177 b
  0737 1190 b
  0742 1204 b
  0747 1219 b
  0753 1235 b
  0759 1252 b
  0766 1271 b
  0773 1292 b
  0780 1314 b
  0787 1338 b
  0795 1365 b
  0803 1394 b
  0812 1425 b
  0821 1460 b
  0831 1498 b
  0841 1539 b
  0852 1584 b
  0863 1632 b
  0875 1684 b
  0887 1738 b
  0901 1795 b
  0914 1855 b
  0929 1915 b
  0944 1975 b
  0960 2035 b
  0977 2092 b
  0995 2145 b
  1014 2193 b
  1034 2235 b
  1055 2268 b
  1077 2292 b
  1100 2306 b
  1124 2310 b
  1150 2302 b
  1177 2283 b
  1205 2254 b
 stroke
 0.0039 setgray
 stroke
00004 w
stroke
n
 1205 2254 moveto
  1235 2215 b
  1267 2168 b
  1300 2112 b
  1335 2053 b
  1371 1987 b
  1410 1921 b
  1451 1852 b
  1493 1784 b
  1538 1717 b
  1586 1652 b
  1636 1590 b
  1688 1532 b
  1744 1477 b
  1802 1426 b
  1863 1378 b
  1928 1336 b
  1996 1296 b
  2067 1260 b
  2142 1228 b
  2186 1212 b
 stroke
 0.0039 setgray
 stroke
00004 w
stroke
n
 0664 2505 moveto
  0665 2388 b
  0666 2238 b
  0668 2108 b
  0670 1993 b
  0671 1892 b
  0673 1802 b
  0675 1723 b
  0677 1652 b
  0679 1601 b
  0681 1562 b
  0683 1524 b
  0686 1489 b
  0688 1456 b
  0691 1425 b
  0694 1396 b
  0697 1369 b
  0700 1344 b
  0703 1320 b
  0707 1298 b
  0710 1277 b
  0714 1257 b
  0718 1239 b
  0723 1222 b
  0727 1206 b
  0732 1191 b
  0737 1177 b
  0742 1164 b
  0747 1151 b
  0753 1140 b
  0759 1129 b
  0766 1119 b
  0773 1109 b
  0780 1100 b
  0787 1092 b
  0795 1084 b
  0803 1076 b
  0812 1069 b
  0821 1063 b
  0831 1057 b
  0841 1051 b
  0852 1045 b
  0863 1040 b
  0875 1035 b
  0887 1031 b
  0901 1026 b
  0914 1022 b
  0929 1018 b
  0944 1015 b
  0960 1011 b
 stroke
 0.0039 setgray
 stroke
00004 w
stroke
n
 0960 1011 moveto
  0977 1008 b
  0995 1005 b
  1014 1002 b
  1034 0999 b
  1055 0997 b
  1077 0995 b
  1100 0992 b
  1124 0990 b
  1150 0989 b
  1177 0987 b
  1205 0985 b
  1235 0984 b
  1267 0982 b
  1300 0981 b
  1335 0980 b
  1371 0979 b
  1410 0978 b
  1451 0977 b
  1493 0976 b
  1538 0975 b
  1586 0974 b
  1636 0973 b
  1688 0973 b
  1744 0972 b
  1802 0972 b
  1863 0971 b
  1928 0971 b
  1996 0970 b
  2067 0970 b
  2142 0969 b
  2186 0969 b
 stroke
 showpage
n
.25 .25 scale
1 setlinejoin
 restore